\documentclass[12pt]{article}
\usepackage{graphicx}
\usepackage{latexsym,amsmath,amsfonts,amssymb}
\newcommand{\be}{\begin{equation}}
\newcommand{\ee}{\end{equation}}
\newcommand{\beq}{\begin{equation}}
\newcommand{\eeq}{\end{equation}}
\newcommand{\bea}{\begin{eqnarray}}
\newcommand{\eea}{\end{eqnarray}}

\newcommand{\ba}{\begin{eqnarray}}
\newcommand{\ea}{\end{eqnarray}}
\begin{document}
\baselineskip=15.5pt
\pagestyle{plain}
\setcounter{page}{1}


\def\del{{\partial}}
\def\vev#1{\left\langle #1 \right\rangle}
\def\cn{{\cal N}}
\def\co{{\cal O}}
\def\IC{{\mathbb C}}
\def\IR{{\mathbb R}}
\def\IZ{{\mathbb Z}}
\def\RP{{\bf RP}}
\def\CP{{\bf CP}}
\def\Poincare{{Poincar\'e }}
\def\tr{{\rm tr}}
\def\tp{{\tilde \Phi}}

\newcommand{\bA}{{\bf A }}
\newcommand{\bN}{{\bf N }}

\def\TL{\hfil$\displaystyle{##}$}
\def\TR{$\displaystyle{{}##}$\hfil}
\def\TC{\hfil$\displaystyle{##}$\hfil}
\def\TT{\hbox{##}}
\def\HLINE{\noalign{\vskip1\jot}\hline\noalign{\vskip1\jot}}
\def\seqalign#1#2{\vcenter{\openup1\jot
  \halign{\strut #1\cr #2 \cr}}}
\def\lbldef#1#2{\expandafter\gdef\csname #1\endcsname {#2}}
\def\eqn#1#2{\lbldef{#1}{(\ref{#1})}%
\begin{equation} #2 \label{#1} \end{equation}}
\def\eqalign#1{\vcenter{\openup1\jot
    \halign{\strut\span\TL & \span\TR\cr #1 \cr
   }}}
\def\eno#1{(\ref{#1})}
\def\href#1#2{#2}
\def\half{{1 \over 2}}



\def\NO{\nonumber}

\def\bea{\begin{eqnarray}}
\def\eea{\end{eqnarray}}

\def\beqx{\begin{displaymath}}
\def\eeqx{\end{displaymath}}

\newcommand{\bmat}{\left(\begin{array}}
\newcommand{\emat}{\end{array}\right)}

\def\half{\frac{1}{2}}


\newtheorem{definition}{Definition}[section]
\newtheorem{theorem}{Theorem}[section]
\newtheorem{lemma}{Lemma}[section]
\newtheorem{corollary}{Corollary}[section]


\def\a{\alpha}
\def\b{\beta}
\def\d{\delta}
\def\e{\epsilon}
\def\f{\phi}
\def\g{\gamma}
\def\h{\eta}
\def\j{\psi}
\def\k{\kappa}
\def\l{\lambda}
\def\m{\mu}
\def\n{\nu}
\def\o{\omega}
    \def\om{\omega}
\def\p{\pi}
\def\q{\theta}
    \def\th{\theta}
\def\r{\rho}
\def\s{\sigma}
\def\t{\tau}
\def\x{\xi}
\def\z{\zeta}
\def\D{\Delta}
\def\F{\Phi}
\def\G{\Gamma}
\def\J{\Psi}
\def\L{\Lambda}
\def\O{\Omega}
    \def\Om{\Omega}
\def\P{\Pi}
\def\Q{\Theta}
    \def\Th{\Theta}
\def\S{\Sigma}
\def\U{\Upsilon}
\def\X{\Xi}


\def\ve{\varepsilon}
\def\vr{\varrho}
\def\vs{\varsigma}
\def\vq{\vartheta}
    \def\vth{\vartheta}
\def\tvf{\tilde{\varphi}}
\def\vf{\varphi}
    \def\vphi{\varphi}


\def\ca{{\cal A}}
\def\cb{{\cal B}}
\def\cc{{\cal C}}
\def\cd{{\cal D}}
\def\ce{{\cal E}}
\def\cf{{\cal F}}
\def\cg{{\cal G}}
\def\ch{{\cal H}}
\def\ci{{\cal I}}
\def\cj{{\cal J}}
\def\ck{{\cal K}}
\def\cl{{\cal L}}
\def\cm{{\cal M}}
\def\cn{{\cal N}}
\def\co{{\cal O}}
\def\cp{{\cal P}}
\def\cq{{\cal Q}}
\def\car{{\cal R}}
\def\cs{{\cal S}}
\def\ct{{\cal T}}
\def\cu{{\cal U}}
\def\cv{{\cal V}}
\def\cw{{\cal W}}
\def\cx{{\cal X}}
\def\cy{{\cal Y}}
\def\cz{{\cal Z}}



\def\Sc#1{{\hbox{\sc #1}}}      
\def\Sf#1{{\hbox{\sf #1}}}      
\def\mb#1{\mbox{\boldmath $#1$}}


\def\slpa{\slash{\pa}}                         
\def\slin{\SLLash{\in}}                                 
\def\bo{{\raise-.3ex\hbox{\large$\Box$}}}               
\def\cbo{\Sc [}                                         
\def\pa{\partial}                                       
\def\de{\nabla}                                         
\def\dell{\nabla}                                       
\def\su{\sum}                                           
\def\pr{\prod}                                          
\def\iff{\leftrightarrow}                               
\def\conj{{\hbox{\large *}}}                            
\def\ltap{\raisebox{-.4ex}{\rlap{$\sim$}} \raisebox{.4ex}{$<$}}   
\def\gtap{\raisebox{-.4ex}{\rlap{$\sim$}} \raisebox{.4ex}{$>$}}   
\def\face{{\raise.2ex\hbox{$\displaystyle \bigodot$}\mskip-2.2mu \llap {$\ddot
        \smile$}}}                                   
\def\dg{\dagger}                                     
\def\ddg{\ddagger}                                   
\def\trans{\mbox{\scri T}}                           
\def\>{\rangle}                                      
\def\<{\langle}                                      


\def\sp#1{{}^{#1}}                                   
\def\sb#1{{}_{#1}}                                   
\newcommand{\sub}[1]{\phantom{}_{(#1)}\phantom{}}    
\newcommand{\supt}[1]{\phantom{}^{(#1)}\phantom{}}    
\def\oldsl#1{\rlap/#1}                               
\def\slash#1{\rlap{\hbox{$\mskip 1 mu /$}}#1}        
\def\Slash#1{\rlap{\hbox{$\mskip 3 mu /$}}#1}        
\def\SLash#1{\rlap{\hbox{$\mskip 4.5 mu /$}}#1}      
\def\SLLash#1{\rlap{\hbox{$\mskip 6 mu /$}}#1}       
\def\wt#1{\widetilde{#1}}                            
\def\Hat#1{\widehat{#1}}                             
\def\lbar#1{\ensuremath{\overline{#1}}}              
\def\VEV#1{\left\langle #1\right\rangle}             
\def\abs#1{\left| #1\right|}                         
\def\leftrightarrowfill{$\mathsurround=0pt \mathord\leftarrow \mkern-6mu
        \cleaders\hbox{$\mkern-2mu \mathord- \mkern-2mu$}\hfill
        \mkern-6mu \mathord\rightarrow$}        
\def\dvec#1{\vbox{\ialign{##\crcr
        \leftrightarrowfill\crcr\noalign{\kern-1pt\nointerlineskip}
        $\hfil\displaystyle{#1}\hfil$\crcr}}}           
\def\dt#1{{\buildrel {\hbox{\LARGE .}} \over {#1}}}     
\def\dtt#1{{\buildrel \bullet \over {#1}}}              
\def\der#1{{\pa \over \pa {#1}}}                        
\def\fder#1{{\d \over \d {#1}}}                         
\def\tr{{\rm tr \,}}                                    
\def\Tr{{\rm Tr \,}}                                    
\def\diag{{\rm diag \,}}                                
\def\preal{{\rm Re\,}}                                  
\def\pim{{\rm Im\,}}                                    


\def\partder#1#2{{\partial #1\over\partial #2}}        
\def\parvar#1#2{{\d #1\over \d #2}}                    
\def\secder#1#2#3{{\partial^2 #1\over\partial #2 \partial #3}}  
\def\on#1#2{\mathop{\null#2}\limits^{#1}}              
\def\bvec#1{\on\leftarrow{#1}}                         
\def\oover#1{\on\circ{#1}}                             


\def\Deq#1{\mbox{$D$=#1}}                               
\def\Neq#1{\mbox{$cn$=#1}}                              
\newcommand{\ampl}[2]{{\cal M}\left( #1 \to #2 \right)} 


\def\NPB#1#2#3{Nucl. Phys. B {\bf #1} (19#2) #3}
\def\PLB#1#2#3{Phys. Lett. B {\bf #1} (19#2) #3}
\def\PLBold#1#2#3{Phys. Lett. {\bf #1}B (19#2) #3}
\def\PRD#1#2#3{Phys. Rev. D {\bf #1} (19#2) #3}
\def\PRL#1#2#3{Phys. Rev. Lett. {\bf #1} (19#2) #3}
\def\PRT#1#2#3{Phys. Rep. {\bf #1} C (19#2) #3}
\def\MODA#1#2#3{Mod. Phys. Lett.  {\bf #1} (19#2) #3}


\def\norder{\raisebox{-.13cm}{\ensuremath{\circ}}\hspace{-.174cm}\raisebox{.13cm}{\ensuremath{\circ}}}
\def\bz{\bar{z}}
\def\bw{\bar{w}}
\def\-{\hphantom{-}}
\newcommand{\dd}{\mbox{d}}
\newcommand{\scr}{\scriptscriptstyle}
\newcommand{\scri}{\scriptsize}
\def\rand#1{\marginpar{\tiny #1}}               
\newcommand{\rstar}{\rand{\bf\large *}}
\newcommand{\rup}{\rand{$\uparrow$}}
\newcommand{\rdown}{\rand{$\downarrow$}}


\def\ads{{\it AdS}}
\def\adsp{{\it AdS}$_{p+2}$}
\def\cft{{\it CFT}}

\newcommand{\ber}{\begin{eqnarray}}
\newcommand{\eer}{\end{eqnarray}}

\newcommand{\beqar}{\begin{eqnarray}}
\newcommand{\cN}{{\cal N}}
\newcommand{\cO}{{\cal O}}
\newcommand{\cA}{{\cal A}}
\newcommand{\cT}{{\cal T}}
\newcommand{\cF}{{\cal F}}
\newcommand{\cC}{{\cal C}}
\newcommand{\cR}{{\cal R}}
\newcommand{\cW}{{\cal W}}
\newcommand{\eeqar}{\end{eqnarray}}
\newcommand{\tht}{\thteta}
\newcommand{\lm}{\lambda}\newcommand{\Lm}{\Lambda}
\newcommand{\eps}{\epsilon}


\newcommand{\nonu}{\nonumber}
\newcommand{\oh}{\displaystyle{\frac{1}{2}}}
\newcommand{\dsl}
  {\kern.06em\hbox{\raise.15ex\hbox{$/$}\kern-.56em\hbox{$\partial$}}}
\newcommand{\id}{i\!\!\not\!\partial}
\newcommand{\as}{\not\!\! A}
\newcommand{\ps}{\not\! p}
\newcommand{\ks}{\not\! k}
\newcommand{\dv}{d^2x}
\newcommand{\Z}{{\cal Z}}
\newcommand{\N}{{\cal N}}
\newcommand{\Dsl}{\not\!\! D}
\newcommand{\Bsl}{\not\!\! B}
\newcommand{\Psl}{\not\!\! P}
\newcommand{\eeqarr}{\end{eqnarray}}
\newcommand{\ZZ}{{\rm \kern 0.275em Z \kern -0.92em Z}\;}

                                                                                                    
\def\del{{\delta^{\hbox{\sevenrm B}}}} \def\ex{{\hbox{\rm e}}}
\def\azb{A_{\bar z}} \def\az{A_z} \def\bzb{B_{\bar z}} \def\bz{B_z}
\def\czb{C_{\bar z}} \def\cz{C_z} \def\dzb{D_{\bar z}} \def\dz{D_z}
\def\im{{\hbox{\rm Im}}} \def\mod{{\hbox{\rm mod}}} \def\tr{{\hbox{\rm Tr}}}
\def\ch{{\hbox{\rm ch}}} \def\imp{{\hbox{\sevenrm Im}}}
\def\trp{{\hbox{\sevenrm Tr}}} \def\vol{{\hbox{\rm Vol}}}
\def\rl{\Lambda_{\hbox{\sevenrm R}}} \def\wl{\Lambda_{\hbox{\sevenrm W}}}
\def\fc{{\cal F}_{k+\cox}} \def\vev{vacuum expectation value}
\def\nodiv{\mid{\hbox{\hskip-7.8pt/}}}
\def\ie{{\em i.e.}}
\def\ie{\hbox{\it i.e.}}

\def\CC{{\mathchoice
{\rm C\mkern-8mu\vrule height1.45ex depth-.05ex
width.05em\mkern9mu\kern-.05em}
{\rm C\mkern-8mu\vrule height1.45ex depth-.05ex
width.05em\mkern9mu\kern-.05em}
{\rm C\mkern-8mu\vrule height1ex depth-.07ex
width.035em\mkern9mu\kern-.035em}
{\rm C\mkern-8mu\vrule height.65ex depth-.1ex
width.025em\mkern8mu\kern-.025em}}}
                                                                                                    
\def\RR{{\rm I\kern-1.6pt {\rm R}}}
\def\NN{{\rm I\!N}}
\def\ZZ{{\rm Z}\kern-3.8pt {\rm Z} \kern2pt}
\def\IB{\relax{\rm I\kern-.18em B}}
\def\ID{\relax{\rm I\kern-.18em D}}
\def\II{\relax{\rm I\kern-.18em I}}
\def\IP{\relax{\rm I\kern-.18em P}}
\newcommand{\CS}{{\scriptstyle {\rm CS}}}
\newcommand{\CSs}{{\scriptscriptstyle {\rm CS}}}
\newcommand{\rc}{\nonumber\\}
\newcommand{\bear}{\begin{eqnarray}}
\newcommand{\eear}{\end{eqnarray}}
\newcommand{\W}{{\cal W}}
\newcommand{\LL}{{\cal L}}
                                                                                                    
\def\mani{{\cal M}}
\def\calo{{\cal O}}
\def\calb{{\cal B}}
\def\calw{{\cal W}}
\def\calz{{\cal Z}}
\def\cald{{\cal D}}
\def\calc{{\cal C}}
\def\to{\rightarrow}
\def\ele{{\hbox{\sevenrm L}}}
\def\ere{{\hbox{\sevenrm R}}}
\def\zb{{\bar z}}
\def\wb{{\bar w}}
\def\nodiv{\mid{\hbox{\hskip-7.8pt/}}}
\def\menos{\hbox{\hskip-2.9pt}}
\def\dr{\dot R_}
\def\drr{\dot r_}
\def\ds{\dot s_}
\def\da{\dot A_}
\def\dga{\dot \gamma_}
\def\ga{\gamma_}
\def\dal{\dot\alpha_}
\def\al{\alpha_}
\def\cls{{closing}}
\def\vev{vacuum expectation value}
\def\tr{{\rm Tr}}
\def\to{\rightarrow}
\def\too{\longrightarrow}


\def\a{\alpha}
\def\b{\beta}
\def\d{\delta}
\def\e{\epsilon}           
\def\f{\phi}               
\def\vf{\varphi}  \def\tvf{\tilde{\varphi}}
\def\vp{\varphi}
\def\g{\gamma}
\def\h{\eta}
\def\j{\psi}
\def\k{\kappa}                    
\def\l{\lambda}
\def\m{\mu}
\def\n{\nu}
\def\o{\omega}  \def\w{\omega}
\def\q{\theta}  \def\th{\theta}                  
\def\r{\rho}                                     
\def\s{\sigma}                                   
\def\t{\tau}
\def\u{\upsilon}
\def\x{\xi}
\def\z{\zeta}
\def\pt{\tilde{\varphi}}
\def\tt{\tilde{\theta}}
\def\lab{\label}  
\def\6{\partial}
\def\wg{\wedge}
\def\atanh{{\rm arctanh}}
\def\bpsi{\bar{\psi}}
\def\bt{\bar{\theta}}
\def\bvf{\bar{\varphi}}

%
                                                                                                    
\newfont{\namefont}{cmr10}
\newfont{\addfont}{cmti7 scaled 1440}
\newfont{\boldmathfont}{cmbx10}
\newfont{\headfontb}{cmbx10 scaled 1728}
\renewcommand{\theequation}{{\rm\thesection.\arabic{equation}}}
\begin{titlepage}

\begin{center} \Large \bf Unquenched flavor in the gauge/gravity correspondence

\end{center}

\vskip 0.3truein
\begin{center}
Carlos 
N\'u\~nez${}^{\dagger}$ \footnote{c.nunez@swansea.ac.uk}, \'Angel 
Paredes${}^{\,\ddagger}$\footnote{aparedes@ffn.ub.es} and Alfonso V. 
Ramallo${}^{*}$\footnote{alfonso@fpaxp1.usc.es}
\vspace{0.2in}\\
\vskip 0.2truein
${}^{\dagger}$ \it{Department of Physics\\ University of Swansea, Singleton
Park\\
Swansea SA2 8PP, United Kingdom}\\
\vspace{0.2in}
\vskip 0.1truein

${}^{\,\ddagger}$\it{Departament de F\'\i sica Fonamental
and ICCUB Institut de
Ci\`encies del Cosmos, Universitat de Barcelona, Mart\'\i\ i Franqu\`es, 1,
E-08028, Barcelona, Spain
}
\vspace{0.2in}
\vskip 0.1truein
${}^{*}$\it{
Departamento de  F\'\i sica de Part\'\i culas, Universidade
de Santiago de
Compostela\\and\\Instituto Galego de F\'\i sica de Altas
Enerx\'\i as (IGFAE)\\E-15782, Santiago de Compostela, Spain
}

\vspace{0.3in}
\end{center}
\vskip 1cm
\centerline{{\bf Abstract}}
Within the AdS/CFT correspondence, we review the 
studies of field theories with a large number of adjoint and 
fundamental fields, in the Veneziano limit. 
We concentrate in set-ups where the fundamentals are introduced
by a smeared set of D-branes.
We make emphasis on the general ideas and then 
in subsequent chapters that can be read independently, describe 
particular considerations in various different models. 
Some new material is presented along the various sections.


\vskip1truecm
\vspace{0.1in}
\end{titlepage}
\setcounter{footnote}{0}
\tableofcontents
\section{Introduction, General Idea and Outline}
\label{sec: intro}

\subsection{Introduction and outline}

The AdS/CFT conjecture originally proposed by Maldacena \cite{Maldacena:1997re},
refined in \cite{Gubser:1998bc, Witten:1998qj} 
and reviewed in \cite{Aharony:1999ti} has been one of the most interesting
developments in theoretical physics of the last decades.
It has become one of the most 
powerful analytic tools to deal with strong coupling effects of some 
particular
gauge  theories in the planar limit $N_c \to \infty$. 
The most studied and best understood case corresponds to $SU(N_c)$
${\cal N}=4$ SYM
which is a highly supersymmetric conformal theory and which only contains
matter in the adjoint representation of the gauge group.
Certainly, there are many interesting field theories which do not share these 
properties and this fact has lead to an enormous amount of effort devoted to
extending the duality along  different paths. Consequently, people have constructed
gravity duals of non-supersymmetric, non-conformal gauge theories,
in different vacua and
with diverse
matter contents. 
One can mention the attempt of building a dual as close as possible to QCD as an
aim for these generalizations. However, one should keep in mind that this 
is just one
among many desirable motivations, since understanding gauge theories at strong coupling
(or using gauge theories to understand gravity) is a very relevant problem {\it per se},
with both theoretical and phenomenological possible implications.

An important development of AdS/CFT has been to generalise the matter content of the gauge theories
under consideration and, in particular, to include fields which transform in the
fundamental representation of the gauge group, as the QCD quarks 
do\footnote{With an abuse of language, we will use 
throughout this review
the words {\it quark} or {\it flavor}
to refer to any field, fermionic or bosonic, 
transforming in the fundamental representation of the gauge group. Accordingly,  by {\it mesons} we will mean bound states of {\it quarks}. }.
A first possibility is to add the flavors in the {\it quenched} approximation.
The word {\it quenched} comes from the lattice literature
and, in that context, it amounts to setting
the quark fermion determinant to one. In more physical terms, {\it quenching} corresponds
to discarding the quark dynamical effects. This means that quantum effects produced by 
the fundamentals are neglected, the quarks are considered as external non-dynamical objects
in the sense that they do not run in the loops\footnote{In the lattice, usually, quenching is
thought to be a good approximation for heavy quarks whereas for the gauge-gravity examples
the relevance of the quenched approximation comes from having parametrically less fundamental than
adjoint fields $N_f \ll N_c$.}. From the string side, adding quenched quarks to a given
gauge theory
corresponds to incorporating a set of brane probes in the dual background, which is not
modified with respect to the quark-less case. By analysing the
worldvolume physics of these flavor branes (typically using the Dirac-Born-Infeld + 
Wess-Zumino action) a lot of physically interesting questions can be understood.
For instance and just to mention a few, 
chiral symmetry breaking can be neatly described, phase diagrams can be
constructed and meson spectra can be exactly computed.
It is hard to do justice to the huge literature in the subject so let us just
mention the seminal papers \cite{Karch:2002sh}, \cite{Kruczenski:2003be} and a recent review 
\cite{Erdmenger:2007cm}.

Thus, it is fair to say that the study of quenched flavor within the gauge-gravity
correspondence
has been a very fruitful program. Nonetheless, there are physical features which 
are intimately related to the quantum effects of the quarks.
Examples are the consequences of the presence of fundamentals
on the running couplings, which may ultimately lead to conformal points, conformal windows
\cite{Banks:1981nn} or
Seiberg-like dualities \cite{Seiberg:1994pq}. More phenomenologically, multihadron production,
the screening of the color charge or the large mass of the $\eta'$ meson are spin-offs of
these quantum effects. Let us also mention that the most successful 
application of string duals towards phenomenology has been the construction of solutions that
can be used as toy models for the experimental quark-gluon plasma. 
Thus, a very interesting program is to build black hole solutions with unquenched flavor which
are really dual to {\it quark}-gluon plasmas, {\it i.e.} such that the effect of
the dynamical
quarks affects the plasma physics, as is expected to be the case in the real world.

These observations largely motivate the study 
of theories with unquenched quarks
from the string theory dual point of view.
Unquenching the flavors of the gauge theory has a very precise implication for the 
dual  theory: the gravity background has to be modified by the inclusion of the quarks,
namely, one needs to take into account the back-reaction on the geometry produced by the
flavor branes. The main goal in the following will be to present methods to compute such
back-reacted solutions. This will be done by presenting different examples
 that, hopefully, will help the reader to gain insight in both the physical questions and
 the technical tools used to address them.

In this review, we will focus on a specific family of unquenched constructions. 
Namely, we will discuss at length just 
solutions of type IIA or type IIB string theory in which the fundamentals come from a
{\it smeared} set of flavor branes. In section \ref{heuristiczzz}, we will try to provide
a general understanding of this notion of smearing the flavor and argue why we find it
a case of particular interest. 
As we will see, by considering the case of smeared D-branes we can build a systematic approach 
applicable to different situations and which typically results in large simplifications as compared
to other kind of flavor D-brane distributions.
This smearing procedure referring to flavors
was first introduced in \cite{Bigazzi:2005md} in a non-critical string framework and
in  \cite{Casero:2006pt}  in a well-controlled ten-dimensional context.

It is important to remark that this smearing is by no means the only possibility to introduce
unquenched fundamentals in gauge-gravity duals. Many important works have 
followed alternative
paths to construct different models. We are not able to review them here, but we provide a survey
of the literature in section \ref{sec:localized}.

\subsubsection*{Outline}

We will devote the rest of  section \ref{sec: intro} to 
further clarifying the kind of physical problems
we want to address and to give the general methods and notions which are common
to all the constructions we will present later.

Then, sections \ref{AdS5X5}-\ref{models2}
 will analyse different models that
are ordered in increasing order of complexity. Each section can be 
 read mostly independently
from the rest.
The discussion of each model can always be regarded as a two step process.
First, one has to solve the equations for finding 
the back-reacted solutions of type II supergravity
coupled to a set of D-brane sources. Second, one can use these solutions to extract
the physics of the conjectured gauge theory duals with unquenched flavors.
Readers interested in different aspects of the problem can consult the
different parts independently. We would like to stress that, even without making any
reference to the gauge-gravity correspondence, the string theory solutions and methods developed
to find them are interesting by themselves. 

Section \ref{AdS5X5} deals with the backreaction of D7-branes on $AdS_5 \times X^5$
spaces, where $X^5$ stands for a Sasaki-Einstein space. As a matter of fact,
a large part of the discussion can
be carried out without specifying the $X^5$. Notwithstanding, the two most interesting
cases correspond to $X^5=S^5$ and $X^5=T^{1,1}$. At different points during section
\ref{AdS5X5}, we will refer to these particular examples in order to explain concrete
features. We will present supersymmetric solutions for massless and massive quarks, and 
also non-supersymmetric black hole
solutions which are dual to theories at finite temperature,
in a deconfined plasma phase. All these solutions share the
property of having a singularity, associated with a UV Landau pole in the field theory
(when quarks are massless and the temperature is zero, there is also a naked IR singularity).
We will show how to make well-defined IR predictions from the geometry, even
in the presence of the UV singularity
(in much the same spirit as in field theory renormalization).

In section \ref{sec:D5D5}, we will discuss a model in which both the color and flavor
branes are D5's. It is dual to a (3+1)-dimensional ${\cal N}=1$ theory 
with a UV completion.
Among several nice features of the model that will be presented, we would like to 
remark here that it incorporates a geometrical description of a Seiberg-like duality.
Section \ref{sec:2+1} is also built from a D5-D5 intersection and in fact shares
several similarities with the previous model. The construction corresponds to color D5's
wrapping a compact 3-cycle and therefore the dual field theory is (2+1)-dimensional.

In section \ref{KS} we examine the addition of D7-branes to the
conspicuous Klebanov-Strassler model \cite{Klebanov:2000hb}. From the physical point of view, how
the unquenched flavors affect a duality cascade is particularly interesting.
From a technical point of view, the system is slightly more involved than the
previous ones because different RR and NSNS forms are turned on. 
However, despite this complication, it is remarkable that almost all functions of the ansatz
can be integrated in a closed form.

Section \ref{models2} reviews  a different class of models. The dual gauge theories
are built on wrapped supersymmetric D-branes 
with the peculiarity that some of the adjoint scalars
remain massless. As we will explain, it is not sensible to smear the branes in
all the transverse directions.
The associated main technical difficulty will be
the fact that one has to solve 
partial differential equations to find the background.

Profiting  from the experience gained by discussing these examples one by one,
in section \ref{mathviewpointzz}, we will give a more mathematical viewpoint
of the constructions. In particular, we will take some tools of differential
geometry to describe in a concise and compact way the distributions of mass and
charge due to the presence of the flavor branes.

Finally, in section \ref{outlook} we will conclude by summarizing the whole topic
trying to give a general perspective of the results obtained and by 
also providing an outlook of the subject.

\subsection{Presentation of the problem}\label{section1}

As anticipated in the introduction, we will discuss the addition of 
flavors to field theories (mostly focusing on SUSY examples, but this is 
not mandatory) using AdS/CFT or more generally, gauge-strings duality.

We hope it is clear to the reader that the addition of flavors (fields 
transforming in the fundamental representation) is a very interesting 
exercise from a  dynamical point of view. Indeed, in a theory with adjoint 
fields (let us for the sake of this discussion, consider the case of a 
confining field theory) the presence 
of fields transforming in the fundamental will produce the breaking of the 
``QCD-string'' or screening. Of course, the fundamentals will add a new 
symmetry, that can be $SU(N_f)$ or, like in massless QCD, $SU(N_f)\times 
SU(N_f)$;  a baryonic $U(1)_B$ symmetry should also appear. Obviously the 
presence of global symmetries (and their possible spontaneous or explicit 
breaking) 
will directly reflect in 
the spectrum.  Apart from this, it will happen that the states before the 
addition of fundamentals, that is the glueballs, will interact and mix 
with the mesons, giving place to new diagonal combinations that will 
be the observed states.
Moreover, anomalies will be modified, as fermions that transform in 
the fundamental will run in the triangles. Also gauge couplings will run 
differently and finally new dualities (Seiberg-like \cite{Seiberg:1994pq}) 
may appear.
In the rest of this article, we will discuss how all of the above 
mentioned features are encoded in string backgrounds.

It is clear that we need to add new objects to our string background. 
These new objects are D-branes, on which a gauge field propagates,  
encoding 
the 
presence of a $U(N_f)$ gauge symmetry (in the bulk), dual to the global 
$U(N_f)$ in the dual QFT. Also, it is on these D-branes that the meson 
fields, represented by excitations of the branes, propagate and interact.
Following a nomenclature that by now became standard, we will call these 
D-branes ``flavor branes''.

It is then clear that to add flavors to a field theory whose dual we know, 
we should consider the original (unflavored) string background and add 
flavor-branes. Now, the point is how to proceed technically to add these 
new branes.
It may be useful to consider  two developments of the 1970's, that will 
turn to illuminate on the answer to this question.

In the papers \cite{'t Hooft:1973jz} and \cite{Veneziano:1976wm}
't Hooft and Veneziano respectively considered the influence of 
fundamentals when the following scaling is taken
\beq
N_c \to \infty, \;\;\;\lambda=g_{YM}^2 N_c=fixed,
\eeq
and considered the two possible cases ('t Hooft and Veneziano 
respectively)
\bea
& & N_f=fixed,\;\;\; x=\frac{N_f}{N_c}\to 0,\nonumber\\
& & N_f \to \infty, \;\;\; x=\frac{N_f}{N_c}=fixed\,\,.\nonumber
\eea
It is very illuminating to see how different diagrams contributing to the 
same physical process (to fix ideas, an n-point correlator of mesonic 
currents) scale in these two cases.
In this respect, 
a formula for the kinematical factor  of  the scattering of $n$ mesons was
produced in \cite{Capella:1992yb},
considering diagrams with $w$ internal fermion loops (windows), $h$
non-planar handles and $b$ boundaries,
\beq
<B_1....B_n> \sim \Big(\,\frac{N_f}{N_c}\,\Big)^w \, N_c^{(2-\frac{n}{2} -2h -b)}.
\label{scatk}
\eeq
Consider the case of scattering of two mesons $n=2$. We see that diagrams
like the first one in the figure ($w=h=0, b=1,n=2$) scale like a constant
$N_c^0\sim 1$, the
second diagram (with $w=1,h=0,b=1,n=2$) scales like $\frac{N_f}{N_c}$,
while the third one (with $w=0, h=0,b=2,n=2$, that is non-planar) goes
like $N_c^{-1}$.

\begin{figure}[htb!]
\centering
\includegraphics[scale=0.5]{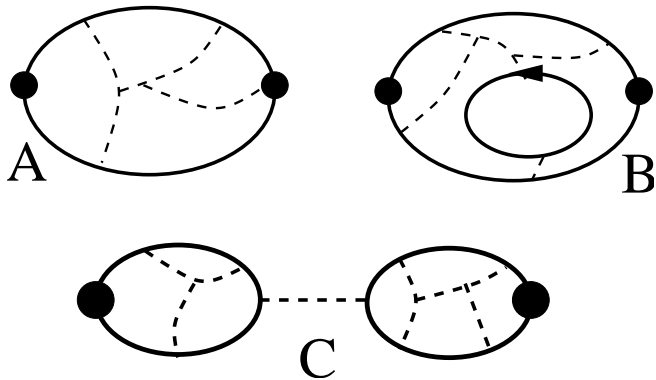}
\caption{\small Diagrams for a meson propagator, with two
insertions of the meson operator ($n=2$) shown as thick
points on the boundaries. The dashed lines are gluons
that fill the diagram in the large $N_c$ limit and the
thick lines are quarks. A) Planar diagram with no internal quark loops
($h=0$, $w=0$), the scaling is $\sim 1$.
B) Planar diagram with an internal quark loop $h=0$, $w=1$,
the scaling is $\sim N_f/N_c$.
C) Non-planar diagram with no internal quark loops $h=0$, $w=0$, $b=2$,
the scaling is $\sim 1/N_c $.}
\label{htb!}\end{figure}
So, we see that from this view point, the Veneziano scaling captures more 
physics, represented here by diagrams like $B$ in Fig. \ref{htb!}.
Nevertheless, there may be some particular problems for which studying 
things in the 't Hooft scaling may be enough.

From the view point of a lattice theorist, 
working in the 't Hooft scaling, hence neglecting the effects of fundamentals running inside loops,
is the same as working in what they would call the 
``quenched approximation''.
We can think of the field theory as being quenched when the fundamental 
fields do not propagate inside the loops. One natural way of doing this is 
to consider the case of very massive quarks.
Indeed, when quenching, 
one considers an expansion of the fermionic determinant (for massive quarks) of the form:
\beq
\log \det [\gamma^\mu D_\mu +m]= Tr \log(\gamma^\mu D_\mu +m)= Tr \log (m[1 + \frac{\gamma^\mu D_\mu }{m}]) \sim 
Tr \log(m) + O(1/m)\,\,.
\eeq
Keeping only the constant term (or considering a very large mass) is equivalent to saying
 that fundamentals are very difficult to pair-produce, hence 
their presence inside loops will be very suppressed. Another way to quench 
in the field theory is to consider the case in which the quotient 
$x=\frac{N_f}{N_c}$ is very small. Notice that 
the quenched theory is not equivalent to a theory with only adjoints, as 
fundamentals can occur 
in external lines, like in a correlator of two mesonic currents as 
exemplified in the diagram A of Fig.\ref{htb!}. 
Needless to say, lattice theorists developed techniques to quench 
fundamental fields with arbitrary mass.
Also, while at first sight the quenching as described above is not a good operation as it breaks unitarity (not including all possible diagrams)
this kind of troubles 
will be avoided when working in the 't Hooft scaling, 
where unitarity problems will be suppressed in $1/N_c$ (but of course 
will be present in a lattice version of theories with finite $N_c$).

The interesting point to be taken from this, by a physicist working in 
gauge-string duality is 
that both scalings ('t Hooft's and Veneziano's) can be realized with D-branes.
Indeed, in both cases we must add D-branes 
(to realize symmetries and new states as discussed above), but we can add these flavor branes in two ways:
\begin{itemize}
\item{We can add $N_f$ flavor branes in such a way that we will only 
{\it probe} the geometry produced by the $N_c$ color branes. In this case the 
dynamics of the probe-flavor branes (the mesons) will be influenced by the 
dynamics of the color branes (the glueballs) but not vice-versa. This is a 
good 
approximation if $x=\frac{N_f}{N_c}\to 0$, which immediately sets us in the 't Hooft scaling limit.
Notice however, that when the $N_c$ contribution  to some 
particular quantity vanishes, the flavor effects may
be the leading ones even when $N_f \ll N_c$.}

\item{We can add $N_f$ flavor branes, in such a way that they will deform the already existing geometry, in other words {\it backreacting} on the 
original ``color'' geometry. In field theory language, we would say that the dynamics of the glueballs and that of the mesons influence each other, 
leading to new states that will be a mixture of mesons and glueballs. This is surely what we need to do if $x=\frac{N_f}{N_c}=fixed$ and doing this 
will set us in the Veneziano scaling limit. }
\end{itemize}
More technically, in the 't Hooft scaling 
limit we are studying the Born-Infeld-Wess-Zumino dynamics for a D$k$ flavor 
brane in a background created by 
$N_c$ color D$p$-branes (we will always work in Einstein frame in the 
following):
\beq
S_{BIWZ}= -T_{k}\int d^{k+1} x \, e^{\frac{k-3}{4}\phi}
 \sqrt{\det[\hat g_{ab} + 2\pi \alpha'e^{-\frac{\phi}{2}} {\cal F}_{ab}]}+ T_{k}\int  C \wedge e^{{\cal F}}\,\,,
\label{BIWZaction}
\eeq
where $\hat g_{ab}, {\cal F}_{ab}$
are fields {\it induced} by the color branes background on the (few) flavor branes.
The `shape' of the flavor branes (induced metric)
will then influence the mass spectrum and interactions of the fluctuations of the flavor branes (the mesons), explicitly realizing the picture 
advocated above. This line of research was initiated 
by Karch and Katz \cite{Karch:2002sh}
and substantially clarified in subsequent papers 
\cite{Kruczenski:2003be}, \cite{Kruczenski:2003uq}-\cite{Sakai:2005yt}. This
line kept on growing in the 
last few years, finding numerous 
applications. See the paper \cite{Erdmenger:2007cm}
for a comprehensive review.

On the other hand, working in the Veneziano scaling limit 
implies that we will need to study the action
\beq
S= S_{IIA/IIB} + S_{BIWZ}.
\label{iibbiwz}
\eeq
There will be new equations of motion, 
encoding explicitly the numbers $N_f, N_c$. 
As discussed above, it is now very clear that proceeding like this will be 
the only 
possibility when the number of flavors is comparable with the number of 
colors.
Also, it makes manifest the fact that the 
dynamics of glueballs (represented on the string side by $S_{IIA/IIB}$
which is of order $g_s^{-2}\sim N_c^2$) 
is influenced and influences back on 
the dynamics of fundamentals 
(represented by $S_{BIWZ}$, of order $g_s^{-1} N_f \sim N_c N_f$).
The rest of this review will focus on this second scaling (Veneziano).

Notice that (in both scalings) we are making an explicit difference 
between the color $SU(N_c)$ {\it gauged}
symmetry and the flavor $SU(N_f)$ {\it global} symmetry on the field 
theory side. 
From the string theory construction, this qualitative difference is connected to the 
fact that the volume of
the flavor branes is infinite, as compared to the volume of the color branes.
Indeed, in the bulk, we only need to realize the field theory
global symmetry, and we do it with the gauge field present in the Born-Infeld-Wess-Zumino action. Searching for solutions of $D_p$ color branes in 
interaction with $D_k$ flavor branes in pure IIA/IIB supergravity is an interesting problem, but will not 
represent the physical system we are after, as only flavor singlet states would be included in the dynamics.

Before we proceed studying the 
formalism and examples to clarify the details, some comments are in order.

\subsection{The string action and the scaling limit in $N_c$ and $N_f$}

Let us study a bit more the expression of eq. (\ref{iibbiwz}), being 
careful about coefficients. We will consider the case of a set of $N_c$ 
``color'' D$p$-branes and $N_f$ ``flavor'' D$k$-branes. The action for 
this system will be, in Einstein frame:
\bea
& & S= \frac{1}{2\kappa_{10}^2}\int d^{10} x \sqrt{g_{10}} 
\Big[R-\frac{1}{2}(\partial\phi)^2 -\frac{e^{-\phi}}{12}H_3^2 -\sum_l 
\frac{e^{ \frac{5-l}{2} \phi}}{2 \times l!} F_{l}^2   \Big] + \int CS-terms 
\nonumber\\
& &\,\,\,
 - N_f T_{k} \int d^{k+1}x \,e^{\frac{k-3}{4}\phi}\sqrt{\det[\hat g_{ab} + 2\pi\alpha' e^{-\frac{\phi}{2}}
{\cal F}_{ab}]} + N_f T_{k} \int_{k+1} C\wedge e^{\cal F}\,\,,\nonumber\\\rc
& & S= \frac{1}{2\kappa_{10}^2} \Big[ \int L(IIA/IIB) - 2\kappa_{10}^2 
N_f T_{k} \int L_{BIWZ} \Big],
\label{lagrangianzz}
\eea
where by $F_l$ we have denoted the various RR fields and with $CS-terms$ the 
possible Chern-Simons terms. We have taken the simplification of writing the action for
the set of flavor branes as $N_f$ times that of a single D-brane, which is enough for
the large $N$ counting we want to undertake here.
The gravitational constant and D-brane tension are:
\beq
2\kappa_{10}^2= (2\pi)^7 g_s^2 \alpha'^{\,4},\;\;\;\qquad T_{k}=\frac{1}{(2\pi)^k g_s 
(\alpha')^{\frac{k+1}{2}}} \,\,.
\label{stringthconsts}
\eeq
The typical quantization condition for the color branes reads:
\be
\frac{1}{2\kappa_{10}^2} \int F_{8-p} = N_c T_p\,\,.
\label{quantDp}
\ee
 As a 
consequence of (\ref{lagrangianzz}), we will have equations of 
motion, that generically will read for the metric and dilaton:
\bea
& & R_{\mu\nu}-\frac{1}{2}g_{\mu\nu}R= T_{\mu\nu}[IIA/B] - 
2\kappa_{10}^2 N_f T_{k} T_{\mu\nu}[brane],\nonumber\\\rc
& & \nabla^2 \phi = \frac{\partial}{\partial \phi}\Big[L[IIA/B] - 
2\kappa_{10}^2 N_f T_{k} L[BIWZ]   \Big],
\label{verygeneraleom}
\eea
together with the modified (by the CS-terms) Maxwell equations and 
importantly, the 
Bianchi identity for the (magnetic) Ramond-Ramond field $F_{8-k}$ that couples to the 
flavor D$k$-branes:
\beq
d F_{8-k}= 2\kappa_{10}^2 N_f T_{k} \delta^{9-k}(\vec{r})\,\,,
\eeq
indicating that the flavor branes are localized (all together) at the 
position 
$\vec{r}=0$. Similarly the $T_{\mu\nu}[brane]$ contains delta functions with support on
the position of the flavor branes. In principle, one will need to solve 
second order, nonlinear, partial differential equations.

Instead of directly dealing with the above equations, we want to present
here an  argument to understand which parameter controls the size of the flavor
effects on the action and, therefore, on the solution.
We remark that
the reasoning below is qualitative and in particular we
will just write a background for flat D$p$-branes
as considered for instance in \cite{Itzhaki:1998dd}. 
This will be enough for understanding the
scaling with the parameters, at least in the
cases studied in this review. 
In the following, we just focus on the behaviour with respect
to $N_f$, $N_c$, $g_{YM}^2$ and do not care about numerical prefactors.
We will use notation similar to  \cite{Itzhaki:1998dd}. 
Consider the
background associated to a stack of D$p$ color branes (in Einstein frame):
\bea
ds^2&=& e^{-\frac{\phi}{2}}\alpha' \left[
\frac{(\sqrt{\alpha'}U)^{(7-p)/2} }{\alpha'\, c_p \sqrt{g_s N_c}}dx_{1,p}^2+
\frac{\alpha'\,c_p \sqrt{g_s N_c}}{(\sqrt{\alpha'}U)^{(7-p)/2}}dU^2
+ c_p \sqrt{g_s N_c} (\sqrt{\alpha'}U)^{(p-3)/2}d\Omega_{8-p}^2\right]\,\,,\rc
e^\phi &\sim& \left( \frac{g_s N_c}{(\sqrt{\alpha'}U)^{7-p}}\right)^{\frac{3-p}{4}}\,\,,
\label{itzhakimetric}
\eea
where $c_p$ is a known
numerical constant and $U$ is an energy scale. 
On a background of this kind, we want to introduce
$N_f$ D$k$-flavor branes and to know which is the relative importance of the associated
terms in the action (\ref{lagrangianzz}) and equations of motion (\ref{verygeneraleom}).
With that aim, let us start by computing the coefficient in
front of  the term coming from the RR-form sourced
by the color branes in (\ref{lagrangianzz}), namely
$(2\kappa_{10}^2)^{-1} \sqrt{g_{10}} e^{\frac{p-3}{2}\phi} F_{8-p}^2$.
Using (\ref{quantDp}) and (\ref{itzhakimetric}), we find that the lagrangian density associated
to the color branes goes as:
\be
{\cal L}_{color} \sim (2\kappa_{10}^2)^{-1} \sqrt{g_{10}}\,\,
e^{\frac{5\phi}{2}} (\alpha'^{-1}) (g_s N_c)^{(p-4)/2} (\sqrt{\alpha'} U )^{(p-3)(p-8)/2}\,\,.
\label{LLcolor}
\ee
Let us now look at the DBI term. We assume that the flavor D$k$-branes are extended
along the Minkowski directions, the radial direction $U$ and $k-p-1$ directions within the
sphere. We find:
\be
{\cal L}_{flavor,DBI}  
\sim  N_f T_k e^{\frac{k-3}{4}\phi}
\sqrt{\hat g_{k+1}} \sim \,\, \frac{N_f}{N_c} \,\,g_{eff}^{\frac{k-p}{2}}\,\, {\cal L}_{color}\,\,,
\ee
where in order to get the last expression we have used 
(\ref{stringthconsts}),
(\ref{itzhakimetric}),
(\ref{LLcolor}) and defined a dimensionless effective coupling as in \cite{Itzhaki:1998dd}:
\be
g_{eff}^2 \sim g_{YM}^2 N_c U^{p-3}\sim g_s N_c (\sqrt{\alpha'} U)^{p-3}\,\,.
\label{geff}
\ee
Thus, parametrically, the action from the flavor branes as compared with
that from the color background 
is weighed by $\frac{N_f}{N_c}\,\, g_{eff}^{\frac{k-p}{2}}$.
We now want to take a low energy decoupling limit as in \cite{Itzhaki:1998dd}
(see also \cite{Boonstra:1998mp}), 
namely the dimensionless effective coupling
 $g_{eff}$ and $U$ are fixed as $\alpha'\to 0$.
Thus, the Veneziano scaling limit in this framework amounts to:
\be
N_c,N_f\to \infty\,\,,\qquad\quad g_{eff} \ \ \textrm{fixed},
\qquad\quad \frac{N_f}{N_c}\,\, g_{eff}^{\frac{k-p}{2}} \ \ \textrm{fixed},
\label{ncnfscaling}
\ee
where the last relation comes from demanding that the flavor effects
are also fixed. 
Staying in the supergravity limit requires:
\be
1 \ll g_{eff}^2 \ll N_c^{\frac{4}{7-p}}\,\,,
\label{validitz}
\ee
a constraint that limits the range of energy scales $U$ for which the supergravity
description is valid \cite{Itzhaki:1998dd}. Notice that if we further require that the flavor 
terms do not  parametrically
dominate over the color ones, this can further restrict $U$, depending on $p$ and $k$.

The probe limit, in which the flavor action is negligible as compared to the
gravity action, comes from making the last quantity
in (\ref{ncnfscaling})
 vanishingly small\footnote{In the literature, it is 
usually written that the probe limit is good when $N_f \ll N_c$. That is not strictly
correct. For instance in the D3-D7 case, the probe approximation is valid when,
parametrically, $\lambda\, N_f \ll N_c$.}. 
As expected, that term is strictly zero in the 't Hooft limit.
We now comment on the values of $p$, $k$ that will appear in the following sections:

For the  D3-D7 case of section
\ref{AdS5X5}, the parameter weighing the flavor effects
is $(g_{YM}^2 N_c)\frac{N_f}{N_c} \sim \lambda\frac{N_f}{N_c}$.
For the cascading case of section \ref{KS}, the result is similar but one has to replace
$N_c$ by the number of fractional branes $M$. 
Getting ahead of the discussion of upcoming sections we notice that,
in these cases, 
it is not enough to take this parameter
fixed, but it should also be small. This will be
due to positive beta functions, as will be thoroughly discussed.

From (\ref{ncnfscaling}), we see that  $k=p$ (sections
\ref{sec:D5D5},\ref{sec:2+1} and \ref{models2})
is particularly interesting\footnote{
Even if in all these sections we
will deal with wrapped branes and therefore the backgrounds are not that similar
to (\ref{itzhakimetric}), the argument above still yields the correct result.}, since it is really $\frac{N_f}{N_c}$
what has to be taken fixed. 
For this reason, only in these cases one
can hope to describe - within gravity - phenomena as Seiberg-like dualities
(see section \ref{sec:Seib}). Loosely  speaking, the Klebanov-Strassler duality
cascade \cite{Klebanov:2000hb} lies in this class of $k=p$ theories, since it 
can be understood as the interplay between two
sets of D5-branes wrapping vanishing two-cycles.

We close this section by mentioning other brane intersections 
that will not be discussed further in later sections. In a D2-D6 system,
the effective coupling (\ref{geff}) decreases at large $U$ and therefore the
flavor backreaction on the glue fades away in the UV - see 
(\ref{ncnfscaling}) - as expected in 
a superrenormalizable theory. This was observed in \cite{cherkishash} when studying a solution
with localized D6-branes. In a D4-D8 intersection, the opposite happens. The probe brane approach can be
valid in an intermediate regime $1 \ll g_{eff}^2 \ll \frac{N_c}{N_f}$ but at a given $U$
the fundamentals eventually take over and dominate. Notice that the value of $U$ for which
the D4-D8 theory loses its validity is parametrically smaller than that for which the 
unflavored D4-brane theory becomes pathological, which is set by (\ref{validitz}).

\subsection{The method}

Looking back at eq. (\ref{lagrangianzz}), one 
can appreciate that in general finding the solution describing 
the 
backreaction between the type II closed strings and the open strings described 
by the Born-Infeld 
action is quite a challenging problem. 
Indeed, the fact that the flavor branes (BIWZ) are localized in the 
ten-dimensional space 
implies that we will have to solve 
second order, nonlinear, coupled and partial 
differential equations with localized sources. 
Basically what makes things so difficult 
are the presence of delta function sources and the fact that the 
differential equations describing the 
dynamics are ``partial'' (in 
principle depending on all the variables 
describing the space transverse to the 
flavor branes). In order to get some intuition of the answer, we may consider the case in which 
we will ``erase'' the dependence on these 
transverse coordinates (this is like considering the 
``s-wave'' of the putative multipole decomposition of the full solution in 
this transverse space) and delocalize the sources. 
To achieve this, we will propose to {\it smear} the flavor branes 
over their perpendicular space.

On the field theory side, this will amount to considering systems where the addition of the degrees 
of freedom  transforming in the fundamental does not break any of the global symmetries of the unflavored 
QFT. Also, it may happen that the original $U(N_f)$ is explicitly broken to $U(1)^{N_f}$ as we are 
separating the flavor branes - see the discussion in section 7 of 
\cite{HoyosBadajoz:2008fw} and in section 2 of \cite{Bigazzi:2008zt}. 
An intuitive 
understanding of the smearing procedure will be discussed in 
section \ref{heuristiczzz}, while a more formal approach will be 
treated in section \ref{mathviewpointzz}.
For technical reasons, this procedure is cleaner in examples preserving some amount of SUSY, since the 
force between flavor branes is cancelled and the smearing is at no cost of energy.

In the examples described in the following sections, we will proceed like this:
\begin{itemize}
\item{Consider an unflavored string background and find the embedding of flavor branes that will preserve 
some SUSY, or (in non-SUSY examples) that will be stable and solve the equations of motion for the brane. 
In the SUSY cases,
this can be achieved by 
considering kappa-symmetric embeddings, that we review 
generically below.}
\item{Consider now $N_f$ flavor branes in that particular 
embedding and smear them, getting an action in ten dimensions, as will 
be explained with all generality in section \ref{mathviewpointzz}.}
\item{Solve the equations derived from (\ref{lagrangianzz}), 
that will now contain smeared branes and will 
be ordinary differential 
equations. In SUSY cases, 
there will be a set of BPS equations to be solved. In non-SUSY examples 
one might manage to get a fake superpotential and fake-BPS equations 
\cite{Freedman:2003ax}.   } 
\end{itemize}
Moreover, one has to check that the flavor embeddings considered are still a solution
in the backreacted geometry. 

Let us review briefly the main technical points collected above.

\subsubsection{BPS equations, Kappa symmetry (SUSY probes) and smearing}
\label{sec: kappa}

Let us consider the case of a supersymmetric background, namely a solution
of type II sugra
for which the supersymmetry variations of the gravitino and the dilatino
 vanish $\delta \psi_\mu = \delta \lambda =0$.
We will not give here details on the form of these expressions, which can
be found elsewhere. For instance,  the string frame
SUSY transformations of both type IIA and type IIB 
are written down in appendix A of \cite{Martucci:2005rb}. 

Given a background that preserves some amount of SUSY, the idea is to find 
the hyper-surfaces in which to extend the flavor branes (in other words, 
finding the embeddings for flavor branes) so that these will preserve all  
(or a fraction) of the SUSY of the background.

One then writes an eigenvalue problem, imposing that the preserved spinors 
of the background are eigenspinors of the kappa-symmetry matrix
\beq
\Gamma_\kappa\,\, \epsilon=\epsilon\,\,.
\eeq
See the papers \cite{Cederwall:1996ri}
for the definition of $\Gamma_\kappa$.

Once we have the kappa-symmetric embeddings as described above, we now 
proceed to write an action describing the dynamics of the closed and open 
strings, as in eq. (\ref{lagrangianzz}). We then realize that the problem 
will lead (unless we are adding D9-branes) to a system of partial 
differential equations. As discussed above, we then proceed to smear these 
flavor branes. For this we propose an ansatz for the metric, where the 
embedding of the flavor branes is clear and distribute them along the 
directions of their transverse space. This distribution of the flavor 
branes can be done in a uniform way. In some sense, we are `deconstructing' 
the transverse space to the flavor branes by adding at each point one of 
the many $N_f$ flavor branes.

The key point is that once  the BPS equations and kappa-symmetry conditions
are simultaneously satisfied,
the problem is solved. In fact, it is a general result \cite{Koerber:2007hd} that
the SUSY equations $\delta \psi_\mu = \delta \lambda =0$, together with the 
Bianchi identities
 - and equations of motion -
for the different forms modified by calibrated (namely, kappa-symmetric) sources
imply the full set of equations of motion. 

In the following, we will discuss first an intuitive way of 
understanding this smearing.  Then 
we will apply this to different examples in sections \ref{AdS5X5}-\ref{models2}.
Finally, in section \ref{mathviewpointzz}, we will present
a formal way of implementing the backreaction from smeared sources.

\subsection{A heuristic viewpoint}
\label{heuristiczzz}

In the following sections, we will introduce the necessary mathematical machinery to consistently
compute solutions of string theory in which smeared backreacting flavor branes are present.
Before that, it is worth to make a digression in order to explain the 
general set-up in
simple, heuristic terms.

\begin{figure}[htbp]
\centerline{
\includegraphics[width=.3\textwidth]{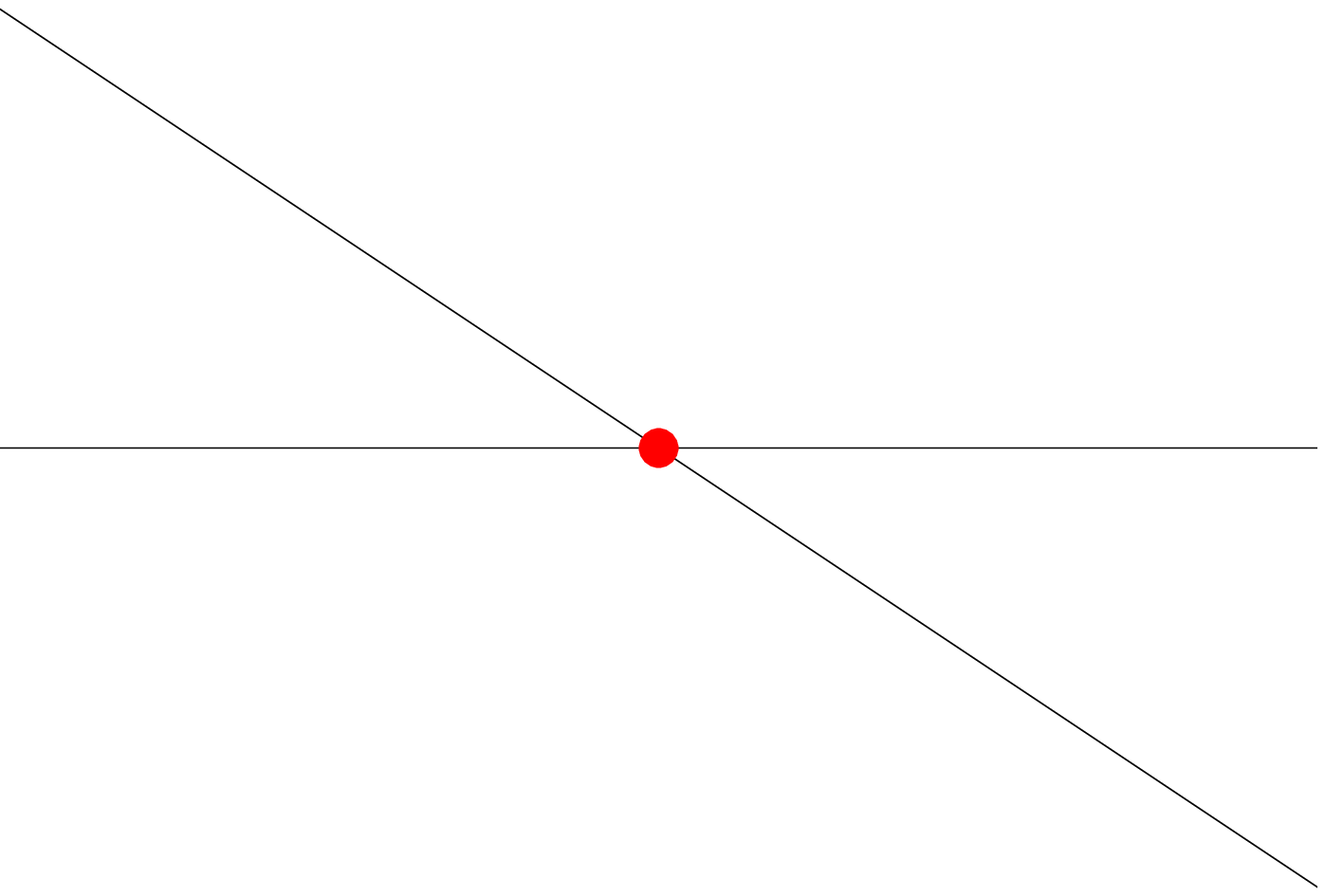}
\qquad\qquad\qquad
\includegraphics[width=.3\textwidth]{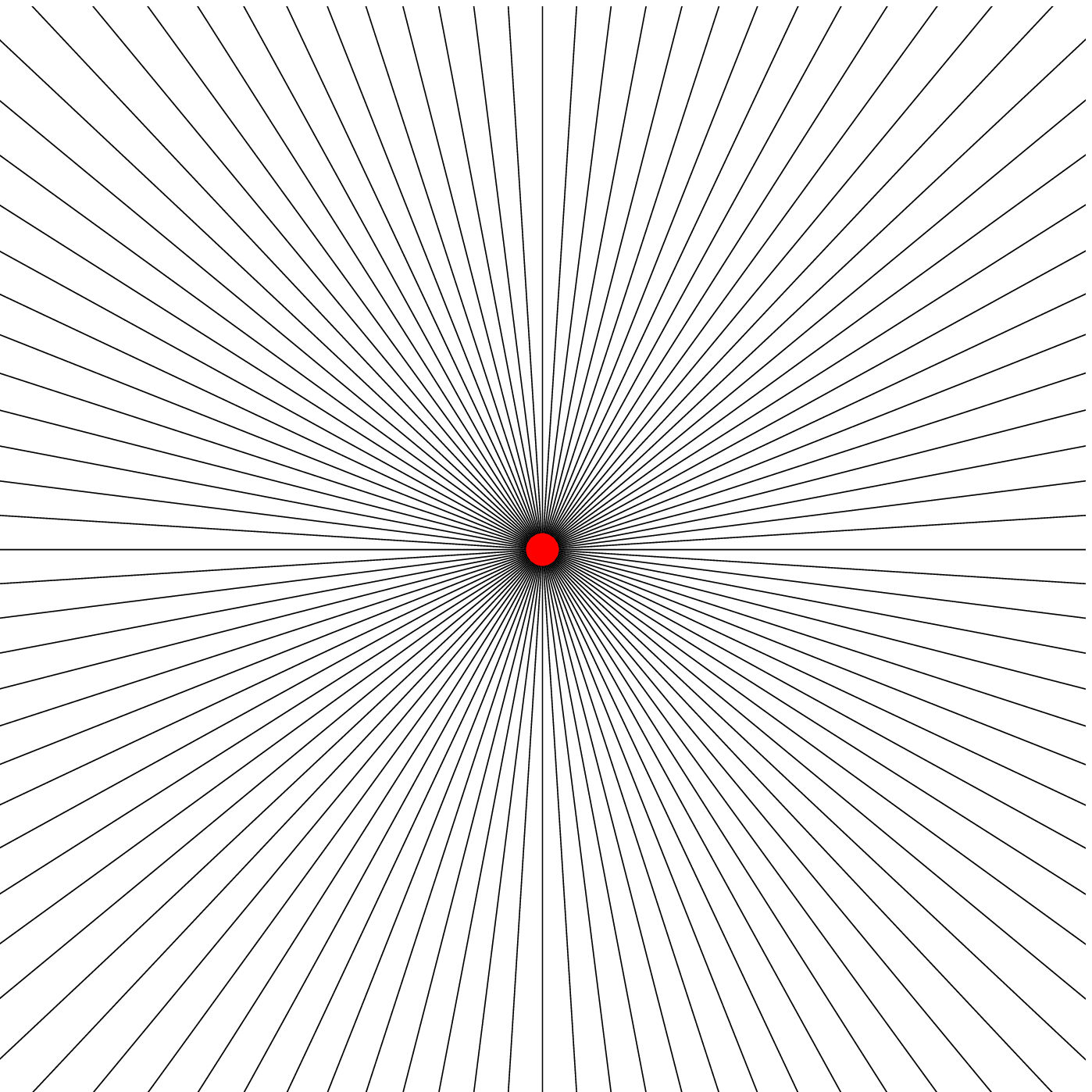}} 
\caption{On the left, a point-like charge (in red) and two lines of charge at different angles. 
On the right, a configuration with many lines of charge. In the 
asymptotic limit of an infinite
number of lines, they just correspond to a radial charge density. This picture depicts an 
analogous situation to the case of smeared flavored branes, when the fundamental fields are
massless.}
\label{massless_plot}
\end{figure}

Let us make an analogy with electrostatics. Suppose we want to compute the electric field
generated by a point-like charge and a couple of lines of charge, as depicted on the left
of Figure \ref{massless_plot}.  In order to depict the situation, we show 
dimension 1 lines
of charge in a total space of dimension 2, but clearly the situation can be generalized by changing
such dimensions. Since in the left plot there is no particular symmetry in the configuration, the
resulting electric field will have a not so simple expression. But let us imagine that we consider a huge
number of lines of charge as in the plot of the right, and homogeneously distribute them in the angle
they form with the horizontal axis.
In the limit of many lines, radial symmetry is recovered, the charge density
is ``smeared" and
 will be just given by a single (monotonically decreasing) function
$\rho(r)$. The electric field, accordingly, will also be radially symmetric.
Notice that this process of describing a large number of discrete objects by a continuous distribution
is ubiquitous in physics: for instance a ``homogeneous" gas is a collection of atoms or the
``homogeneous" Universe considered in cosmological models contains a collection of galaxy clusters.
Also, solutions with different kinds of 
smeared sources have been considered many times 
in string theory contexts not necessarily related to gauge-gravity duality, see for
instance \cite{Grana:2006kf}, \cite{DeWolfe:2008zy}.

When comparing to the string theory set-up, the point-like
charge in the center corresponds to the color branes and the lines to the flavor branes
(which have to extend to infinity). The limiting, radially symmetric configuration, corresponds to
the kind of smeared situations that we will discuss in this review.\footnote{More precisely, it corresponds
to the situations analysed in sections \ref{AdS5X5}-\ref{KS}. For the 
cohomogeneity 2 cases analysed in section
\ref{models2}, the different functions depend on two radial variables. A heuristic picture for such
situations is presented in section \ref{models2}.} 
All functions of the ansatz can then be considered to depend on a single radial variable.
For flavor branes, the different 
``angles" correspond to adding fundamental matter which couples differently to the rest of the fields. 
In some of the cases discussed in the following, we will see how this is reflected in the field theory
superpotential (section \ref{masslessKW}).

We can still get further intuition from this simple analogy. In the right plot of figure
\ref{massless_plot}, we see that all lines intersect at the center. From the string point of view,
that means that the flavor branes are stretched down to the bottom of the geometry and the quarks
are massless.
In this situation, the charge density
$\rho(r)$ is highly peaked at $r=0$. Essentially, that is the reason why for the 
solutions with massless
quarks described in the following sections there is a curvature singularity at the origin, where
all flavor branes meet.

\begin{figure}[htbp]
\centerline{
\includegraphics[width=.3\textwidth]{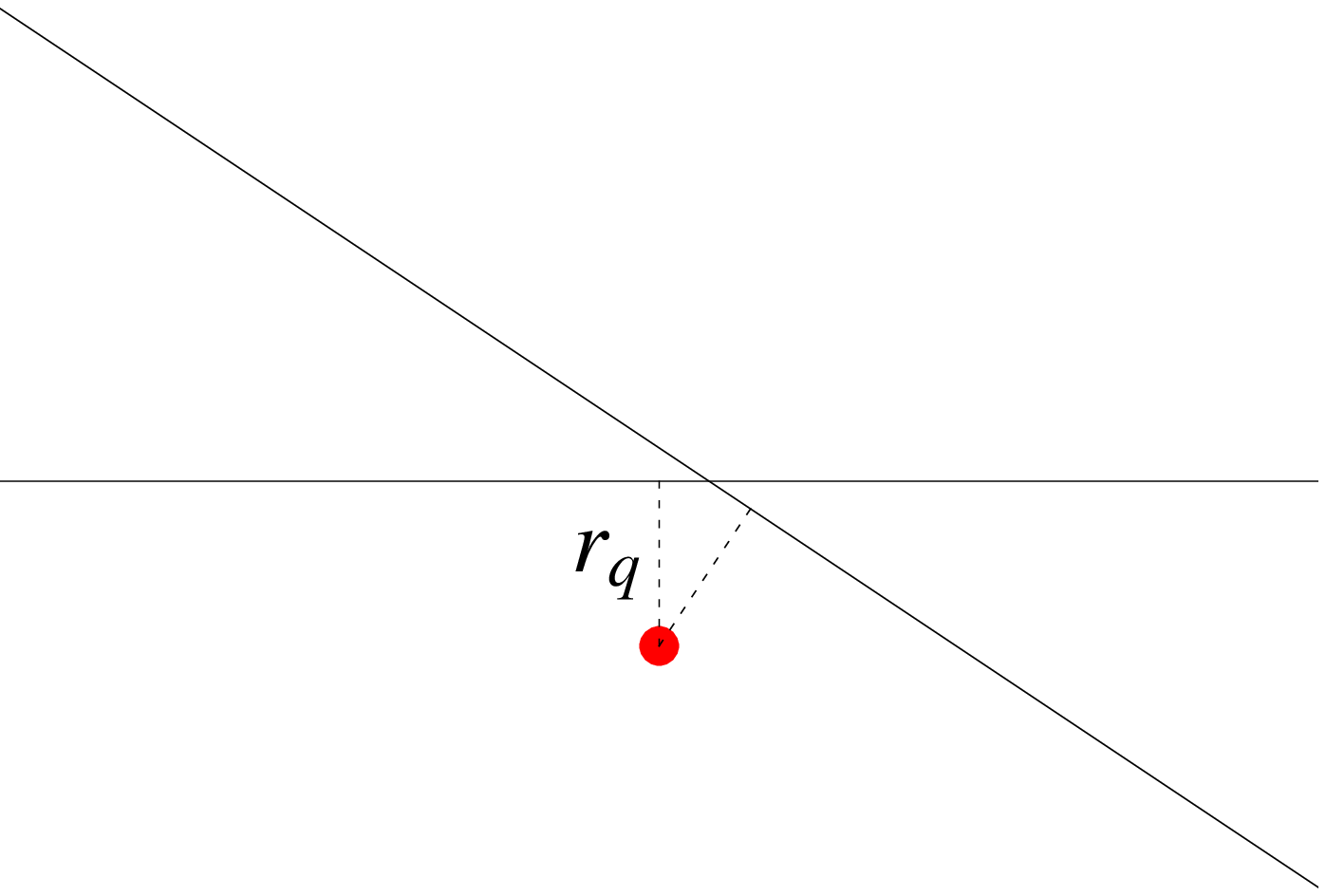}
\qquad\qquad\qquad
\includegraphics[width=.3\textwidth]{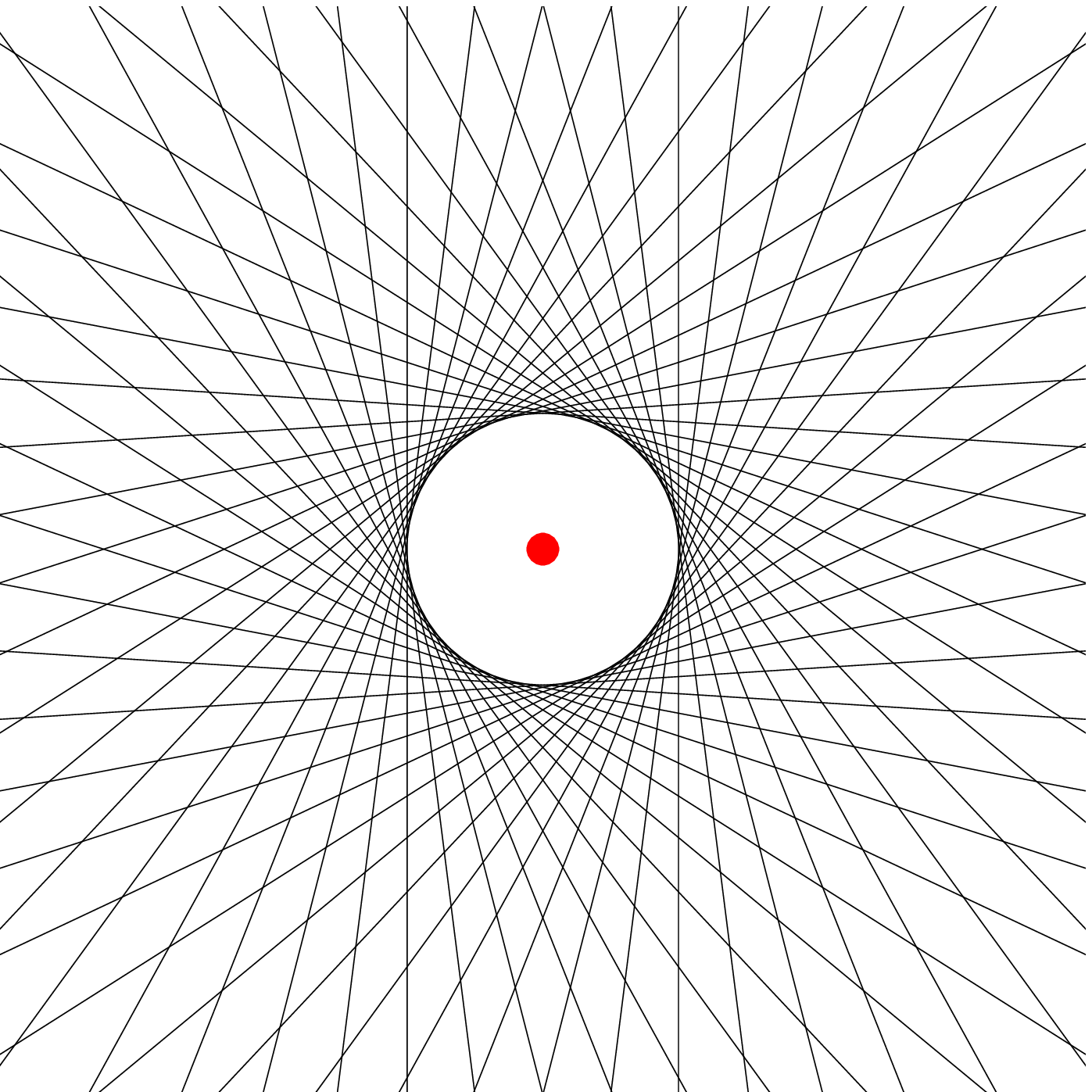}}
\caption{These pictures depict  
analogous situations to the case of flavored branes, when the fundamental fields are
massive. On the right, again, we add a large number of lines, such that in the limit radial
symmetry is recovered.}
\label{massive_plot}
\end{figure}

Then, a simple way of getting rid of such a singularity is to displace the lines of charge from the origin,
while still keeping the radial symmetry. This is depicted in figure \ref{massive_plot}. 
If we dub the distance from any of the lines to the center as $r_q$, the density of charge
$\rho(r)$ will vanish for $0<r<r_q$, while it will asymptote to the ''massless" $r_q=0$ one as $r\gg r_q$.
From the brane 
construction, this displacement
typically corresponds to giving a mass to the fundamentals (or, in particular cases, it could
correspond to a non-trivial vacuum expectation value).
The solution of section \ref{massiveKW}, which indeed is regular in the IR, is a neat example of this notion.
Another possibility is to add temperature and to hide the singularity behind a horizon,
see section \ref{D3D7plasma}.

Going back to electrostatics for the right plot of figure \ref{massive_plot}, we know from Gauss' law
 that the charge density outside does not affect the central region. The corresponding statement in the
 field theory is that the massive fields decouple from the IR physics below the scale given by their mass.
We find it interesting that, through this heuristic reasoning, Gauss' law is 
connected to the decoupling of heavy
particles (or holomorphic decoupling in the SUSY cases).

Even if the example of electrostatics is useful to qualitatively
picture what we will do in the following, the analogy is
by no means perfect. We note two differences: 
first, we will be working with gravity, which is non-linear and, thus, one cannot find the final
solution by superposing the fields generated by different sources (which in the case of electrostatics
would make it rather trivial to find the electric field for the configurations depicted on the left
of figures \ref{massless_plot} and \ref{massive_plot}). Second, our ``lines of charge" (the flavor
branes) are dynamical. This means that is it not enough to compute the background fields generated from the
sources but one also has to check that the sources are stably embedded in the geometry.

We end this section by summarizing the pros and cons of looking 
for duals of unquenched theories for which the string solutions include smeared flavor
branes, many of which can be inferred from the heuristic discussion above. On the positive side:

\begin{itemize}

\item The smearing simplifies the situation allowing us to write ans\"atze depending on a single radial coordinate,
and therefore the problem is eventually reduced to a set of ODEs. (For the cases of section \ref{models2}, they
depend on two radial coordinates and thus one finds PDEs in terms of two 
independent variables, but again, without delta-function localized 
sources.)

\item Possible issues related to singularities and strong coupling are ameliorated in the same sense as 
they are washed out in electrostatics when
considering a smooth charge density rather than a sum of delta-functions over a large number of electrons.

\item It allows a simple application  of the powerful mathematical tools of  calibrated geometry \cite{Gaillard:2008wt},
see section \ref{mathviewpointzz}.

\end{itemize}

On the negative side:
\begin{itemize}

\item Obviously, if we require the flavor branes to be smeared, we are limiting ourselves to considering a very
particular subset of all the possible flavored theories. In particular, we require the superpotentials to 
effectively recover (some of) the global symmetries of the theory without flavors.

\item Related to the previous point,
one cannot realize, in general, theories with $U(N_f)$ flavor groups. Since the flavor branes are required to
sit at different points in the internal space, the typical string connecting different flavor branes is heavy
and the flavor symmetry is typically broken to $U(1)^{N_f}$ (one may also interpret the 
solutions as having flavor symmetry $U(k)^{N_f/k}$ for some $k\ll N_f$).
From the point of view of the field theory, this amounts to having a Veneziano expansion
with ``one window graphs'', as pointed out in \cite{Bigazzi:2008zt}.
In principle, this fact can hinder the realization of some interesting physical features in the dual 
set-ups considered. 

\end{itemize}

\subsection{Localized sources and other approaches}
\label{sec:localized}

As already remarked, this review focuses on solutions of string theory for which 
there are D-brane sources homogeneously smeared over a given family of possible 
embeddings, and that can be interpreted as duals of strongly coupled gauge
theories in the Veneziano limit. As stressed above, this is a very particular
subset of all the possible duals of theories with flavor. In a generic 
case, one
should consider the sources to be localized at certain positions, such that the
density of charge is given by a sum over Dirac delta functions. Such generic case
is  technically more challenging. 
However, remarkable works along these lines have appeared, pursuing solutions
with the flavor branes localized at a single point of space (notice this is not
the most general case either). 
We will not review them
in any detail here, but the goal of this section is to provide a brief guide
to the literature on the subject.

The main ingredient of this approach consists of finding solutions of 
supergravity which can be interpreted as intersections of branes of different
dimension, with each stack of branes localized at a fixed position of space-time.
In the context of gauge-gravity duality, the search for such solutions was initiated
in \cite{Kehagias:1998gn}, \cite{Aharony:1998xz}. These papers discussed D3-D7 intersections,
which have been the most studied in the literature (see below for
different set-ups). A lot of progress was reported in subsequent years
\cite{Grana:2001xn}, \cite{Bertolini:2001qa}, \cite{Bertolini:2002xu}.
Among other aspects, these papers presented a better understanding of the 
D3-D7 solutions,
the inclusion of fractional branes and clear matchings with field theory issues
such as the running of couplings and anomalies.
Further work on the D3-D7 localized system was performed in \cite{Burrington:2004id}
(where the conifold was also addressed),
\cite{Liu:2004ru} (where D7 brane backreaction on bubbling geometries was considered) and
\cite{Kirsch:2005uy}, where the  solution corresponding to D3-D7 in flat space
 was completed by providing an expression for 
the warp factor in closed form.  
It is also worth mentioning \cite{Ouyang:2003df} where a flavor D7-brane
in a cascading theory was considered and its backreaction introduced 
as a perturbation. The finite temperature generalization of
the background of \cite{Ouyang:2003df} was discussed in  \cite{Mia:2009wj}.

Let us now outline the literature on D2-D6 localized intersections, which can be interpreted
as duals of 2+1 supersymmetric gauge theories coupled to fundamentals introduced
by the D6-branes. The construction of the type IIA solutions (and their relation
to M-theory) was carried out in \cite{Itzhaki:1998uz}, \cite{Cherkis:2002ir}, \cite{cherkishash}.
In \cite{Erdmenger:2004dk}, meson excitations of this background were discussed and,
in particular,
the holographic dictionary relating meson-like operators
to certain (closed string) supergravity modes was presented.
On the other hand, the authors of \cite{GomezReino:2004pw} found a finite temperature
version of the solution, which was used to discuss the thermodynamics of the system.
Very recent progress in the D2-D6 systems, their M-theory uplifts and the detailed
relation to Chern-Simons theories with flavor has been reported in
\cite{gaiottoetal}.

Regarding D4-D8 intersections,
localized solutions  in that
set-up were constructed in \cite{Nastase:2003dd} in an early attempt to build a QCD dual.
In the context of the Sakai-Sugimoto model \cite{Sakai:2004cn}, backreaction from
localized D8-$\bar{\rm {D}8}$ branes was analysed in \cite{Burrington:2007qd}.

It is also worth mentioning recent solutions in heterotic string theory which
were argued to be related to flavored theories \cite{Carlevaro:2009jx}.

\vskip.25cm

Interestingly, there are a few papers in which 
similar situations were considered in subcritical string theory
and therefore defined in dimensions lower than ten.
In many of these cases,
each flavor brane fills the
whole space-time (therefore they 
are not localized, neither
smeared).
Some physics can then be extracted by using exact string theory methods
but what these models have in common is that it is not possible to handle 
them
within a well-controlled gravity description: 
gravity-like actions with just two derivatives suffer curvature corrections which cannot 
be neglected, nor consistently
computed. However, 
there is the hope that the two-derivative actions
 can nevertheless provide additional non-trivial insights in the
physics of the system.
This idea was put forward by Klebanov and Maldacena in \cite{Klebanov:2004ya},
who considered a D3-D5 system in a six dimensional background (the cigar).
Such a system is dual to 4D ${\cal N}=1$ SQCD as was shown using exact worldsheet methods
in \cite{Fotopoulos:2005cn,Murthy:2006xt}. 
For a recent discussion on the dual to the flavor singlet sector of
${\cal N}=2$ superconformal QCD in a subcritical string framework, see
\cite{Gadde:2009dj}.
The set-up of \cite{Klebanov:2004ya} was generalized to different situations in \cite{Alishahiha:2004yv},
\cite{Bigazzi:2005md},
\cite{Casero:2005se}.
The finite temperature physics of a model in \cite{Bigazzi:2005md} was
analysed in \cite{Bertoldi:2007sf}.
Bottom-up approaches
(in which a high-dimensional gravity theory is proposed to describe
some specific features of QCD) with space-time filling flavor branes
have been discussed in \cite{Gursoy:2007cb}, \cite{Sin:2007ze}.
Recently, a bottom-up approach to the 
conformal window along these lines has
appeared 
\cite{Jarvinen:2009fe}.

\vskip.25cm

Finally, let us mention a recent contribution by Armoni
\cite{Armoni:2008jy}, in which a way of departing from the quenched approximation
was proposed. The fermion determinant is expanded in terms of Wilson loops. It then
turns out that a sum of correlators of an observable with the Wilson loops boils 
down to an expansion in $\frac{N_f}{N_c}$, which can in principle be computed.
It would be nice to further develop possible implications of this 
observation in holographic set-ups.

\setcounter{equation}{0}
\section{Flavor deformations of $AdS_5\times X^5$}
\label{AdS5X5}

Our first concrete application of the procedure described 
will be the flavor deformation of $AdS_5\times S^5$. 
This is the simplest possible case and,
hopefully, it will neatly illustrate the comments of section \ref{section1}. 
In fact, for most of the discussion,
the formalism applies to any $AdS_5\times X^5$ geometry, $X^5$ being a five-dimensional 
compact Sasaki-Einstein (SE) space, so we will refer to this more general case during this
whole section. At some points, we will use the two notable examples $X^5 = S^5$ or $T_{1,1}$
to clarify particular issues. 

Let us start with a general comment. Since the $AdS_5\times S^5$ 
theories without flavor are conformal, we expect that once we include extra matter, a positive
beta function is generated. This is in fact the case and leads to the appearance of a Landau 
pole. Nevertheless, as in QED,
the theory renders meaningful IR physics 
even if the UV is ill-defined, as long as the IR and UV are well separated scales. 
However, this separation does not allow to have $N_f$ and $N_c$ of the same order.
As we will see, one can define a parameter $\epsilon \sim \lambda N_f/N_c$ 
which weighs the internal
flavor
loops and that has to be kept small. The effect of the unquenched quarks can then be computed as
an expansion in $\epsilon$.

After introducing the framework in sections \ref{sec21}, \ref{sec22},
we present the unquenched supersymmetric 
(${\cal N}=1$ in 4d) solutions in \ref{sec23}. In section \ref{sec:screening},
we present an instance of the effects of the unquenched flavors on a physical quantity,
namely on the mass of a particular meson tower.
Then, in section \ref{D3D7plasma}, we break supersymmetry by turning on temperature and analyse
the physics of the dual quark-gluon plasma. We end in section \ref{sec:rangeval} by discussing the range
of validity for the solutions and approximations used.

\subsection{The geometries and field theories without flavors}
\label{sec21}

 The models we discuss here
 are obtained by placing a stack of $N_c$ D3-branes at the origin of the six-dimensional cone over $X^5$. The corresponding type IIB background reads:
\bear
&&ds^2\,=\,\big[\,h(r)\,\big]^{-1/2}\,dx^2_{1,3}\,+\,\big[\,h(r)\,\big]^{1/2}\,\big[\,dr^2\,+\,r^2\,ds^2_{X^5}\,\big]\,\,,\rc\rc
&&F_5\,=\,dh^{-1}\,dx^0\wedge\cdots\wedge dx^3\,+\,{\rm Hodge\,\, dual}\,\,,\rc\rc
&&h(r)\,=\,\frac{Q_c}{ 4r^4}\,\,,\qquad\qquad Q_c\,\equiv\,\frac{(2\pi)^4\,g_s\,\alpha'^2\,N_c}{{\rm Vol}(X^5)}\,\,,
\label{AdS5X5back}
\eear
where we have taken the near horizon limit. The dilaton is
 constant  and all the other fields of type IIB supergravity vanish. 
 In general the metric of the SE space $X^5$ can be written as a Hopf fibration over a four-dimensional K\"ahler-Einstein (KE) manifold:
\beq
ds^2_{X^5}\,=\,ds^2_{KE}\,+\,\big(\,d\tau\,+\,A_{KE}\,\big)^2\,\,,
\label{SEmetric}
\eeq
where $\tau$ is the fiber and $A_{KE}$ is the connection one-form whose exterior derivative gives the K\"ahler form $J_{KE}$ of the KE base:
\beq
dA_{KE}\,=\,2J_{KE}\,\,.
\label{A-J-KE}
\eeq

Let us first consider  the particular case in which $X^5$ is the five-sphere $S^5$. In this case the KE base is the manifold $CP^2$ (with the Fubini-Study metric)  and the space transverse to the color branes, with metric $dr^2+r^2\,ds^2_{S^5}$, is just 
$\mathbb{R}^6$.  When $X^5=S^5$ the coefficient $Q_c$ appearing in (\ref{AdS5X5back}) is just $Q_c\,=\,16\pi g_s\alpha'^2\,N_c$. Moreover, as is well-known, the field theory dual to the $AdS_5\times S^5$ background is 
${\cal N}=4$ SYM in 4d, which, in ${\cal N}=1$ language,  can be written  in terms of a vector multiplet and of  three chiral superfields $\Phi_i$ ($i=1,2,3$) transforming in the adjoint representation of the gauge group and interacting by means of the cubic superpotential:
\beq
W_{{\cal N}=4}\,=\,{\rm Tr}\,\Big[\,\Phi_1\,[\Phi_2,\Phi_3]\Big]\,\,.
\eeq
If we represent the transverse $\mathbb{R}^6$ of the  $AdS_5\times S^5$ solution in terms of three complex variables $Z_i$ ($i=1,2,3$), one can regard the $Z_i$'s as the geometric realization of the adjoint superfields $\Phi_i$. 

The second prominent example which we will analyze in detail is the one in which $X^5$ is the $T^{1,1}$ space with metric:
\beq
ds^2_{T^{1,1}} = \frac16\sum_{i=1}^2[d\theta_i^2+ \sin^2\theta_id\varphi_i^2] + \frac19 [d\psi+\sum_{i=1}^2\cos\theta_id\varphi_i]^2\,,
\label{T11metric}
\eeq
where the range of the angles is $\psi \in [0,4\pi)$,
$\varphi_i \in [0,2\pi)$, $\theta_i \in [0,\pi]$.  Since ${\rm Vol}(T^{1,1})=16\pi^3/27$, the coefficient $Q_c$ for this solution is $Q_c=27\pi g_s\alpha'^2\,N_c$. In this case the space transverse to the color branes is the conifold, which is a 6d Calabi-Yau cone which can also be described as the locus of the solutions of the algebraic equation:
\be
z_1z_2=z_3z_4\,,
\label{conifold-rel}
\ee
where the $z_i$ are four complex coordinates. The relation between these variables and the coordinates used in (\ref{T11metric}) is the following:
\bear
z_1 = r^\frac32 e^{\frac{i}{2}(\psi - \varphi_1 - \varphi_2)}
\sin\frac{\theta_1}{2}\sin\frac{\theta_2}{2}\,\,,\qquad
z_2 = r^\frac32 e^{\frac{i}{2}(\psi + \varphi_1 + \varphi_2)}
\cos\frac{\theta_1}{2}\cos\frac{\theta_2}{2}\,\,,\rc
z_3 = r^\frac32 e^{\frac{i}{2}(\psi + \varphi_1 - \varphi_2)}
\cos\frac{\theta_1}{2}\sin\frac{\theta_2}{2}\,\,,\qquad
z_4 = r^\frac32 e^{\frac{i}{2}(\psi - \varphi_1 + \varphi_2)}
\sin\frac{\theta_1}{2}\cos\frac{\theta_2}{2}\,\,.
\label{zetas}
\eear
Notice also that the metric written in (\ref{T11metric}) is of the form (\ref{SEmetric}) where the KE base is just the $S^2\times S^2$ space parameterized by the angles $(\theta_i, \varphi_i)$ and  one should make the following identifications:
\bear
&&\tau\,=\,\psi/3\,\,,\qquad\qquad
A_{T^{1,1}}\,=\,\frac13\,\big(\cos\theta_1\,d\varphi_1\,+\,\cos\theta_2\,d\varphi_2\,\big)\,\,,\rc\rc
&&J_{T^{1,1}}\,=\,\frac{dA_{T^{1,1}}}{ 2}\,=\,-\frac16\,
\big(\,\sin\theta_1\,d\theta_1\wedge d\varphi_1\,+
\,\sin\theta_2\,d\theta_2\wedge d\varphi_2\,\big)\,\,.
\label{Kahler-conifold}
\eear
The field theory dual to the $AdS_5\times T^{1,1}$ background is 
the  ${\cal N}=1$ superconformal quiver gauge theory with gauge group $SU(N_c)\times SU(N_c)$ and bifundamental matter fields $A_1, A_2$ and $B_1, B_2$ transforming respectively in the ({\bf $N_c,\bar N_c$}) and in the ({\bf$\bar N_c, N_c$}) representations of the gauge group \cite{kw}, \ie\ the so-called Klebanov-Witten (KW) model. The matter fields form two $SU(2)$ doublets and interact through a quartic superpotential
\be
W_{KW} = \hat h\,\epsilon^{ij}\epsilon^{kl}\,{\rm Tr}\, [A_iB_kA_jB_l]\,.
\label{KW-superpotential}
\ee
The fields $A_i$ and $B_i$ can be related to the coordinates $z_i$ by means of the following  relations:
\be
z_1 = A_1 B_1, \quad z_2 = A_2 B_2, \quad z_3 = A_1B_2, \quad z_4 = A_2B_1 \,,
\label{maps}
\ee
which automatically solve the defining conifold equation (\ref{conifold-rel}). 

\subsection{Flavor branes and smeared charge distribution}
\label{sec22}

The flavor branes for the  $AdS_5\times X^5$ backgrounds just described are D7-branes extended along the four Minkowski directions as well as along a non-compact submanifold of the cone over $X^5$. The type of flavor that the D7-branes add depends both on the space $X^5$ and on the submanifold they wrap in the transverse space. We first illustrate the situation
with the two examples of $X^5 = S^5$ and $T_{1,1}$ and at the end display the general expressions.

The first instance is
the case in which $X^5=S^5$. In this case a simple kappa symmetry analysis shows that, in order to preserve eight supersymmetries,  the D7-branes must be extended along a codimension two hyperplane in $\mathbb{R}^6$ which, in terms of the complex coordinates $Z^i$ can be written as:
\beq
a_1\,Z^1\,+\,a_2\,Z^2\,+\,a_3\,Z^3\,=\,\mu\,\,,
\label{hol-embedding-S5}
\eeq
with the  $a_i$ and $\mu$ being complex constants satisfying $\sum_1^3 |a_i|^2 = 1$. 
On the field theory side these flavor branes introduce ${\cal N}=2$ fundamental hypermultiplets $(Q^r, \tilde Q_r)$ ($r=1,\cdots N_f$) - nonetheless,
a generic collection of branes
within the family (\ref{hol-embedding-S5})  retains just ${\cal N}=1$ susy.
The corresponding superpotential for an embedding such as the one in (\ref{hol-embedding-S5}) can be written as:
\beq
W\,=\,W_{{\cal N}=4}\,+\,\tilde Q_r\,\big[\,\sum_j a_j\,\Phi_j\,+\,m\,\big]\,Q^r\,\,,
\eeq
where the mass $m$ is related to the constant $\mu$ in (\ref{hol-embedding-S5}). 
Notice that since the embeddings are holomorphic, it is not possible to smear them in a way
in which the full $SO(6)$ isometry is realised.
After smearing over the embeddings (\ref{hol-embedding-S5}), one can recover, at most, $SU(3)\times U(1)$,
as will be seen directly from the dual solution.

In the case of the $AdS_5\times T^{1,1}$ background there are two classes of holomorphic embeddings which correspond to different types of flavors in the KW theory. In terms of the $z_i$ coordinates of (\ref{zetas}) the representative embedding of the first class is given by the equation $z_1=\mu$. This is the so-called Ouyang embedding 
\cite{Ouyang:2003df}, which has two branches in the massless limit $\mu=0$. 
In  each of these branches the D7-brane adds fundamental matter to one of the two nodes of the KW quiver and antifundamental matter to the second. The corresponding superpotential contains cubic couplings between the quark fields $q_i$ and $\tilde q_i$ ($i=1,2$) and the bifundamental fields $A_i$ and $B_i$.  For example, for the massless embedding $z_1=0$ the superpotential (\ref{KW-superpotential}) is modified as:
\be
W_{z_1=0} = W_{KW} + h_1{\tilde q_1}A_1q_2 + h_2 {\tilde q_2} B_1 q_1\,,
\ee
where, here and in the following, traces over color indices and sums over the $N_f$ flavor indices are implied. The second class of D7-brane embeddings is the one giving rise to non-chiral flavors, whose representative element is given by the equation 
$z_1-z_2=\mu$.  In this case every D7-brane adds fundamental and antifundamental flavor to one node of the KW quiver and the flavor mass terms do not break the classical symmetry of the massless theory. The corresponding superpotential contains only 
mass terms and
quartic couplings, namely:
\begin{equation}
W = W_{KW} +\hat h_1\, \tilde q_1 [A_1B_1-A_2B_2 ]q_1 + \hat h_2\,\tilde q_2 [B_1A_1-B_2A_2]q_2 + k_i\,(\tilde q_i q_i)^2 + m\,(\tilde q_i q_i)\,.
\label{KWsuperpquark}
\end{equation}

In order to develop our program and construct backreacted gravity solutions for smeared distributions of flavor branes  following ref. \cite{Benini:2006hh}, we should be able to find a family of equivalent embeddings for each type of configuration described above.  In the case of the $AdS_5\times S^5$ background eq. (\ref{hol-embedding-S5}) provides such a family. 
Notice that, even if each individual embedding of the form (\ref{hol-embedding-S5}) preserves ${\cal N}=2$,
which supersymmetries are preserved depends on the $a_i$'s. 
Nevertheless, one can check that all the holomorphic embeddings
 of the type (\ref{hol-embedding-S5})  are mutually supersymmetric and, 
 due to the holomorphic nature of the linear equation (\ref{hol-embedding-S5}), they preserve the same common four  supersymmetries (${\cal N}=1$)
for all values of the constants $a_i$.  Thus, we can use these constants to parameterize the family of different planes that constitute our continuous distribution of flavor branes.

In the case of the $AdS_5\times T^{1,1}$ background one can generalize the chiral embedding $z_1=\mu$ by acting with the $SU(2) \times SU(2)$ symmetry of the conifold. The corresponding family of embeddings takes the form:
\be
\sum_{i=1}^4 \alpha_i z_i = \mu\,, 
\label{general_chiral}
\ee
with the complex constants $\alpha_i$ spanning a conifold (up to overall complex rescalings)
\be
\alpha_1 \alpha_2 - \alpha_3 \alpha_4 =0\,.
\label{ouyang_eq}
\ee
Notice that embeddings like $z_1-z_2=\mu$ are not in this family. Indeed, the non-chiral embeddings $z_1-z_2=\mu$ can be generalized as:
\be
\bar p z_1 - p z_2 + \bar q z_3 + q z_4 = \mu\,,
\label{genembeddingks}
\ee
where $p, q$ span a unit 3-sphere, \ie\ they satisfy $|p|^2+|q|^2\,=1$. 

In spite of the differences among the cases presented above, the charge distribution generated by these families of embeddings can be written in a common form. The reason for this universality is the underlying Sasaki-Einstein structure. In order to illustrate this fact, let us consider the chiral embeddings (\ref{general_chiral}) in the case in which the mass parameter $\mu$ is zero. Without loss of generality we can rescale the $\alpha_i$ coefficients and fix $\alpha_1=1$. Then, (\ref{ouyang_eq}) fixes $\alpha_2=\alpha_3\alpha_4$ and, after using (\ref{zetas}), the massless embedding  equation
\be
 z_1 + \alpha_3 \alpha_4 z_2 + \alpha_3 z_3 + \alpha_4 z_4 =0\,,
\label{genmless}
\ee
nicely factorizes as:
\be
\left( \sin\frac{\theta_1}{2}
+ \alpha_3 e^{i \varphi_1} \cos \frac{\theta_1}{2}\right)
\left( \sin\frac{\theta_2}{2}
+ \alpha_4 e^{i \varphi_2} \cos \frac{\theta_2}{2}\right)=0\,.
\label{genmlessex}
\ee
Notice that the vanishing of each of the factors in (\ref{genmlessex}) determines a branch in which the branes sit at a fixed point of one of the two two-spheres parameterized by the angles $(\theta_i, \varphi_i)$. The constants $\alpha_3$ and $\alpha_4$ determine the particular point at which each brane is sitting in each $S^2$. Indeed, if $\xi^\alpha_1$ and $\xi^\alpha_2$ are systems of worldvolume coordinates for the D7-branes, these two branches can be written as:
\bear
&&\xi^\alpha_1 = \{x^0,x^1,x^2,x^3,r,\theta_2,\varphi_2,\psi\}\,\,, \qquad
\theta_1=\theta_1^*\,=\,\text{const.} \,\,,\qquad \varphi_1=\varphi_1^*=\text{const.}, \qquad\rc\rc
&&\xi^\alpha_2 = \{x^0,x^1,x^2,x^3,r,\theta_1,\varphi_1,\psi\} \,\,,\qquad
\theta_2=\theta_2^*=\text{const.}\,\,, \qquad \varphi_2=\varphi_2^*=\text{const.}
\label{branches}
\eear
\begin{figure}[ht]
\begin{center}
\includegraphics[width=0.95\textwidth]{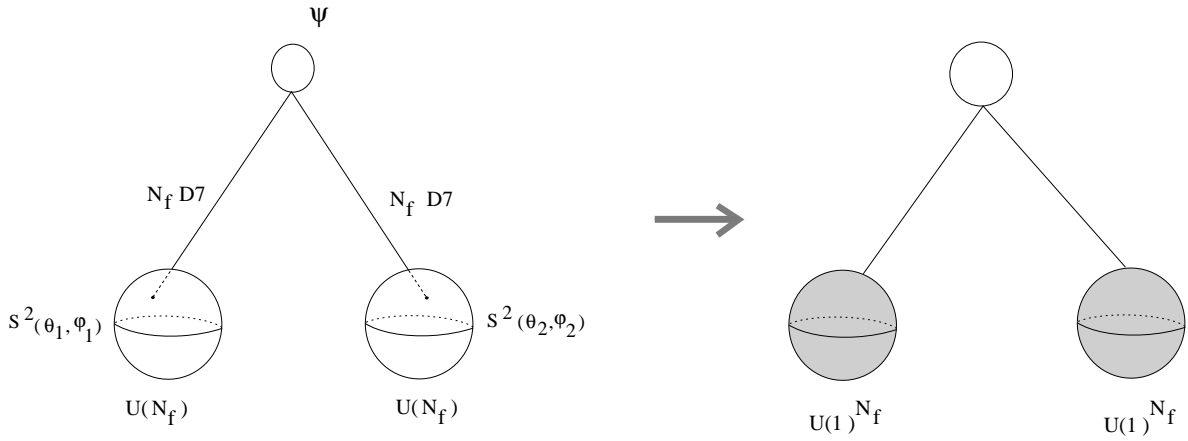}
\end{center}
\caption[smearingfig]{We see on the left side the two stacks of $N_f$ flavor-branes localized on
each of their respective $S^2$'s (they wrap the other $S^2$). The flavor
group is clearly $U(N_f) \times U(N_f)$. After the smearing on the right
side of the figure, this global symmetry is broken to $U(1)^{N_f - 1}\times
U(1)^{N_f -1} \times U(1)_B \times U(1)_A$. \label{smearingfig}} 
\end{figure}
In figure \ref{smearingfig} we have represented the two branches 
for the embedding (\ref{genmless}). 
From the field theory side, which particular embedding we choose determines the coupling 
between the associated
quarks and the bifundamentals. Roughly speaking, the contribution to the superpotential 
of an embedding determined by some $\alpha_3$, $\alpha_4$ is
$h_1{\tilde q_1}(A_1 + \alpha_4 A_2) q_2 + h_2 {\tilde q_2} 
(B_1 + \alpha_3 B_2) q_1$. Thus, when we {\it smear} and sum over all the possible $\alpha_3$ and $\alpha_4$,
both $SU(2)$'s (the one rotating the $A_i$'s and the one rotating the $B_i$'s) is effectively
recovered. Figure \ref{smearingfig} is the geometric interpretation of this effect.

It is straightforward to compute the charge density produced by this localized D7-brane configuration. Indeed, taking into account the contribution of the two branches, one gets:
\beq
\Omega^{loc}\,=\,\delta^{(2)}\,(\theta_1-\theta_1^{*}\,,\,\varphi_1-\varphi_1^{*})\,
d\theta_1\wedge d\varphi_1\,+\,
\delta^{(2)}\,(\theta_2-\theta_2^{*}\,,\,\varphi_2-\varphi_2^{*})\,
d\theta_2\wedge d\varphi_2\,\,.
\eeq
To produce an homogeneous configuration of $N_f$ D7-branes we should distribute in every branch the branes homogeneously along their transverse two-sphere. In the continuum limit $N_f\to\infty$ this procedure amounts to performing
 an integration over each $S^2$ with the corresponding volume element, namely:
\bear
&&\Omega\,=\, \Big[\,\int{N_f\over 4\pi}\,\sin\theta_1^*\,\,
\delta^{(2)}\,(\theta_1-\theta_1^{*}\,,\,\varphi_1-\varphi_1^{*})\,
d\theta_1^{*}\, d\varphi_1^{*}\,\Big] d\theta_1\wedge d\varphi_1\,+\,\rc\rc
&&\qquad
+\,\Big[\,\int{N_f\over 4\pi}\,\sin\theta_2^*\,\,
\delta^{(2)}\,(\theta_2-\theta_2^{*}\,,\,\varphi_2-\varphi_2^{*})\,
d\theta_2^{*}\, d\varphi_2^{*}\,\Big] d\theta_2\wedge d\varphi_2\,\,.
\label{Omega-micro}
\eear
The integrations over $\theta_i^{*}$ and $\varphi_i^{*}$ in (\ref{Omega-micro}) can be immediately performed, yielding the following expression for the smeared charge distribution of D7-branes:
\beq
\Omega\,=\,{N_f\over 4\pi}\,\,\Big(\,\sin\theta_1\,d\theta_1\wedge d\varphi_1\,+\,
\sin\theta_2\,d\theta_2\wedge d\varphi_2\,\Big)\,\,.
\label{Omega-massless-conifold}
\eeq
Notice that in (\ref{Omega-micro}) we have included  the normalization factor $N_f/ 4\pi$ in such a way that the resulting distribution densities $\sin\theta_i^*\,N_f/ 4\pi\,$ are normalized to $N_f$ when they are integrated over $S^2$. Notice that, as already pointed out above, the flavor symmetry of the smeared configuration is $U(1)^{N_f}$ rather than $U(N_f)$, since the branes are not placed on top of each other. Interestingly, a similar calculation for the embeddings (\ref{genembeddingks}) in the massless case $\mu=0$ gives rise to the same charge density for the smeared configuration as in (\ref{Omega-massless-conifold}) \cite{Benini:2007gx}. 
This is because the form of $\Omega$ in (\ref{Omega-massless-conifold}) is determined by the 
$SU(2)\times SU(2) \times Z_2$ global symmetry which we want to recover after smearing.

Actually, one can rewrite (\ref{Omega-massless-conifold}) in a form which can be easily generalized to any continuous family of equivalent D7-brane massless embeddings in an arbitrary Sasaki-Einstein manifold. Indeed, by using (\ref{Kahler-conifold}) one can rewrite the right-hand side of  (\ref{Omega-massless-conifold}) in terms of the K\"ahler form of $T^{1,1}$ as:
\beq
\Omega\,=\,-{3N_f\over 2\pi}\,\,J_{T^{1,1}}\,\,.
\label{Omega-T11}
\eeq
For an arbitrary Sasaki-Einstein space $X^5$, the expression (\ref{Omega-T11}) generalizes to:
\beq
g_s\,\Omega\,=\,-2\,Q_f\,J_{KE}\,\,,
\label{Omega-X5}
\eeq
where $Q_f$ is the following constant coefficient:
\beq
Q_f\,=\,{{\rm Vol}(X^3)\,g_s\,N_f\over 4\,{\rm Vol}(X^5)}\,\,.
\label{Qf}
\eeq
In (\ref{Qf}) $X^3$ is the compact submanifold of $X^5$ wrapped by the D7-brane in a massless embedding . Notice that in this case the D7-brane worldvolume along the space transverse to the color branes is always of the form $I\times X^3$, where $I$ is a non-compact interval along the holographic radial direction.  It is worth noticing that the factor ${\rm Vol}(X^5)/{\rm Vol}(X^3)$ appearing on the right-hand side of (\ref{Qf}) is just the volume transverse to any individual flavor brane,  over which we are distributing the D7-branes. For the massless chiral embeddings in the conifold one can readily check, after taking into account the contribution of both branches in (\ref{branches}), that 
${\rm Vol}(X^3)=16\pi^2/9$.  Since ${\rm Vol}(T^{1,1})=16\pi^3/27$, one can easily prove that (\ref{Omega-X5}) reduces to (\ref{Omega-T11}). In the case $X^5=S^5$ the three-manifold $X^3$ is just a unit $S^3$ and ${\rm Vol}(X^3)=2\pi^2$. Therefore, we obtain the following values of $Q_f$  for $X^5=S^5, T^{1,1}$:
\beq
Q_f\,=
\begin{cases}
{g_s\,N_f\over 2\pi}\,\, &{\rm for} \,\,\,\,X^5=S^5\,\,,\cr\cr
{3g_s\,N_f\over 4\pi}\,\,&{\rm for} \,\,\,\,X^5=T^{1,1}\,\,.
\end{cases}
\label{Qfs-cases}
\eeq
The charge density $\Omega$ determines the ansatz of $F_1$ in the backreacted geometry. Indeed, the WZ part of the D7-brane action contains a term in which the RR eight-form potential $C_8$ is coupled to the D7-brane worldvolume. The continuous limit for this term amounts to performing the following substitution:
\beq
S_{WZ} = T_7 \sum_{N_f} \int_{{\cal M}_8} \hat C_8\, \rightarrow\, T_7  \int_{{\cal M}_{10}} \Omega\wedge C_8\,\,,
\eeq
which leads to the following violation of the Bianchi identity for $F_1$:
\beq
dF_1 = - 2\kappa_{(10)}^2\,T_7	\Omega = - g_s \Omega\,\,.
\label{gen_bianchi}
\eeq
Taking into account the general expression of $\Omega$ for a massless embedding written in (\ref{Omega-X5}), as well as the relation (\ref{A-J-KE}) between the one-form $A_{KE}$ and the Sasaki-Einstein K\"ahler form $J_{KE}$, one is led to adopt \cite{Benini:2006hh} the following ansatz for $F_1$:
\beq
F_1\,=\,Q_f\,(d\tau+A_{KE})\,\,.
\label{F1-massless}
\eeq
A simple modification of  (\ref{F1-massless}) for $F_1$  allows us to extend the ansatz to the case in which the quarks are massive \cite{Benini:2006hh}. This modification corresponds to introducing a function $p(\rho)$ of the holographic coordinate $\rho$ and performing the substitution $Q_f\to Q_f\,p(\rho)$ in (\ref{F1-massless}). This function $p(\rho)$ encodes the effects of the non-trivial profile of the D7-branes. Indeed, when the quarks are massive the brane does not extend along the full range of the radial coordinate $\rho$ and, accordingly, $p(\rho)$ must vanish  for $\rho<\rho_q$, where $\rho=\rho_q$ is the radial location of the tip of the D7-brane. Moreover, the function $p(\rho)$ should approach the value $p=1$ 
when
 $\rho \gg \rho_q$ since in this region the quarks are 
 effectively massless. The form of the function $p(\rho)$ is not universal and depends on the particular embedding of the D7-brane. For the three embeddings in  the cases $X^{5}=S^{5}$ and $T^{1,1}$ discussed above, the expressions for $p(\rho)$ are given below. At this point let us simply notice that the charge density $\Omega$ is modified with respect to the massless case as:
\beq
g_s\,\Omega\,=\,-2\,p(\rho)\,Q_f\,J_{KE}\,-\,Q_f\,\dot p(\rho)\,d\rho\wedge (d\tau + A_{KE})\,\,,
\label{gsomega}
\eeq
where the dot denotes derivative with respect to the radial variable $\rho$.

\subsection{Backreacted ansatz and solution}
\label{sec23}

Let us now write an ansatz for the backreacted D3-D7 background for a generic Sasaki-Einstein space $X^5$ \cite{Benini:2006hh}. It is clear from the discussion of the previous subsection that, after performing the smearing, the resulting RR one-form $F_1$ introduces a distinction between the directions of the $U(1)$ fiber and  of the KE base of $X^5$.  Therefore, it seems clear that the effect of the smeared flavor branes on the metric should be  an internal deformation of the $X^5$ in the form of a relative squashing between the KE space and the Hopf 
fiber\footnote{Just in the case when $X^5$ is the sphere $S^5$, this squashing breaks part of the isometry
$SO(6) \to SU(3) \times U(1)$, where $SU(3)$ is the isometry of the K\"ahler-Einstein base $CP^2$.
}. 
Accordingly,  let us adopt the following ansatz  for the metric in Einstein frame:
\beq
ds^2\,=\,\Big[\,h(\rho)\,\Big]^{-{1\over 2}}\,dx^2_{1,3}\,+\,
\Big[\,h(\rho)\,\Big]^{{1\over 2}}\,\Big[\,
e^{2f(\rho)}d\rho^2\,+\,e^{2g(\rho)}\,ds^2_{KE}\,+\,e^{2f(\rho)}\,
\big(\,d\tau+A_{KE})^2\,\Big]\,,
\label{metrictzero}
\eeq
where $g(\rho)$ and $f(\rho)$ are the functions that implement the squashing mentioned above and the function multiplying $d\rho^2$ amounts to choosing a particular radial variable $\rho$ which is convenient for our purposes. Moreover, the dilaton will depend on $\rho$  and the RR forms $F_5$ and $F_1$ have the form:
\be
 \phi=\phi(\rho)\,, \qquad
 F_{5} = Q_c\,(1\,+\,*)\varepsilon(X^5)\,\,,\qquad F_1\,= Q_f\,p(\rho)\,
(d\tau+A_{KE})\,,
\label{dilaton-forms}
\ee
where $\varepsilon(X^5)$ is the volume element of $X^5$ and $Q_c$ and $Q_f$ are written in (\ref{AdS5X5back}) and (\ref{Qf}) respectively. The function $p(\rho)$, whose form depends on the D7-brane embedding, takes into account the effects of massive quarks, as explained above. 

Given the ansatz (\ref{metrictzero})-(\ref{dilaton-forms})  one can easily study the supersymmetric variations of the dilatino and gravitino in type IIB supergravity and find the corresponding first-order BPS equations, which ensure the preservation of four supersymmetries. The resulting equations are \cite{Benini:2006hh}:
\bear
&&\partial_\rho g= e^{2f-2g} \,\,,\qquad
\qquad\qquad\quad\ 
\qquad\partial_\rho f = 3-2 e^{2f-2g} - \frac{Q_f}{2}p(\rho)
e^\phi\,\,,\rc\rc
&&\partial_\rho \phi  = Q_f p(\rho) e^\phi \,\,,\qquad 
\qquad\qquad\qquad
\partial_\rho h=-Q_c e^{-4g}\,\,.
\label{BSPsysKW}
\eear
Remarkably, the system (\ref{BSPsysKW}) can be integrated analytically for any function 
$p(\rho)$. In order to present this solution, let us define the function $\eta(\rho)$ as follows:
\be
\eta(\rho) = Q_f e^\phi \int_{\rho_q}^\rho e^{6\xi} p(\xi) d\xi\,\,,
\label{eta}
\ee
where $\rho_q$ is the value of the radial coordinate at the tip of the flavor brane ($p(\rho<\rho_q)=0$).  Then, we can write down quite simple expressions for $f,g,\phi$, namely:
\bear
e^{-\phi} &=& e^{-\phi_*} -Q_f \int_{\rho_*}^\rho p(\xi) d\xi\,\,,\rc\rc
e^g &=& c_2 e^\rho e^{-\frac{\phi}{6}} \left(1+e^{-6\rho} (c_1 e^\phi + \eta) \right)^{\frac16}\,\,,\rc\rc
e^f &=& c_2 e^\rho e^{-\frac{\phi}{6}} \left(1+e^{-6\rho} (c_1 e^\phi + \eta) \right)^{-\frac13}\,\,,
\label{general-sol}
\eear
where we have introduced a reference scale $\rho_*$ and we have defined $\phi_*=\phi(\rho=\rho_*)$. Notice that the warp factor $h$ can be obtained as the integral of $e^{-4g}$, as follows from the last equation in the BPS system (\ref{BSPsysKW}). In (\ref{general-sol}) $c_1$ and $c_2$ are integration constants that we now fix. 
First, if we demand
IR regularity of the solution, we need $g=f$ when $\rho \leq \rho_q$. 
Since $\eta$ vanishes at 
$\rho= \rho_q$, we need $c_1=0$. Moreover, the  constant $c_2$ is just some overall scale and has no physical meaning. It is natural
to fix it to $\alpha'^{\frac12} e^{\frac{\phi_*}{6}}$ in order to give appropriate dimensions and to recover the usual expression for the metric when $Q_f=0$. Therefore, we find:
\bear
e^{\phi-\phi_*} &=& \frac{1}{1 -e^{\phi_*} Q_f \int_{\rho_*}^\rho p(\xi) d\xi }\,\,,\rc\rc
e^g &=& \sqrt{\alpha'}\,\, e^\rho e^{-\frac{\phi-\phi_*}{6}} \left(1+e^{-6\rho}  \eta \right)^{\frac16}\,\,,\rc\rc
e^f &=&\sqrt{ \alpha'} \,\,e^\rho e^{-\frac{\phi-\phi_*}{6}} \left(1+e^{-6\rho}  \eta \right)^{-\frac13}\,\,.
\label{general-sol-fixed}
\eear
Notice that, when $Q_f=0$, we recover the unflavored $AdS_5\times X^5$ background. Indeed, in this case $\phi=\phi_*$ and $\eta=0$ and, after performing the change of the radial variable $r=\sqrt{\alpha'}\,e^{\rho}$, we get that $e^g=e^f=r$ and the background (\ref{metrictzero})-(\ref{dilaton-forms}) coincides with the one written in (\ref{AdS5X5back}).

Let us now introduce the following parameter
\beq
\epsilon_*\,\equiv\,Q_f\,e^{\phi_*}\,\,,
\label{epsilon_*}
\eeq
which, as we will see in a while, controls the effects of quark loops in the  backreacted supergravity solution. Indeed, the gauge/gravity dictionary for the type of theories we are studying relates the exponential of the dilaton to the Yang-Mills coupling constant. For example, for the (flavored) ${\cal N}=4$ $SU(N_c)$ theory, dual to the deformed $AdS_5\times S^5$ background, the gauge coupling is
$g_{YM}^2=4\pi\,g_s e^\phi$ and, thus, the 't Hooft coupling at the scale $\rho_*$ is given by:
\beq
\lambda_*=4\pi\,g_s N_c e^{\phi_*}\,\,.
\label{lambda_*}
\eeq
For the  quiver theories that correspond to different $X^5$ geometries, the gauge groups are of the form $SU(N_c)^n$. Let us generalize a relation from the orbifold constructions
$\sum_i^n 4\pi g_{YM,i}^{-2}= (g_s e^\phi)^{-1}$ \cite{kw,Lawrence:1998ja}, 
and consider all the gauge couplings
$g_{YM,i}$ to be equal. Then $4\pi\,g_s N_c e^{\phi}$, strictly speaking, gives the 't Hooft coupling at each node of the quiver, divided by $n$. However, with an abuse of language we will simply refer to it as the 't Hooft coupling. Therefore, by using (\ref{lambda_*}) and the definition of $Q_f$ in
(\ref{Qf}) in (\ref{epsilon_*}), we get:
\be
\epsilon_* = \frac{Vol(X^3)}{16\pi\,Vol(X^5)}\lambda_*  \frac{N_f}{N_c}\,\,.
\label{epsstar}
\ee
In particular, when $X^5=S^5$ this relation becomes:
\beq
\epsilon_{*\,(X^5=S^5)} = \frac{1}{8\pi^2}\lambda_*  \frac{N_f}{N_c}\,\,.
\eeq
Notice that the fact that $\phi$ is not constant in the backreacted solution is simply a reflection, in the gauge theory dual,  of the running of the Yang-Mills coupling constant when matter is added to a conformal theory. 

In terms of $\epsilon_*$ the dilaton and the function $\eta$ of (\ref{eta})  take the form:
\beq
e^{\phi-\phi_*}\,=\,{1\over 1-\epsilon_*\,\int_{\rho_*}^{\rho}\,
p(\xi)\,d\xi}\,\,,
\qquad\qquad
\eta\,=\,\epsilon_*\,e^{\phi-\phi_*}\,\,
\int_{\rho_q}^{\rho}\,\,e^{6\xi}\,p(\xi)\,d\xi\,\,.
\label{dilaton-eta-epsilon*}
\eeq
One of the prominent features of our solution is the fact that, for $N_f\not=0$,  the dilaton blows up at some UV scale $\rho=\rho_{LP}$, determined by the condition:
\beq
\int_{\rho_*}^{\rho_{LP}}\,p(\xi)\,d\xi\,=\,\epsilon_*^{-1}\,\,.
\eeq
Clearly, in order to have a well-defined solution, we should restrict the value of the radial coordinate $\rho$ to the range $\rho<\rho_{LP}$. In view of the relation between the Yang-Mills coupling $g_{YM}$ and the dilaton ($g_{YM}^2\sim e^{\phi}$), the divergence of $\phi$ implies that $g_{YM}$ blows up at some UV scale, \ie\ that the gauge theory develops a Landau pole. This UV pathology of our solution was expected on physical grounds since the flavored gauge theory has positive beta function. Indeed, we will check below in some particular case that our solution reproduces the running of the coupling constant of the dual field theory.

\subsubsection{The supersymmetric solution with massless quarks}
\label{masslessKW}

We  now consider the particular case of massless quarks, which corresponds to taking the charge distribution given by (\ref{Omega-X5}) or simply  $p(\rho)=1$. In this case (\ref{dilaton-eta-epsilon*}) simply gives:
\beq
e^{\phi-\phi_*}\,=\,{1\over 1+\epsilon_*\,(\rho_*-\rho)}\,\,,\qquad\qquad
e^{-6\rho}\,\eta\,=\,{\epsilon_*\over 6}\,e^{\phi-\phi_*}\,\,,
\label{masslessSUSY-phi-eta}
\eeq
and the solution written in (\ref{general-sol-fixed}) reduces to:
\bear
e^{g}&=&\sqrt{\alpha'}\,\, e^\rho\,\left(1+\epsilon_* \Big(\frac16 +\rho_*-\rho\Big)\right)^\frac16\,\,,\rc\rc
e^{f}&=&\sqrt{\alpha'}\,\, e^\rho\,\left(1+\epsilon_* (\rho_*-\rho)\right)^\frac12
\left(1+\epsilon_*  \Big(\frac16 +\rho_*-\rho \Big)\right)^{-\frac13}\,\,,\rc\rc
\frac{dh}{d\rho} &=& -{Q_c \over\alpha'^{2}}\,\,\, e^{-4\rho} \left(1+\epsilon_*  \Big(\frac16 +\rho_*-\rho \Big)\right)^{-\frac23}\,\,.
\label{susysol}
\eear
Notice that the location of the Landau pole in this case is just  $\rho_{LP}=\rho_*+\epsilon_*^{-1}$ and that the range of $\rho$ for which the solution (\ref{susysol}) makes sense is $\rho\in (-\infty, \rho_{LP})$.  Moreover, by using the definition of $\epsilon_*$ in (\ref{epsilon_*}) one can immediately show that the dilaton can be written as:
\beq
e^{\phi(\rho)}\,=\,{1\over Q_f\,(\,\rho_{LP}-\rho\,)}\,\,.
\label{dilaton-rhoLP}
\eeq
Let us now verify that the dependence on $\rho$ of $\phi$ in (\ref{dilaton-rhoLP}) matches the expectations from field theory. For concreteness we will consider the case of ${\cal N}=4$ SYM with matter. Similar checks can be done in other cases (see  \cite{Benini:2006hh} for the case of the Klebanov-Witten theory). By using the relation between the Yang-Mills coupling and the dilaton discussed above, as well as the value of $Q_f$ for $X^5=S^5$ written in (\ref{Qfs-cases}), one gets:
\beq
{8\pi^2\over g^2_{YM}}\,=\,N_f\,(\rho_{LP}-\rho)\,\,.
\label{g-rho}
\eeq
In order to read the running of the coupling constant from (\ref{g-rho})  we must convert the dependence on the coordinate $\rho$ in (\ref{g-rho}) into a dependence on the energy scale of the corresponding dual field theory. At an energy scale $\mu$ much lower that the Landau pole scale $\Lambda_{UV}$ (\ie\ for $\rho<<\rho_{LP}$) the scaling dimensions of the adjoints and fundamentals take their canonical values and the  natural radius/energy relation is:
\beq
\rho_{LP}\,-\,\rho\,=\,\log{\Lambda_{UV}\over \mu}\,\,.
\eeq
Plugging this relation in (\ref{g-rho}) we get:
\beq
{8\pi^2\over g^2_{YM}}\,=\, N_f\,\log{\Lambda_{UV}\over \mu}\,.
\eeq
Therefore, we get a logarithmic scaling of the coupling of the type 
$8\pi^2/ g^2_{YM}=b\log E$, with $b=-N_f$, which matches the one-loop field theory 
result\footnote{In principle, one could object that, being strongly coupled, the matter fields
could get large anomalous dimensions making this result suspicious. However, 
since we
are performing a small perturbative (in $\epsilon_*$) deformation of the unflavored 
backgrounds, the anomalous dimensions for the fundamental multiplets cannot differ
much from their quenched values. For the $X^5=S^5$ case, those anomalous dimensions
vanish.
We thank F. Bigazzi for stressing this point to us.}
 in which one has that $b=3N_c-3N_c-N_f$. 

In order to have a clearer understanding of the deformation of the $AdS_5\times S^5$ metric introduced by the flavor, it is very convenient to change to a new radial variable $r$, which is defined by requiring that the warp factor takes the same form as in the unflavored case (see (\ref{AdS5X5back})):
\beq
h\,=\,{R^4\over r^4}\,\,,\qquad\qquad
R^4\,=\,{Q_c\over 4}\,\,.
\label{hR}
\eeq
By integrating the last equation in (\ref{susysol}) we can get $h(\rho)$ and thus $r(\rho)$.  We will perform this integration order by order in a series expansion in powers of $\epsilon_*$. The additive integration constant will be fixed by requiring that $r(\rho_*)\equiv r_*=\sqrt{\alpha'}\,e^{\rho_*}$. One gets:
\bear
r&=&\alpha'^{\frac12} e^\rho\Big[1+\frac{\epsilon_*}{72} \Big(
e^{4\rho-4\rho_*}-1 +12(\rho_*-\rho)\Big)
+\frac{5\epsilon_*^2}{10368}\Bigl(e^{8\rho-8\rho_*} + 6 e^{4\rho-4\rho_*}(3+4(\rho_*-\rho))\rc
&&- (19 -24 (\rho_*-\rho) +144 (\rho_*-\rho)^2) \Bigr) + O(\epsilon_*^3)
\Big]\,\,.
\eear
It is now straightforward to obtain the functions
 $f(r)$, $g(r)$ and the dilaton $\phi(r)$ as expansions in powers of $\epsilon_*$. Up to second order we have:
\bear
e^f&=&r\Big[1-\frac{\epsilon_*}{24}(1+\frac13 \frac{r^4}{r_*^4})+
\frac{\epsilon_*^2}{1152}\left(17-\frac{94}{9}\frac{r^4}{r_*^4}+\frac59\frac{r^8}{r_*^8}
-48 \log(\frac{r}{r_*})\right)+ O(\epsilon_*^3)\Big]\,\,,\rc
e^g&=&r\Big[1+\frac{\epsilon_*}{24}(1-\frac13 \frac{r^4}{r_*^4})+
\frac{\epsilon_*^2}{1152}\left(9-\frac{106}{9}\frac{r^4}{r_*^4}+\frac{5}{9}\frac{r^8}{r_*^8}
+48 \log(\frac{r}{r_*})\right)+ O(\epsilon_*^3)\Big]\,\,,\rc
\phi&=&\phi_*+ \epsilon_* \log\frac{r}{r_*} + \frac{\epsilon_*^2}{72}\left(1-\frac{r^4}{r_*^4}
+12 \log\frac{r}{r_*} + 36 \log^2\frac{r}{r_*}\right)+ O(\epsilon_*^3)\,\,.
\label{susymasslesssol}
\eear
Eq. (\ref{susymasslesssol}) neatly displays the effects of quark loops in the deformation of the geometry and in the running of the dilaton (the latter is related to the running of the gauge coupling, as argued above). It is important to point out that the deformed geometry has a curvature singularity at the origin $r=0$ (or $\rho=-\infty$)
(this singularity is similar to the one that appears at $r=0$ in a 2-dimensional manifold with
metric $ds^2 = dr^2 + r^2 (1+r) d\varphi^2$). In the same IR limit, $e^\phi$ runs to $0$.
 As argued in section \ref{heuristiczzz}, the appearance of this singularity can 
be intuitively understood as due to the fact that, in this massless case, all branes of our smeared distribution pass through the origin and the charge density is highly peaked at that point. 
From the field theory side, one can think of the singularity as appearing because the
theory becomes IR free, as first pointed out in \cite{Aharony:1998xz}.
Consistently with these interpretations and with the heuristic picture of
section \ref{heuristiczzz}, the IR singularity can be easily cured by giving a mass to the quarks
(it is a ``good" singularity according to the criteria of \cite{Maldacena:2000mw},\cite{Gubser:2000nd}). 
We will explicitly verify this fact in the next subsection.

\subsubsection{The supersymmetric solution with massive quarks}
\label{massiveKW}
 
 Let us now find the backreacted supergravity solution for massive quarks. As mentioned above, the function $p(\rho)$ entering the ansatz for $F_1$ in this case is not universal and depends on the particular 
 Sasaki-Einstein space $X^5$ and on the  family of D7-brane embeddings chosen. 
 For concreteness we first concentrate  in discussing the case in which $X^5=S^5$. The calculation of the function $p(\rho)$ in this case was performed in appendix C of \cite{Bigazzi:2009bk}. If 
$|\mu|=e^{\rho_q}$, one has:
 \beq
 p(\rho)\,=\,\Big[\,1\,-\,e^{2(\rho_q-\rho)}\,\Big]^2\,\Theta(\rho-\rho_q)\,\,.
 \eeq
When $\rho\ge \rho_q$ the function $p(\rho)$ is non-vanishing and one has to perform the integrals appearing in (\ref{dilaton-eta-epsilon*}) . These integrals can be straightforwardly done in analytic form and yield the result:
\bear
e^g&=&\sqrt{\alpha'}\,\, e^\rho\,\left(1+\epsilon_* \Big(\frac16 +\rho_*-\rho-\frac16 e^{6\rho_q-6\rho}
-\frac32 e^{2\rho_q-2\rho} + \frac34 e^{4\rho_q-4\rho}-\frac14 e^{4\rho_q-4\rho_*}+ e^{2\rho_q-2\rho_*}
\Big)\right)^\frac16\,\,,
\rc\rc
e^f&=&\sqrt{\alpha'}\,\, e^\rho\,\frac{\left(1+\epsilon_* (\rho_*-\rho-e^{2\rho_q-2\rho}+
\frac14 e^{4\rho_q-4\rho}
+e^{2\rho_q-2\rho_*}-
\frac14 e^{4\rho_q-4\rho_*}
)\right)^\frac12}
{\left(1+\epsilon_* (\frac16 +\rho_*-\rho-\frac16 e^{6\rho_q-6\rho}
-\frac32 e^{2\rho_q-2\rho} + \frac34 e^{4\rho_q-4\rho}-\frac14 e^{4\rho_q-4\rho_*}+ e^{2\rho_q-2\rho_*}
)\right)^{\frac13}}
\,\,,
\rc\rc
\phi&=& \phi_* -\log\Big[1+\epsilon_*\, \Big(\rho_*-\rho-e^{2\rho_q-2\rho}+
\frac14 e^{4\rho_q-4\rho}
+e^{2\rho_q-2\rho_*}-
\frac14 e^{4\rho_q-4\rho_*}\Big)\Big]\,\,.
\label{susysolmassive}
\eear
As a check, notice that setting $\rho_q \to -\infty$ one recovers the massless
solution  of (\ref{masslessSUSY-phi-eta}) and (\ref{susysol}). 
We still have to write the solution for $\rho<\rho_q$. In this case $p(\rho)$ vanishes and
the dilaton is constant and, by continuity, it has the value that can be read from 
(\ref{susysolmassive}) inserting $\rho=\rho_q$:
\be
\phi_{IR}=\phi_q = \phi_* -\log(1+\epsilon_*\, (\rho_*-\rho_q-
\frac34 
+e^{2\rho_q-2\rho_*}-
\frac14 e^{4\rho_q-4\rho_*}))\,\,.
\ee
The functions $f$ and $g$ are equal and given by:
\be
e^{f}=e^{g}=\alpha'^{\frac12}\, e^\rho e^{-\frac16(\Phi_{IR}-\Phi_*)}\,\,,\qquad\qquad
(\rho<\rho_q)\,\,.
\ee
It follows straightforwardly from these results that the IR singularity at $\rho=-\infty$ of the massless case disappears when $\mu\not=0$ since the background reduces to $AdS_5\times S^5$ for $\rho<\rho_q$. Moreover, one can verify that the metric is also regular at $\rho=\rho_q$. Thus, as stressed in section \ref{heuristiczzz}, the smearing of massive flavors allows one to smooth out IR singularities. 

Similar calculations can be done for the conifold theories. In this case we redefine the parameter $\mu$ of the embedding equations (\ref{general_chiral}) and (\ref{genembeddingks}) as $|\mu|=e^{3\rho_q\over 2}$. The charge distribution for the family 
 (\ref{general_chiral}) of chiral embeddings was obtained in ref. \cite{Bigazzi:2008zt}, with the result:
\beq
p(\rho)\,=\,\Big[\,1\,-\,e^{3(\rho_q-\rho)}\,\big(1+3\rho-3\rho_q\big)\,\Big]\,
\Theta(\rho-\rho_q)\,\,.
\eeq
Similarly, 
for the non-chiral embeddings (\ref{genembeddingks}) the function $p(\rho)$ is given by \cite{fkw2}:
\beq
p(\rho)\,=\,\Big[\,1\,-\,e^{3(\rho_q-\rho)}\,\Big]\,\Theta(\rho-\rho_q)\,\,.
\eeq
The corresponding supergravity solutions have been written down in refs. \cite{Bigazzi:2008zt,fkw2}. They are regular in the IR, much in the same way as in the $X^5=S^5$ case detailed above.

\subsection{Screening effects on the meson spectrum}
\label{sec:screening}

The holographic theories with flavors present mesonic excitations, meaning
that there exists a spectrum of colorless physical states created by operators
which are bilinears in the fundamental fields. 
They are associated to normalizable excitations of the flavor branes as was neatly explained
in the seminal paper
\cite{Kruczenski:2003be}. For a  review of this broad subject, see \cite{Erdmenger:2007cm}.
Notice that the notion of ``meson" we use here generalizes that used
in QCD. For instance, the ``mesons" of \cite{Kruczenski:2003be} are excitations of a non-confining
theory and in this case the dimensionful quantity that sets the meson masses is just the 
quark mass (divided by a power of the 't Hooft coupling), not a dynamically generated scale. 

In the present section, we review how the presence of unquenched flavors can affect the discrete
mesonic spectrum. Again, we will restrict ourselves to the smeared set-up and follow 
\cite{unquenchedmesons}. For discussions about screening effects on the spectrum 
 in cases with localized rather than
smeared  backreacting flavor branes, we refer the reader to \cite{Kirsch:2005uy,Erdmenger:2007cm}.
The effect of the smeared flavors on the hydrodynamical transport coefficients (in a finite temperature
setting) was studied in \cite{Bigazzi:2009bk,BCT}.
It is also worth mentioning that,
within the model we will introduce in section \ref{sec:D5D5},
the screening effects on the glueball spectrum have been recently analyzed in
\cite{Elander:2009bm}.

For the sake of briefness, we will just focus on an example and discuss a
 particular
 mesonic excitation in the
backreacted Klebanov-Witten model. 
The analysis and conclusions for different modes and/or different models should
be similar, see \cite{unquenchedmesons} for some other examples.
In particular, we will consider oscillations of a D7-brane which introduces massive
non-chiral flavor \cite{Kuperstein:2004hy}, and just look at the oscillation of
the gauge field that gives rise to a vector mode in the dual field theory.
Thus, we discuss the physics of a meson whose ``constituent quarks" are massive in the 
presence of many dynamical massless flavors.

We write the gauge field along the Minkowski directions as
$A_\mu=a_v(\rho) \,\xi_\mu \, e^{ikx}$, where $\xi_\mu$ is a constant transverse vector.
The equation describing this oscillation was found in \cite{unquenchedmesons},
building on the formalism introduced in \cite{Kuperstein:2004hy}. It reads:
\be
0=\partial_\rho\left(e^{2g-3\r}(e^{3\r}-e^{3\r_Q})\partial_\r a_v \right)
+ M_v^2  h\, e^{2g+2f}
\left(1+ e^{3\r_Q - 3\r} \Big(\frac34 e^{2g-2f} -1\Big) \right) a_v\,\,,
\label{eqforav}
\ee
where $M_v^2 = - k^2$, the constant $\rho_Q$ is the minimal value of $\rho$ reached by the D7-brane 
(related to the quark mass) and
$f,g,h$ are given in (\ref{susysol}).

Notice that for the meson excitation, we just use a D-brane probe, namely, we consider the oscillation
of a single brane in a fixed background. At first sight, this could look contradictory, since our aim is
always to take into account the effect of the flavor branes on the geometry. 
Then, one may think about considering coupled fluctuations of brane and background fields.
Nevertheless, this is not necessary: there are $N_f \gg 1$ flavor branes which are affecting the background
but when we consider a meson, only one (or two) out of this $N_f$ is fluctuating. Therefore, the effect of this
oscillation on the background is suppressed by $N_f^{-1}$ with respect to the contribution of the whole set
of branes and therefore is consistently negligible. On the other hand, the existence of the rest of flavors
- and the associated quantum effects on the spectrum - are taken into account through the deformation they have produced in
the background geometry.

Following the standard procedure \cite{Kruczenski:2003be,Erdmenger:2007cm}, a discrete tower of values for
$M_v$ should be found when selecting solutions of (\ref{eqforav}) which are regular and normalizable.
Since the background has a Landau pole, some prescription is needed for dealing with the UV limit
(large $\rho$). Technically, we will just require that the fluctuation $a_v$ vanishes at $\rho_*$. 
Physically, one can check that this is a consistent procedure if $\rho_Q << \rho_*$: we are interested
in some IR physics which should be independent of the UV completion of the theory at $\rho > \rho_*$, up
to corrections suppressed by powers of the UV scale. Namely, we neglect contributions of  order
$e^{\rho_Q - \rho_*}\sim \frac{\Lambda_{IR}}{\Lambda_{UV}}$ and check that the spectrum can be written
in terms of IR quantities. The value $\rho_*$ disappears from the final result, apart from the quoted
negligible corrections. See \cite{unquenchedmesons} for further discussions on the issue.
In section \ref{D3D7plasma}, we will see similar examples of how to deal with
the Landau pole. In that case, the IR scale, which has to be much smaller that the arbitrary UV scale
at $\rho_*$, is set by the temperature rather than by the quark mass.

In order to estimate the spectrum from (\ref{eqforav}), we can use a WKB approximation.
In \cite{unquenchedmesons}, using a formalism developed in \cite{Russo:1998by}, an
expression for the mass tower in terms of the principal quantum number $n$ was found.
Adapting notation to the one we are using here:
\be
M_v^{(n)} \approx \frac{\pi}{\Sigma_v}\,n\,\,,\qquad\quad
\Sigma_v \equiv \int_{\rho_Q}^{\rho_*} h^{\frac12} e^f \sqrt{\frac
{1+ e^{3\r_Q - 3\r} (\frac34 e^{2g-2f} -1)}{1 -  e^{3\r_Q - 3\r}}}d\rho\,\,.
\ee
Let us evaluate this integral at first order in $\epsilon_*$, by inserting
(\ref{susysol}). We still have to fix the additive constant for $h$, which we can 
do by requiring $h(\r_*)=0$ (in \cite{unquenchedmesons}  $h(\r_{LP})=0$ was used.
It is crucial that both prescriptions give the same result, up to  quantities
in $e^{\rho_Q - \rho_*}\sim \frac{\Lambda_{IR}}{\Lambda_{UV}}$).
We shift
to a coordinate $u$ such that $u\equiv e^{\rho-\rho_Q},\ u_* \equiv e^{\rho_*-\rho_Q}$.
Defining $\lambda_Q$ as the 't Hooft coupling (\ref{lambda_*}) at the quark mass scale, 
inserting the value of $Q_c$ in (\ref{AdS5X5back})
and defining
$T_Q\equiv \left(e^{\phi/2} \sqrt{-g_{tt} g_{xx}}\right)|_{\rho=\rho_Q}$ as the tension of a hypothetical
fundamental string stretched at constant $\rho=\rho_Q$, we can write the estimate for
the meson masses as:
\be
M_v^{(n)} \approx \frac{T_Q^{\frac12}}{\lambda_Q^{\frac14}}\frac{\pi \, n}
{\frac{3^{\frac34}}{4\sqrt{2\pi}} \int_1^{u_*}\left( \frac{\sqrt{4u^3-1}}{u^2 \sqrt{u^3-1}} 
+ \epsilon_Q \frac{7-4u^3 + 4 (4u^3-1)\log u}{24u^2 \sqrt{4u^3-1} \sqrt{u^3-1}} \right) du }\,\,.
\label{WKBest}
\ee
It is important to stress once again that this expression is written only in terms of 
IR quantities, once we discard terms of order $u_*^{-1} = e^{\rho_Q - \rho_*}$,
 namely contributions like $\log u_*$ have cancelled out. Notice that the upper limit of the
 integrals can be taken to infinity if we again insist in discarding $O(u_*^{-1})$ 
 contributions. The expression (\ref{WKBest}) is a neat example of how, even having a Landau pole,
 the holographic set-up is able to consistently obtain IR predictions, in exactly the same spirit as
in field theory. We can perform numerically the integration in (\ref{WKBest}), and
we get \cite{unquenchedmesons}:
\be
M_v^{(n)} \approx \frac{T_Q^{\frac12}}{\lambda_Q^{\frac14}} n\, (5.2 - 6 \times 10^{-3} \frac{N_f\lambda_Q}{N_c}
+ \dots )\,\,,
\label{nummesons}
\ee
where in order to substitute $\epsilon_Q$ as in (\ref{epsstar}) we have used
$Vol(X^3)= \frac{16}{9}\pi^2$, $Vol(X^5)= \frac{16}{27}\pi^3$. The expression (\ref{nummesons}) is the
result quoted in \cite{unquenchedmesons}, apart from a different factor of 2 in the definition
of $\lambda_Q$.

The lesson we want to take from this
section is that there is a well defined method to obtain the shift produced by the flavor quantum effects
on the meson spectrum (or, eventually, on any physical observable) as an expansion in the parameter 
$\epsilon \sim \lambda \,N_f / N_c$
which weighs the flavor loops. Heuristically, it may be useful to think 
of the computation leading to (\ref{nummesons}) as (partially)
a  strong coupling analogue of the Lamb shift corrections of QED.

\subsection{Black hole solutions: D3-D7 quark-gluon plasmas}
\label{D3D7plasma}

In this subsection we will review the results in \cite{Bigazzi:2009bk}.
We start by
showing how one can find a black hole 
solution which includes the backreaction effects due to massless quarks.  To perform this analysis
 it is more convenient to work with a new radial variable $\sigma$ such that the metric takes the form:
\be
ds^2 = h^{-\frac12}\left[-b\ dt^2 + d\vec{x}_3^{\,2}\right] + h^\frac12
\left[\, b\,e^{8g+2f}\, d\sigma^2 + e^{2g}\, ds_{KE}^2 + e^{2f}\, (d\tau + A_{KE})^2\,
\right]\,\,.
\label{black10dmetric}
\ee
Notice that we have introduced a new function $b$ which parameterizes the breaking of Lorentz invariance induced by the non-zero temperature $T$. All functions appearing in the metric (\ref{black10dmetric}), as well as the dilaton $\phi$, depend on $\sigma$. Moreover, the RR field strengths $F_5$ and $F_1$ are given by the ansatz (\ref{dilaton-forms}) with the function $p=1$. We remind the reader that fixing $p=1$ corresponds to taking massless quarks\footnote{In 
reference \cite{Bigazzi:2009bk},
the more involved case of massive quarks $p\neq 1$ 
was also discussed. An extra complication is the necessity of finding the non-trivial
D7-brane embeddings in the backreacted geometry.}.

In this non-supersymmetric case we will not have the first-order BPS equations at our disposal and we will have to deal directly  with the second-order equations of motion. Actually, since all the functions we need to compute depend only on the radial coordinate $\sigma$, it is possible to describe the system in terms of a one-dimensional effective action. One can find this effective action by directly substituting the ansatz in the gravity plus branes action (\ref{lagrangianzz}). One gets:
\bear
S_{eff}&=&\frac{Vol(X^5)V_{1,3}}{2\kappa_{10}^2}\int d\sigma \left(
-\frac12\frac{(\partial_\sigma h)^2}{h^2} +12\,(\partial_\sigma g)^2
 + 8 \,\,\partial_\sigma g\,\,\partial_\sigma f\,-\,
 \frac12 (\partial_\sigma \phi)^2+\right.\rc\rc
&+&\left.
\frac{(\partial_\sigma b)}{2b}
\left( \frac{\partial_\sigma h}{h}+
8 \,\partial_{\sigma}\,g
+ 2 \,\partial_{\sigma}\,f 
\right)+
\right. 
\rc\rc
&+&\left.
24\, b\,e^{2f+6g} - 4 b\,e^{4f+4g}\,-\,{Q_c^2\over 2} \frac{b}{h^2}\,  -\, 
{Q_f^2\over 2}\,b\, e^{2\phi+8g}
\, -\,
4Q_f\,b\,
e^{\phi+6g+2f}\right)\,\,.
\label{effeclagr}
\eear
In (\ref{effeclagr}) $V_{1,3}$  denotes the (infinite) integral over the Minkowski coordinates. 
The second derivatives coming from the Ricci scalar have been integrated by parts and, as is  customary,
only the angular part of $F_{5}$ is inserted in the $F_{5}^2$ term (otherwise
the $Q_c$ would not enter the effective action since, on-shell, 
$F_{5}^2=0$ due to the self-duality condition). 
The last  term in (\ref{effeclagr}),  proportional to $Q_f$,  comes from the DBI contribution in (\ref{BIWZaction}). Notice also that the WZ term does not enter
(\ref{effeclagr})  because it does not depend on the metric or the dilaton.

The equations of motion stemming from the effective action (\ref{effeclagr}) are:
\bear
\partial_\sigma^2(\log b)&=&0\,\,,\rc\rc
\partial_\sigma^2(\log h)&=&-Q_c^2 \frac{b}{h^2}\,\,,\rc\rc
\partial_\sigma^2\,g\,&=& -2 b \,e^{4g+4f} + 6 b\,e^{6g+2f} - Q_f \,b\, \,e^{\phi+6g+2f}\,\,,
\rc
\partial_\sigma^2\,f\,&=& 4b\,e^{4g+4f} - \frac{Q_f^2}{2}\,b\,e^{2\phi+8g}\,\,,
\rc\rc
\partial_\sigma^2\phi&=& Q_f^2\,b\,e^{2\phi+8g}\, +
4 Q_f\,b\, e^{\phi +6g+2f}\,\,.
\label{bhSeqsmassless}
\eear
It is straightforward to check that these equations solve the full set of Einstein equations
provided the following ``zero-energy'' constraint is also satisfied:
\bear
0&=& 
-\frac12\frac{(\partial_\sigma h)^2}{h^2} +12\,(\partial_\sigma g)^2
 + 8 \,\,\partial_\sigma g\,\,\partial_\sigma f\,-\,
 \frac12 (\partial_\sigma \phi)^2+\rc\rc
&+&
\frac{(\partial_\sigma b)}{2b}
\left( \frac{\partial_\sigma h}{h}+
8 \,\partial_{\sigma}\,g
+ 2 \,\partial_{\sigma}\,f 
\right)\,-\,24\, b\,e^{2f+6g} +4 b\,e^{4f+4g}\,+\,
\rc\rc
&+&
{Q_c^2\over 2} \frac{b}{h^2}\,  +\, 
{Q_f^2\over 2}\,b\, e^{2\phi+8g}
\, +\,
4Q_f\,b\,
e^{\phi+6g+2f}\,\,.
\label{constraintmassless}
\eear

This constraint can be thought of as the $\sigma\sigma$ component of the Einstein equations or, alternatively,
as the Gauss law from the gauge fixing of $g_{\sigma\sigma}$ in the ansatz 
(\ref{black10dmetric}).

Let us now find a solution of the system of equations (\ref{bhSeqsmassless}) and of  the ``zero-energy'' constraint (\ref{constraintmassless}) that  corresponds to a black hole for the backreacted D3-D7 system. We will require that such a solution is regular  at the horizon and tends to the supersymmetric one at energy scales much higher than the black hole temperature $T$.  Actually, the biggest advantage of the radial variable $\sigma$ introduced above is that the equations of motion of $b$ and $h$ in (\ref{bhSeqsmassless}) are decoupled from the ones corresponding to the other functions of the ansatz. These decoupled equations can be easily integrated in terms of an integration constant $r_h$ as follows:
\be
b=e^{4r_h^4 \,\sigma}
\,\,,\qquad\qquad
h=\frac{Q_c}{4r_h^4}(1-e^{4r_h^4 \,\sigma})\,\,.
\ee
where $\sigma\in (-\infty,0)$. We  now define a new radial coordinate $r$ by means of the relation:
\be
e^{4r_h^4\,\sigma}=1-\frac{r_h^4}{r^4}\,\,,\qquad\qquad
r\in (r_h, +\infty).
\label{rcoorddef}
\ee
Then, $b$ and $h$ take the form:
\beq
b\,=\,1\,-\,{r_h^4\over r^4}\,\,,\qquad\qquad
h\,=\,{R^4\over r^4}\,\,,
\label{bh-r}
\eeq
with $R^4=Q_c/4$. Notice that $h$ is given by the same expression as in (\ref{hR}). Moreover, it is clear  from (\ref{bh-r}) that $r=r_h$ is the position of the horizon and, thus, the extremal limit is attained by sending $r_h$ to zero. In terms of $r$ the metric takes the form:
\beq
ds^2\,=\,{r^2\over R^2}\,\,\Big[\,\,
\Big(1-{r_h^4\over r^4}\Big)\,dt^2\,+\,d\vec{x}_3^{\,2}\,\Big]\,+\,
{R^2\over r^2}\,\,{e^{8\hat g+2\hat f}\over 1-{r_h^4\over r^4}}\,(dr)^2\,+\,
R^2\,\big[\,e^{2\hat g}\,ds_{KE}^2\,+\,e^{2\hat f}\,
(d\tau+A_{KE})^2\,\big]\,\,,
\eeq
where we have defined the functions $\hat f$ and $\hat g$ as follows:
\beq
e^{\hat f}\,\equiv\,{e^{f}\over r}\,\,,\qquad\qquad
e^{\hat g}\,\equiv\,{e^{g}\over r}\,\,.
\eeq
In order to determine completely the background we still have to solve eqs. (\ref{bhSeqsmassless}) and (\ref{constraintmassless}) for $ f$, $g$ and the dilaton $\phi$. We will find this solution by introducing a reference UV scale $r_*$ and by expanding the functions in terms of the parameter $\epsilon_*$ defined in (\ref{epsilon_*}). We will impose that the functions $f$, $g$ and $\phi$ are equal to the SUSY ones of (\ref{susymasslesssol}) when the extremality parameter $r_h$ vanishes. Moreover, we will also require that these functions coincide with those in (\ref{susymasslesssol}) at the UV scale $r_*$. These conditions fix uniquely a solution of (\ref{bhSeqsmassless}) and (\ref{constraintmassless}). Up to second order in $\epsilon_*$ this solution is given by:
\bear
e^{\hat f}&=&1-\frac{\epsilon_*}{24}\Big(1+ \frac{2r^4-r_h^4}{6r_*^4-3r_h^4}\Big)+
\frac{\epsilon_*^2}{1152}\left(17-\frac{94}{9}\frac{2r^4-r_h^4}{2r_*^4-r_h^4}+
\frac59\frac{(2r^4-r_h^4)^2}{(2r_*^4-r_h^4)^2}+\right.\rc
&&\left.-\frac89 \frac{r_h^8 (r_*^4-r^4)}{(2r_*^4-r_h^4)^3}
-48 \log\Big(\frac{r}{r_*}\Big)\right)+ O(\epsilon_*^3)\,\,,\rc
e^{\hat g}&=&1+\frac{\epsilon_*}{24}\Big(1- \frac{2r^4-r_h^4}{6r_*^4-3r_h^4}\Big)+
\frac{\epsilon_*^2}{1152}\left(9-\frac{106}{9}\frac{2r^4-r_h^4}{2r_*^4-r_h^4}+
\frac59\frac{(2r^4-r_h^4)^2}{(2r_*^4-r_h^4)^2}+\right.\rc
&&\left.-\frac89 \frac{r_h^8 (r_*^4-r^4)}{(2r_*^4-r_h^4)^3}
+48 \log\Big(\frac{r}{r_*}\Big)\right)+ O(\epsilon_*^3)\,\,,\rc
\phi&=& \phi_*+\epsilon_* \log\frac{r}{r_*} + \frac{\epsilon_*^2}{72}\left(1-\frac{2r^4-r_h^4}{2r_*^4-r_h^4}
+12 \log\frac{r}{r_*} + 36 \log^2\frac{r}{r_*}+\right.\rc
&&\left.
+\frac92   
\left(Li_2\Big(1-\frac{r_h^4}{r^4}\Big)-Li_2\Big(1-\frac{r_h^4}{r_*^4}\Big)\right)
\right)+ O(\epsilon_*^3)\,\,,
\label{finiteTsol}
\eear
where $Li_2(u)\equiv \sum_{n=1}^\infty \frac{u^n}{n^2}$ is a polylogarithmic function. The functions written in (\ref{finiteTsol}) determine a geometry that is regular at the horizon $r=r_h$. In the next subsection we will study its thermodynamics and we will extract some consequences for the dual field theory with dynamical quarks at non-zero temperature. 

Let us conclude this section with some comments on the stability of our perturbative non-extremal solutions.
A possible way to check for the latter is to consider worldvolume fluctuations of a D7-brane in the setup. If, as in our cases, the brane corresponds to massless flavors, the related quasi-normal modes on the unflavored background all have frequencies with a negative imaginary part of the order of the temperature, 
signaling stability \cite{mesonmelting}, \cite{myers}. 
This result cannot be changed in the flavored case when a perturbative 
expansion in $\epsilon_*$ is done. Thus, in our regime of approximations, stability with respect to those fluctuations is guaranteed.

\subsubsection{Thermodynamics of the solution}
\label{sec:thermo}

In the previous subsections we have defined the backreacted background in terms of an arbitrary UV scale $r_*$ as an expansion in powers of  the parameter $\epsilon_*$ written in (\ref{epsstar}). This scale $r_*$ should be well separated from the Landau pole scale in order to avoid having the pathologies of the latter. Moreover, we are interested in analyzing the physical consequences of this background at energies much lower than the UV scale $r_*$.  In a black hole background dual to a quark gluon plasma the natural IR scale is the location $r_h$ of the horizon, which should be related to temperature $T$ of the plasma. Accordingly, we define $\epsilon_h$ as:
\be
\epsilon_h = \frac{\lambda_h\,Vol(X^3)}{16\pi\,Vol(X^5)}\frac{N_f}{N_c}\,\,,
\ee
where, in what follows,  the subscript $h$ means that the quantities are evaluated at the horizon $r=r_h$.
Thus, $\lambda_h$ is
naturally identified with the 't Hooft coupling at the scale of the plasma temperature.
We therefore have:
\be
\epsilon_h = \epsilon_* \frac{e^{\Phi_h}}{e^{\Phi_*}} = 
\epsilon_* + \epsilon_*^2 \log\frac{r_h}{r_*} + O(\epsilon_*^3)\,\,.
\label{epsirelation}
\ee
We will use this relation to recast the expansions in powers of $\epsilon_*$ as series in $\epsilon_h$. We will assume in what follows that $r_h$ is well below the reference UV scale $r_*$ to ensure that the IR physics does not depend on the UV completion of the theory. In a Wilsonian sense of the renormalization group flow,  the UV details of the theory should not affect the IR physics. Moreover, since, as we will see below, $r_h$ is proportional to the temperature
(at leading order), we have:
\be
\frac{d\epsilon_h}{dT}=\frac{\epsilon_h^2}{T}+ O(\epsilon_h^3)\,\,,
\label{epsilonrun}
\ee
and $T(d\lambda_h/dT)=\epsilon_h\lambda_h$ at leading order. These relations reflect the running of the gauge coupling induced by the dynamical flavors. 

The thermodynamic properties of the  black hole solution are determined by the metric functions at the horizon. After neglecting terms suppressed in powers of $r_h^4/r_*^4$, the values of the functions $\hat f$ and $\hat g$ at $r=r_h$ can be obtained from (\ref{finiteTsol}). One gets:
\be
e^{\hat f_h}\,=\, 1 - \frac{\epsilon_h}{24} + \frac{17}{1152} \epsilon_h^2 + O(\epsilon_h^3)\,\,,\qquad
e^{\hat g_h}\,= \,1 + \frac{\epsilon_h}{24} + \frac{1}{128} \epsilon_h^2+ O(\epsilon_h^3)\,\,.
\label{FSphi}
\ee

The black hole temperature can be obtained by requiring regularity of the euclideanized metric and by
identifying the temperature with the
inverse of the period of the euclideanized time.
A simple computation yields:
\be
T=\frac{2r_h}{2\pi R^2 \,e^{4\hat g_h+\hat f_h}}
=\frac{r_h}{\pi R^2}\left[1-\frac18 \epsilon_h
-\frac{13}{384}\epsilon_h^2 + O(\epsilon_h^3) \right]\,\,,
\label{temperature}
\ee
where in the last step we have used the values of $\hat f_h$ and $\hat g_h$ written in (\ref{FSphi}).

The entropy density $s$ is proportional to $A_8$, the volume at the horizon of the eight dimensional part of the space orthogonal to the $\hat t,r$ 
plane (where $\hat t$ is the Euclidean time), divided by the infinite constant volume of the 3d space directions
$V_3$. From the general form of the metric  we get that:
\be
s=\frac{2\pi\,\,A_8}{\kappa_{10}^2\,V_3} =
\frac{r_h^3 \,R^2 \,e^{4\hat g_h+\hat f_h}\,
Vol(X^5)}{2^5 \pi^6 g_s^2 \alpha'^4}
=\frac{\pi^5}{2Vol(X^5)} N_c^2 \frac{r_h^3}{\pi^3 R^6}\left[1+\frac18 \epsilon_h
+\frac{19}{384}\epsilon_h^2 + O(\epsilon_h^3) \right]\,\,,
\ee
which in terms of the temperature reads:
\be
s=
\frac{\pi^5}{2Vol(X^5)} N_c^2 T^3 \left[1+\frac12 \epsilon_h
+\frac{7}{24}\epsilon_h^2 + O(\epsilon_h^3) \right]\,\,.
\label{entropy}
\ee

As for the other thermodynamic quantities which will follow, the leading term of this formula is the well-known unflavored result.
The $O(\epsilon_h)$ term was already calculated in \cite{myers} with the probe brane technique,
in the $X^5=S^5$ case. Here we have re-obtained this result in a quite standard way, 
by computing the increase of the horizon area produced by the flavor branes. This can be considered
as a crosscheck of the validity of the whole construction.
Finally, the order $\epsilon_h^2$ was first obtained in \cite{Bigazzi:2009bk}.

The ADM energy of the solution  can be computed as an integral of the extrinsic curvature of the eight-dimensional hypersurface of constant time and radius. This calculation is straightforward and has been done in the appendix B of \cite{Bigazzi:2009bk}, with the result:
\be
\varepsilon= \frac{E_{ADM}}{V_3}=\frac38 \frac{\pi^5}{Vol(X^5)} N_c^2 T^4
\left[1+\frac12 \epsilon_h(T) + \frac13 \epsilon_h(T)^2 +O(\epsilon_h(T)^3)\right]\,\,.
\label{ADMresult}
\ee
Again, terms suppressed as powers of $\frac{r_h}{r_*}$ have been neglected. Moreover,
since in the following derivatives with respect to $T$ are going to be taken, we find
it convenient to make explicit that $\epsilon_h$ depends on $T$ (see (\ref{epsilonrun})).
Eq.  (\ref{ADMresult}) yields the energy density of the plasma and, thus, it
allows us  to study the full thermodynamics. Indeed, 
from (\ref{epsilonrun}) and (\ref{ADMresult}) we get immediately the heat capacity (density):
\be
{c_V} = \partial_T \varepsilon = \frac32 \frac{\pi^5}{Vol(X^5)}
N_c^2 T^3 \left[1+\frac12 \epsilon_h(T) +\frac{11}{24} \epsilon_h(T)^2+O(\epsilon_h(T)^3)\right] ~.
\label{heatcapacity}
\ee
The free energy density, and so (minus) the pressure, reads:
\be\label{freeen}
\frac{F}{V_3} =-p = \varepsilon - T s= -\frac18 \frac{\pi^5}{Vol(X^5)}
N_c^2 T^4 \left[1+\frac12 \epsilon_h(T) +\frac16 \epsilon_h(T)^2+O(\epsilon_h(T)^3)\right] ~.
\ee
Notice that, consistently, this satisfies the relation $s=\partial_T p$ (where it is crucial
to take (\ref{epsilonrun}) into account).
This result is confirmed by the direct computation of $F$ from the renormalized euclidean action (see, again, the appendix B of \cite{Bigazzi:2009bk}) and also by  the calculation in \cite{BCT} of the correlator of the tensorial mode in the hydrodynamical approximation.  

The speed of sound $v_s$ is obtained by combining (\ref{entropy}) and (\ref{heatcapacity}), namely: 
\be
v_s^2 = \frac{s}{c_V} = \frac13 \left[1-\frac{1}{6} \epsilon_h(T)^2+O(\epsilon_h(T)^3)\right]~.
\label{vs2}
\ee
Note that the correction to the speed of sound, which measures the deviation from conformality,   only appears at second order and that the sign of the correction is consistent with  the bound  $v_s^2\le \frac13$ conjectured in \cite{Cherman:2009tw}. 
It is also interesting to point out that the solution 
provides a direct measure of the breaking of conformality at second order, namely the so-called interaction measure, given by:
\be
\frac{\varepsilon-3p}{T^4}=\frac{\pi^5 N_c^2}{16Vol(X^5)}\epsilon_h(T)^2\,\,.
\ee

Let us now analyze the viscosity of the plasma predicted by the flavored black hole. Since we are not introducing higher derivatives of the metric in the action, the usual theorems apply and the shear viscosity $\eta$ saturates the Kovtun-Son-Starinets bound
\cite{Kovtun:2004de}, \ie\ $\eta/s=1/4\pi$. Therefore, the shear viscosity $\eta$ can be obtained by dividing by $4\pi$ the entropy density written in (\ref{entropy}). Again, the first-order term coincides with the one calculated in the probe approach in \cite{Mateos:2006yd} while the second-order result was first computed in \cite{Bigazzi:2009bk}. On the other hand, one can also compute the bulk viscosity $\zeta$ for this model, with the result \cite{BCT}:
\be
\zeta=\frac{\pi^4}{72 Vol(X^5)}N_c^2T^3 \left[\epsilon_h(T)^2+O(\epsilon_h(T)^3)\right]\, .
\label{bulk-vis}
\ee
Interestingly, the value of $\zeta$ written in  (\ref{bulk-vis}) saturates the bound proposed in \cite{buchel}:
\be
\frac{\zeta}{\eta}\geq 2\left(\frac13 - v_s^2\right)\, .
\label{bbound}
\ee
For the computation of other transport coefficients, we refer the reader to \cite{BCT}.

\subsubsection{Energy loss of partons}

One of the main phenomenological applications of holography is the analysis of the energy loss of a parton that moves through a quark-gluon plasma. One of the measures of this energy degradation is the so-called jet quenching parameter $\hat q$, which is a transport coefficient that measures the bremsstrahlung experienced by a parton probe due to its interactions with the quarks and gluons of the plasma \cite{Baier}.  At very high energy, and using  the eikonal approximation, the authors of \cite{liu} found a non-perturbative prescription for calculating $\hat q$ 
as the coefficient of $L^2$ in an almost  light-like Wilson loop with dimensions $L^{-}\gg L$. By using 
 the generic formula in \cite{aredmas} (and cutting the integral at
$r_*$), we can write:
\be
\hat q^{-1}= \pi\, \alpha' \int_{r_h}^{r_*} e^{-\frac{\phi}{2}}
\frac{\sqrt{g_{rr}}}{g_{xx}\sqrt{g_{xx}+g_{tt}}}dr
= \frac{\pi\, \alpha'\,R^4}{r_h^2} e^{-\frac{\phi_h}{2}}\int_{r_h}^{r_*} e^{-\frac{(\phi-\phi_h)}{2}}
\frac{e^{4\hat g+\hat f}}{\sqrt{r^4-r_h^4}}\,\, dr
\,\,.
\label{qhat-general}
\ee
The dilaton enters the formula because we are considering the Einstein frame metric. By plugging in (\ref{qhat-general}) the expressions of $\hat f$, $\hat g$ and $\phi$ written in (\ref{finiteTsol}), and by performing the corresponding integrals in $r$, one gets $\hat q$ as a power series expansion in $\epsilon_*$. In the course of this calculation we will neglect terms that are suppressed by powers of $r_h/r_*$ and we will write the result  in a series in $\epsilon_h$ rather than in $\epsilon_*$. 
In terms of  gauge theory quantities one gets \cite{Bigazzi:2009bk}:
\be
\hat q=\frac{\pi^3\sqrt{\lambda_h}\Gamma(\frac34)}{\sqrt{Vol(X^5)}\,\Gamma(\frac54)}T^3
\left[1 + \frac18(2+\pi) \epsilon_h + \gamma\, \epsilon_h^2+ O(\epsilon_h^3)
\right]\,\,,
\label{jetq}
\ee
where we have introduced a constant $\gamma$:
\be
\gamma= \frac{11}{96}+\frac{\pi}{48} + \frac{3\pi^2}{128} + \frac18 {\cal C}+
\frac{1}{48}\ {}_4 F_3\left(
1,1,1,\frac32;\frac74,2,2;1\right)\approx 0.5565\,\,,
\ee
with ${\cal C}\sim 0.91597$ being the Catalan constant.  Notice that the flavor correction to $\hat q$ is positive, \ie\ fundamentals enhance the jet quenching. Actually, eq.  (\ref{jetq}) can be used to estimate this enhancement in the
extrapolation to the realistic RHIC regime. Let us take 
$X^5=S^5$, $N_c=N_f=3$ and $\alpha_s\,=\,g^2_{YM}/4\pi\sim 1/2$. Then, $\lambda_{h}\sim 6\pi$ and $\epsilon_h\sim {N_f\over 4 \pi}\sim 0.24$. Using this value in (\ref{jetq}) we have that $\hat q$ is increased by $20\%$. For example, at 
$T=300$ MeV we obtain $\hat q \sim 5.3$ (Gev)$^2$/fm, to be compared with the value \cite{liu} $\hat q \sim 4.5$ (Gev)$^2$/fm of the unflavored plasma (the RHIC values are
$\hat q \sim 5-15$ (Gev)$^2$/fm). It is also interesting to rewrite (\ref{jetq}) in terms of the entropy density $s$. One gets:
\be
\hat q = c \,\sqrt{\lambda_h} \sqrt{\frac{s}{N_c^2}}T^\frac32
\left[1+\frac{\pi}{8}\epsilon_h+(\gamma-\frac{11}{96}-\frac{\pi}{32})\epsilon_h^2+
O(\epsilon_h^3) \right]\,\,,
\qquad
c=\sqrt{2\pi}\frac{\Gamma(\frac34)}{\Gamma(\frac54)}\,\,,
\ee
which shows a deviation (driven by $\epsilon_h$) from the general expression
 put forward in \cite{lrw2}. In this setting, the presence of fundamentals and the breaking of
 conformality are inevitably mingled. It would be interesting to have the dual of a conformal theory
 with fundamentals to check whether the conjecture \cite{lrw2} holds in such situation.
That was analyzed in \cite{Bertoldi:2007sf} in a non-critical string framework
and, interestingly, the result differs from \cite{lrw2}. The caveat is that
the model studied in \cite{Bertoldi:2007sf} suffers from the usual 
problem of gravity-like approaches to non-critical strings, namely
there are uncontrolled approximations.

Another way of characterizing the energy loss of a parton probe in the plasma is by modeling it as a macroscopic string attached to a probe flavor brane. The string is dragged by a constant force $f$ which keeps its velocity $v$ fixed and transfers to the parton energy and momentum, which is then lost in the plasma at a constant rate.  This energy loss is measured by the drag coefficient $\mu$, which relates the force $f$ and the parton momentum $p$: $f=\mu p$. To compute this drag force one can apply the general procedure of refs.  \cite{herzog1,gubserdrag,herzog2}. By using the Nambu-Goto action for a string in the black hole background one gets that the rate  of momentum transferred to the medium is given by \cite{herzog2}:
\be
\frac{dp}{dt} = -\frac{1}{2\pi\alpha'}C = -
\frac{r_h^2}{2\pi\alpha'\,R^2}e^\frac{\Phi(r_c)}{2}\frac{v}{\sqrt{1-v^2}}=-
\mu\,M_{kin}\,\frac{v}{\sqrt{1-v^2}}\,\,,
\label{drag1}
\ee
where $C$ is the  constant 
determined from the equation $g_{xx}(r_c)g_{tt}(r_c)+C^2=0$ with the point $r_c$ given by $g_{tt}(r_c)+g_{xx}(r_c)v^2=0$, namely $r_c=r_h(1-v^2)^{-\frac14}$.
In (\ref{drag1}) we have introduced, following  ref.  \cite{herzog1}, 
 the kinematical mass $M_{kin}$ such that $p=M_{kin}\frac{v}{\sqrt{1-v^2}}$.
From (\ref{drag1}), using  (\ref{finiteTsol}), (\ref{epsirelation}), (\ref{temperature}), 
we find:
\bear\label{muemme}
\mu\,M_{kin} &=& \frac{\pi^{5/2}}{2} \frac{\sqrt{\lambda_h}}{\sqrt{Vol(X^5)}}\, T^2 \left[
1+\frac18 (2-\log(1-v^2)) \epsilon_h +
\right. \\
&& \left.+
\frac{1}{384}\left[44-20 \log(1-v^2)
+9 \log ^2(1-v^2)+12 Li_2(v^2) \right] \epsilon_h^2 + O(\epsilon_h^3)\right]\,\,.\nonumber
\eear
As happens with the jet quenching, the energy loss (at fixed $v$) is enhanced by the presence of fundamental matter. The quantity $\mu\,M_{kin}$ grows when increasing the velocity. From (\ref{muemme}), formally,
it would diverge as $v\to 1$. However, 
(\ref{muemme}) is not applicable in that limit since 
we have to require $\epsilon_h \log(1-v^2) \ll 1$ for the expansions to be valid.

\subsection{A discussion on the range of validity}
\label{sec:rangeval}

We now discuss, following \cite{Bigazzi:2009bk}, the restriction on the physical
parameters needed for the deconfined flavored plasma 
solution to be physically meaningful. Before we go on, two comments
are in order: first, notice that, even if we will use here the plasma
temperature as the IR scale at
which the relevant physics takes place, this can be substituted by any other IR 
scale, depending on what one wants to study. Thus, for instance, when computing meson
masses at zero temperature as in section \ref{sec:screening}, the discussion below holds,
just taking into account the IR scale there is set by the quark mass.
Second, notice that the restriction
of small $\epsilon_*$ (which leads to $N_f \ll N_c$) 
of the D3-D7 case at hand comes from the existence of a Landau pole.
In holographic theories in which 
there is no Landau pole in the geometry (sections \ref{sec:D5D5}, \ref{sec:2+1}),
there is in principle no 
restriction to $N_f$. In particular, it is possible to consider in those theories
$N_f$ to be of the same order as $N_c$.

As we have already remarked, 
having a pathological UV means that
there must exist a separation of scales between IR and UV.
Concretely, there must exist a hierarchy, which
in terms of the $r$ radial coordinate reads:
\be
r_h \ll r_* \ll r_a < r_{LP}\,\,.
\ee
The quantity $r_h$ sets the scale of the plasma temperature $\frac{r_h}{R^2}\sim \Lambda_{IR}\sim T$,
which is the scale at which we want to analyze the physics.
The point $r_{LP}$ is where the dilaton diverges, signaling a Landau pole in the dual theory.
At a scale $r_a$ the string solution starts presenting subtler 
pathologies, whose
 discussion we delay until the end of this section.
Finally, $r_*$ sets an (arbitrary) UV cutoff scale $\frac{r_*}{R^2}\sim \Lambda_{UV}$.
The solution (\ref{finiteTsol}) will only be used for $r<r_*$.
In a Wilsonian sense of a renormalization group flow, the UV details
should not affect the IR physical predictions.
This feature is reflected in the fact that physical quantities do not depend (up to suppressed contributions)
on $r_*$ or functions evaluated at that point, but only on IR parameters.
Even if the precise value of $r_*$ is arbitrary, we have to make sure that it is possible
to choose it such that it is well above the IR scale (so that the UV
completion 
only has negligible effects on the IR physics) and well below the pathological $r_a, r_{LP}$ scales
(so that the solution we use is meaningful and the expansions do not break down).
To this we turn now.

Let us start by  computing the hierarchy between $r_*$ and $r_{LP}$. Since at
$r_*$ we can approximate the solution by the supersymmetric one, we can read the
position of the Landau pole from (\ref{susysol}). If we insert 
the approximate relation between radial coordinates
$r \approx \sqrt{\alpha'} e^{\rho}$, we find:
\be
\frac{r_*}{r_{LP}}\approx e^{-\frac {1} {\epsilon_*}} \ll 1\,\,,
\ee
as long as $\epsilon_* \ll 1$.

Moreover, one has to make sure that the Taylor expansions (\ref{finiteTsol})
are valid in the region $r_h<r<r_*$. This of course requires $\epsilon_* \ll 1$,
but also that $\epsilon_* \left|\log\frac{r_h}{r_*}\right|\ll 1$ (notice that the absolute value of the
logarithm can be big because $r_h \ll r_*$). This means that
$\frac{r_h}{r_*} \gg e^{-\frac {1} {\epsilon_*}}$.
On the other hand, when computing physical quantities in the previous sections, we always neglect
quantities suppressed as powers of $\frac{r_h}{r_*} \sim \frac{T}{\Lambda_{UV}}$.
 This is the order of magnitude
of the corrections due to the eventual UV completion of the theory at $r_*$. One has to make sure that
the corrections in $\epsilon_*$ we are keeping are much larger than the neglected ones,
namely $\epsilon_* \gg \frac{r_h}{r_*}$. In summary, we have the 
following hierarchy of
parameters (in the following, in order to avoid overly messy expressions, we insert the value of $\epsilon_*$ for the
$X^5=S^5$ case, remembering that for a generic $X^5$, its value is given by
(\ref{epsstar})):
\be
e^{-\frac {1} {\epsilon_*}}\sim 
e^{-\frac {8\pi^2\,N_c} {\lambda_*\,N_f}}
\ll \frac{r_h}{r_*} \sim \frac{T}{\Lambda_{UV}}
\ll \epsilon_* \sim \frac{\lambda_*\,N_f}{8\pi^2\,N_c} \ll 1\,\,.
\ee
As long as $\epsilon_* \sim \frac{\lambda_*\,N_f}{8\pi^2\,N_c} \ll 1$, there always
exists a range of $r_*$ such that this inequality is satisfied. Since we focus on the IR physics of the plasmas, at the scale set by their temperature, the actual physical constraint on the parameters will be
$
\frac{\lambda_h}{8\pi^2}\frac{N_f}{N_c} \ll 1\,\,,
$
which we have written in terms of  the coupling at the scale of the horizon,
$\lambda_h = \lambda_* (1 + O(\epsilon_*))$.

On top of this, we have to make sure that the SUGRA+DBI+WZ action we are using is
valid. As usual, the suppression of
closed string loops requires
$N_c \gg  1$ whereas the suppression of $\alpha'$-corrections
is guaranteed by $\lambda_h \gg 1$. 
We have written the D7-brane worldvolume contribution to the action as a sum of $N_f$ single brane contributions.
This is justified if the typical energy of a string connecting two different branes is large
(in $\alpha'$ units). Since the branes are distributed on a space whose size is controlled
by $R \sim \lambda_h^\frac14 \sqrt{\alpha'}$, we again need $\lambda_h \gg 1$. 
The smearing approximation will be good if
 the distribution of D7-branes on the transverse space is dense, i.e. $N_f\gg 1$.
The discussion up to now is summarized in the following validity regime:
\be\label{validity}
N_c \gg  1\,\,,\qquad  \lambda_h \gg 1\,\,,\qquad N_f \gg 1 \,\,,\qquad 
\epsilon_h= \frac{\lambda_h}{8\pi^2}\frac{N_f}{N_c} \ll 1\,\, .
\ee
Finally, we want to find the regime of parameters in which the flavor corrections are not
only {\it valid} but are also the {\it leading} ones. With this aim,
we ought to demand that the leading $\alpha'$-corrections to the supergravity action
(which typically scale as $\lambda_h^{-\frac32}$
due to terms of the type $\alpha'^3 {\cal R}^4$) are smaller than the flavor
ones, controlled by $\epsilon_h$, namely:
\be
\lambda_h^{-\frac32} \ll \epsilon_h\,\,.
\label{leadingcond}
\ee
Demanding that corrections to the D7-branes contributions
(for instance curvature corrections to the worldvolume action itself or corrections produced by possible
modifications of the brane embeddings due to  curvature
corrections to the background metric)
are subleading does not impose any further restriction. The reason is that their contribution
is typically of order $\epsilon_h \lambda_h^{-c}$ for some $c>0$ which is always subleading
with respect to $\epsilon_h$ as long as (\ref{validity}) is satisfied.

\subsubsection*{The holographic a-function}
As discussed in \cite{unquenchedmesons} and mentioned above, the string solution starts presenting
pathologies at a scale $r_a < r_{LP}$, where the
holographic $a$-function is singular. The utility of the solution for $r>r_a$ is doubtful, but
since we have only used the solutions up to $r_* \ll r_a$ in order to derive the IR physics, this
subtlety does not affect the physical results. We now briefly review the argument in
\cite{unquenchedmesons}, which used the backreacted Klebanov-Witten solution at zero temperature.
The qualitative picture holds for the rest of the cases addressed in the present section and for the case of section \ref{KS} too.

\begin{figure}
 \centering
\includegraphics[width=.4\textwidth]{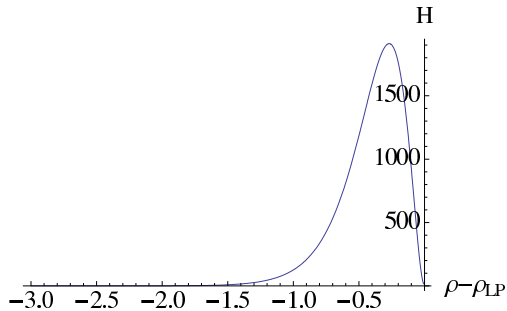}
\caption{ The function $H$ in the massless-flavored KW model at zero temperature.}
\label{Hflav}
\end{figure}

Let us start by considering the metric of a generic dimensional  reduction to 
five dimensions, giving a 5d Einstein frame metric of the form (the $u$ here is, obviously, a redefined
holographic coordinate, namely $u=u(r)$):
\be
ds_5^2 = H(u)^{1/3}\left[dx_{1,3}^2 + \beta(u) du^2\right]\,.
\ee
In standard set-ups, the function $H(u)^{1/6}$, which can be roughly identified with the dual field theory energy scale, monotonically varies with the radial coordinate. This is also required in order for the ``holographic $a$-function'' \cite{holaf}: 
\be
a(u)\sim \beta(u)^{3/2} H(u)^{7/2} [H'(u)]^{-3}\,,
\label{holadef}
\ee
to be finite.\footnote{The monotonicity of $H(u)$ also plays a crucial role in holographic computations of the entanglement entropy, see \cite{kkm}. The notations of that paper are used in the equations above.} 
Instead, the function $H(u)$ is not monotonic here: it increases with $u$ from zero up to a maximum
 at a point $u_a$ and then it decreases back to zero where $h$ vanishes\footnote{
 For the present discussion and in particular for figure \ref{Hflav}
 we will choose the additive integration constant of $h$ such that
 $h$ is zero at the Landau pole. The specific point $u_a$ (namely $r_a$) at which this UV pathology
 sets in depends on this choice. Again, we stress that the important point is the IR results do not
 depend on this choice (modulo suppressed contributions) as long as $r_a \gg r_*$. What we show here is
 that the integration constant can be naturally chosen such that this condition is satisfied.
 }. In the flavored supersymmetric KW case, the $H$ and $a$ functions 
 read:
 \be
 H(\rho)\,\sim\,h\,e^{2f+8g}\,\,,\qquad\quad
a(\rho) \,\sim\, h^{3/2}\,e^{3f}\, H^{7/2} [\partial_\rho H]^{-3}\,,
\label{holafkw}
\ee
where we have not written unimportant overall factors.
 A representative plot is given in figure \ref{Hflav}. 
 The non-monotonic behavior of $H$
 implies that the holographic $a$-function is singular and discontinuous at the ``$a$-scale".
From the plot, we see that  $r_a \sim e^{\r_a}$ is below, but not parametrically separated from $r_{LP}$.

\setcounter{equation}{0}

\section{A dual to ${\cal N}=1 $ SQCD-like theories}
\label{sec:D5D5}

In the following section, we will study a system that  in some sense is qualitatively different from those of the previous sections, though the procedure to deal with the addition of flavors is identical. The main qualitative difference will be that there need not be a hierarchical difference between the number of flavors and the number of colors. The case treated here will represent the addition of fundamental matter to
a field theory that is originally confining and four dimensional at low energies, but that gets some higher dimensional completion in the UV (in principle this allows one to extend the range of the radial coordinate to arbitrarily large values). Some of the qualitative changes that observables of a confining theory undergo when fundamentals are added will be discussed. The developments described  in the present section were
applied to model possible aspects that could appear in
physics beyond the standard model, as we will briefly mention below.

More concretely,   in this section, we will study a dual
to a version of ${\cal N}=1$ SQCD. The model is 
based 
on  D5-branes wrapped on two-cycles inside the resolved conifold-leading 
to a 
geometry related to the deformed conifold. We will first briefly present 
the model without flavors, then study the addition of flavors following 
the 
ideas described in the first sections of this paper: kappa symmetric 
embeddings, smearing, backreaction, system of BPS equations, particular
solutions to this system and finally present a set of checks that the 
correspondence we are proposing is valid and robust, we also explain some 
predictions about the field theory obtained with the string background.
\subsection{The model without flavors}
The proposal is to construct a dual to a field theory with minimal SUSY in 
four dimensions using wrapped branes. Ideas of this kind were first 
explored by Witten in the early days of AdS/CFT. In the paper 
\cite{Witten:1998zw}, Witten presented a model dual to a  version of 
 Yang-Mills theory 
(with an extra massless scalar and UV-completed by an infinite tower of 
massive vectors, scalars and fermions), by wrapping 
a set of $N_c$ D4 branes on a  
circle with SUSY-breaking boundary 
conditions.

The idea here is very similar, only that we will work with D5-branes and 
we will preserve some amount of SUSY. We will compactify the five branes 
in a very subtle way (involving a twisting of the 6-d 
theory) so that only four supercharges will be  preserved in the 
compactified theory for all energies \cite{Maldacena:2000yy}\footnote{The 
fact that a 
twisting procedure (see 
\cite{Witten:1994ev} for a very nice presentation of this idea) is at 
work, implies that even in the far UV, the theory is still preserving only 
four supercharges .}, in other words, the partial SUSY breaking is {\it 
not} due 
to the presence of relevant operators, like mass terms.
This kind of compactification of the six dimensional theory living on a  
stack of D5-branes (when the D5's wrap a two-cycle inside the resolved 
conifold) was well studied in various papers, see \cite{Bertolini:2003iv} 
for various reviews. We will follow mostly the detailed study of 
\cite{Andrews:2005cv}.

One can show that a very generic string background describing a stack of 
$N_c$ D5-branes wrapping a two cycle and preserving four supercharges 
includes a metric, RR-three form $F_3=dC_2$, a dilaton $\phi(\r)$ and  is 
given by,
\bea
ds^2 &=& \alpha' g_s  e^{ \frac{\phi(\rho)}{2}} \Big[dx_{1,3}^2 + 
e^{2k(\rho)}d\rho^2
+ e^{2 h(\rho)}
(d\theta^2 + \sin^2\theta d\varphi^2) +\nonumber\\
&+&\frac{e^{2 g(\rho)}}{4}
\left((\tilde{\omega}_1+a(\rho)d\theta)^2
+ (\tilde{\omega}_2-a(\rho)\sin\theta d\varphi)^2\right)
 + \frac{e^{2 k(\rho)}}{4}
(\tilde{\omega}_3 + \cos\theta d\varphi)^2\Big], \nonumber\\
F_{(3)} &=&\frac{g_s\alpha' N_c}{4}\Bigg[-(\tilde{\omega}_1+b(\rho) 
d\theta)\wedge
(\tilde{\omega}_2-b(\rho) \sin\theta d\varphi)\wedge
(\tilde{\omega}_3 + \cos\theta d\varphi)+\nonumber\\
& & b'd\rho \wedge (-d\theta \wedge \tilde{\omega}_1  +
\sin\theta d\varphi
\wedge
\tilde{\omega}_2) + (1-b(\rho)^2) \sin\theta d\theta\wedge d\varphi \wedge
\tilde{\omega}_3\Bigg],
\label{nonabmetric424}
\eea
where $\tilde\omega_i$ are the left-invariant forms of $SU(2)$
\bea\lab{su2}
&&\tilde{\omega}_1= \cos\psi d\tilde\theta\,+\,\sin\psi\sin\tilde\theta
d\tilde\varphi\,\,,\rc
&&\tilde{\omega}_2=-\sin\psi d\tilde\theta\,+\,\cos\psi\sin\tilde\theta
d\tilde\varphi\,\,,\rc
&&\tilde{\omega}_3=d\psi\,+\,\cos\tilde\theta d\tilde\varphi.
\eea
For convenience, below we will set the parameters $g_s=\alpha'=1$. The 
presence of the $N_c$ color D5-branes is indicated in $F_3$ that 
satisfies
the quantization condition:
\beq
\frac{1}{2\kappa_{(10)}^2}\int_{S^3} F_{(3)} = N_c T_5\,\,.
\label{quantcond}
\eeq
The $S^3$ on which we integrate is parameterized by $\tilde\theta,
\tilde\varphi,\psi$.

We then impose that a fraction of SUSY is preserved, hence we need to 
impose some projections on the Type IIB spinors and a set of BPS 
equations reflecting this arise (see the appendix B in the paper 
\cite{Casero:2006pt} for generous details). The BPS equations are 
non-linear, first order and  coupled for the functions of the background 
in eq. (\ref{nonabmetric424})-for details see appendix B in 
\cite{Casero:2006pt}. Certainly,  solving first order equations is simpler 
than solving the second order Einstein equations; nevertheless the BPS 
equations for the functions $(\phi,h,g,k,a,b)$
are non-linear and coupled, rendering the problem 
complicated.

It is technically convenient to make a `change of basis' to another set of 
functions, so that the BPS equations become first order and nonlinear (of 
course) but can be decoupled, and then solved independently. A change of 
variables that does the job partially was obtained in 
\cite{HoyosBadajoz:2008fw}. The change of basis is from the set of 
functions $[\phi,h,g,k,a,b]$ into the functions 
$[P,Q,\tau,\Phi,Y,\sigma]$. The map reads \cite{HoyosBadajoz:2008fw},
\beq
e^{2h}=\frac14\left(\frac{P^2-Q^2}{P\cosh\t-Q}\right),
\quad e^{2g}=P\cosh\t-Q,\quad e^{2k}=4Y,\quad
a=\frac{P\sinh\t}{P\cosh\t-Q},\quad b=\frac{\s}{N_c}.
\label{changevariables}
\eeq
and 
\beq
\Phi= (P^2-Q^2)\sqrt{Y} e^{2\phi}.
\eeq
As explained in detail in \cite{HoyosBadajoz:2008fw} (see section 3 of 
that paper), the BPS equations can be solved one by one for these new 
functions, obtaining:
\bea
& & Q(\rho)=(Q_o+ N_c)\cosh\tau + N_c (2\rho \cosh\tau -1),\nonumber\\
& & \sinh\tau(\rho)=\frac{1}{\sinh(2\rho-2\rho_o)},\quad 
\cosh\t(\r)=\coth(2\r-2\r_o),\nonumber\\
& & Y(\rho)=\frac{P'}{8},\nonumber\\
& & e^{4\phi}=\frac{e^{4\phi_o} \cosh(2\rho_o)^2}{(P^2-Q^2) Y
\sinh^2\tau},\nonumber\\
& & \sigma=\tanh\tau (Q+N_c)= \frac{(2N_c\rho + Q_o + N_c)}{\sinh(2\rho
-2\rho_o)}.
\label{BPSeqs}
\eea
Note that both $Q$ and the dilaton are given {\em algebraically} in terms 
of the rest of the functions
parametrizing the backgrounds. 
Here $Q_o$ and $\f_o$ are constants of 
integration and we have chosen the integration constant in the dilaton 
field equation $\phi_0$
such that it admits a smooth 
limit as $\r_o\to -\infty$ (this limit gives $\t=\s=0$ and so corresponds 
to what in the paper \cite{Casero:2007jj}  
were called  type {\bf A} backgrounds).

The function $P$ satisfies the following second order equation:
\beq
P'' + P'\Big(\frac{P'+Q'}{P-Q} +\frac{P'-Q'}{P+Q} - 4 \coth(2\rho-2\rho_o)
\Big)=0.
\label{master}
\eeq
We will refer to this equation as the `master' equation, since once we 
have a solution of (\ref{master}) all other functions 
are determined via (\ref{BPSeqs}).
\subsection{Some solutions}
There are many solutions to the  master equation 
(\ref{master}).
A very simple one is given by
\beq
P=2 N_c \r ,\;\;\; Q_o=-N_c.
\label{cv}
\eeq
Once processed back, one computes the functions in the original 
background of 
eq.  (\ref{nonabmetric424}) and one recovers an old  solution 
\cite{Chamseddine:1997nm}. To avoid nasty singular behaviors, in the 
following, we will choose the value 
of the integration constant $Q_o = - N_c$, so that the first term in the 
expression for $Q(\r)$, namely $(Q_o+N_c)$, vanishes\footnote{If we do not make 
this choice, the space ends before $\r=\r_0$, since $Q>P$ possibly giving 
place to 
geodesically incomplete spaces and a divergent dilaton. Hence, we will 
choose the term 
proportional to $\cosh\tau(\r)$ in $Q(\r)$ to vanish.}. 

Aside from the simple 
solution presented above, there are a variety of very 
interesting solutions. For example when the function $P(\r)$ is given near 
$\r=0$ by the following Taylor series:
\be
P= h_1 \r+ \frac{4 h_1}{15}\left(1-\frac{4 N_c^2}{h_1^2}\right)\r^3
+\frac{16 h_1}{525}\left(1-\frac{4N_c^2}{3h_1^2}
-\frac{32N_c^4}{3h_1^4}\right)\r^5+\co(\r^7),
\label{solutionnear0xx}\ee
where $h_1$ is again an arbitrary constant (notice that for $h_1=2N_c$ 
we get back to the solution in eq. (\ref{cv}); we will also assume that 
$h_1>2N_c$). It is 
interesting that 
this solution can be numerically connected in a smooth way with a solution 
for large values of the radial coordinate ($\r\to \infty$) that 
differs greatly from the linear behavior of the solution in eq.  (\ref{cv}). In 
this case, 
it is given by 
\beq
P\sim e^{\frac{4}{3}\r}\Big[c (1- \frac{8}{3} \r e^{-4\r}) +\frac{1}{64 
c}( 256 \r^2 +256 Q_o\r +144 N_c^2 + 64 Q_o^2 
)e^{-\frac{8}{3}\r}+O(e^{-4\r})       \Big]. 
\label{Pinfinity}
\eeq
These solutions were studied explicitly in section 8 of the paper 
\cite{Casero:2006pt} and have a variety of interesting applications that 
we will briefly mention in the following sections.
\subsubsection{An exact recursive solution}
There is one recursive way of obtaining solutions, described in 
\cite{Nunez:2008wi}, that basically uses the fact that the master equation 
(\ref{master}) can be written as (we choose here and in the following 
$\r_0=0$)
\be\label{master-c+}
\pa_\r\left(s(P^2-Q^2)P'\right)+4sP'QQ'=0,\;\;\;\; 
s(\r)=\sinh^2\t=\frac{1}{\sinh^2(2\r)}.
\ee
Integrating eq. (\ref{master-c+}) twice we obtain
\be\label{master-c+-integrated}
P^3-3Q^2P+6\int_{\r_2}^\r d\r'QQ'P+12\int_{\r_2}^\r d\r' 
s^{-1}\int^{\r'}_{\r_1} d\r'' s P'QQ'=c^3 R(\r)^3,
\ee
where
\be
R(\r)\equiv\left(\cos^3\a+\sin^3\a(\sinh(4\r)-4\r)\right)^{1/3},
\ee
being $(c,\alpha)$ the two integration constants of the master 
equation.

Following 
\cite{HoyosBadajoz:2008fw} we write $P$ in a formal expansion in 
inverse powers of $c$ - the integration constant encountered above - as:
\be\label{solution}
P=\sum_{n=0}^\infty c^{1-n} P_{1-n}.
\ee
Inserting this expansion in eq. (\ref{master-c+-integrated}) we obtain 
recursively
\bea
&&P_1=R,\NO\\
&&P_0=0,\NO\\
&&P_{-1}=-\frac13 P_1^{-2}\left(-3Q^2P_1+6\int_{\r_2}^\r d\r' QQ' 
P_1+12\int_{\r_2}^\r d\r' s^{-1}\int^{\r'}_{\r_1} d\r'' 
sQQ'P_1'\right),\NO\\\NO\\
&&P_{-2}=0,\NO\\
&&P_{-n-2}=-\frac13P_1^{-2}\left\{\sum_{m=1}^{n+2}\left(2P_1P_{1-m}P_{m-n-2}+\sum_{k=1}^{n-m+3}P_{1-m}P_{1-k}
P_{m+k-n-2}\right)-3Q^2P_{-n}\right.\NO\\
&&\left.+6\int_{\r_2}^\r d\r' QQ' P_{-n}+12\int_{\r_2}^\r d\r' 
s^{-1}\int^{\r'}_{\r_1} d\r'' sQQ'P_{-n}'\right\},\quad n\geq 1.
\eea
It follows by induction that $P_k=0$ for all even $k$. The large $\rho$ 
expansion of these solutions coincide with that described in 
eq. (\ref{Pinfinity}) Once again, solutions written in this form have 
interesting applications to the physics of cascading quivers on the 
baryonic branch \cite{Maldacena:2009mw}, \cite{gmnp}.
We will not study the physics encoded in the solutions 
described above, suggesting the interested reader to consult the papers 
\cite{Bertolini:2003iv}, 
\cite{Casero:2006pt}, \cite{Nunez:2008wi}.

There is another set of solutions, proposed in \cite{Nunez:2008wi}
and whose physics content was developed in \cite{Nunez:2009da},\cite{Elander:2009pk}
that correspond to what are called ``walking solutions''. The idea here
is to construct string backgrounds such that the dual QFT has a gauge 
coupling with very slow running 
(or ``walking'' coupling). See the papers  
\cite{Nunez:2008wi},\cite{Nunez:2009da},\cite{Elander:2009pk} for detailed explanations on 
the physical implications of these solutions.
\subsubsection{A comment about the dual field theory}\label{qftwof}
The system of wrapped D5-branes has a field theory realized on its 
worldvolume, whose dual background and various 
solutions were described above. The field theory is a version of 
minimally SUSY Yang-Mills. Again, some UV completion takes over at 
high energies \footnote{We are not saying that Super-Yang-Mills needs a 
UV completion, just that the system of D5-branes realizes a theory with 
these characteristics.}. The field theory is minimally SUSY (${\cal N}=1$) and its 
{\it perturbative} spectrum, aside from a massless vector multiplet contains a 
tower of massive vector and chiral multiplets. A careful study of the perturbative dual 
field theory obtained by compactification and twisting of the 
six-dimensional theory living on (unwrapped) D5-branes was done in 
\cite{Andrews:2005cv}. In that paper, the degeneracies and masses of the 
(perturbative) states in the tower mentioned above are given. More 
interestingly, the authors of \cite{Andrews:2005cv} showed that the theory 
is equivalent to ${\cal N}=1^*$ Yang-Mills in a particular Higgs vacuum, where 
the 
extra dimensions appear by deconstruction. In this sense, we will think of 
the theory without flavors either as a six-dimensional theory compactified 
or as a four-dimensional theory with an infinite set of fields.

For our purposes, it will be enough to use the fact  that the lagrangian of the 
field theory reads
\beq
L= Tr[-\frac{1}{4}F_{\mu\nu}^2 -i \bar{\lambda}\gamma^\mu D_\mu \lambda + 
L(\Phi_k, W_k,W)]\,\,,
\eeq
where $\Phi_k$ and $W_k$  represent the infinite number of 
massive chiral and vector multiplets and $W$ denotes the massless vector 
multiplet. The term $L(\Phi_k, W_k,W)$, represents all the kinetic terms 
and interactions that can be deduced from \cite{Andrews:2005cv}. More 
comments about this field theory can be found in the appendix A of 
\cite{HoyosBadajoz:2008fw}.

In what follows, we will summarize the procedure of adding flavors to this 
field theory. The flavor branes, in this particular case are D5-branes.
\subsection{Addition of flavors}
The study of supersymmetric embeddings in backgrounds of the form of 
eq. (\ref{nonabmetric424}), more precisely for the solution given in 
eq.  (\ref{cv}) was initiated in the paper \cite{Nunez:2003cf}. There the 
eigenspinors of the kappa symmetry matrix were found to be the spinors 
preserved by the background for a variety of D5-brane embeddings. For the 
purposes of this review, we will focus on the ``cylinder embeddings'' 
described in section 6.3 of \cite{Nunez:2003cf} and in more detail in 
section 6.5.3 of the third paper in reference \cite{Bertolini:2003iv}.
In this case the flavor D5-branes are extended along the $R^{1,3}$ 
Minkowski directions, on the radial direction $\r$ and also wrap the 
R-symmetry direction $\psi$. Intuitively, the flavor branes are localized 
in the directions $(\theta,\tilde{\theta},\varphi,\tilde{\varphi})$, but 
interestingly enough, any constant value of these coordinates ensures that 
we have a kappa symmetric configuration. This is a very important fact, as 
we can put one flavor brane ``at each point'' of the four manifold 
$\Sigma[\theta,\tilde{\theta},\varphi,\tilde{\varphi}]$ and still have a 
SUSY configuration.

This is precisely what we will take advantage of when smearing. Let us see 
this in more detail: if, as discussed in the first section, we 
write the action describing the closed strings (IIB) and the open strings 
(BIWZ), we will have
\bea
& & S=\frac{1}{2\kappa_{(10)}^2}
\int d^{10}x \sqrt{-g} \left[R-\frac12 (\partial_\mu \phi)
(\partial^\mu \phi)-\frac{1}{12}e^{\phi}F_{(3)}^2\right]\qquad\qquad
\nonumber\\\rc
&& \qquad\qquad-T_5 \sum^{N_f} 
\int_{{\cal M}_6} d^6x e^{\frac{\phi}{2}}\sqrt {-\hat g_{(6)}}\,+\, T_5 \sum^{N_f} 
 \int_{{\cal M}_6} P[C_{6}]\,\,,
\label{gravaction}
\eea
where the integrals are taken over the six-dimensional
worldvolume of the flavor
branes ${\cal M}_6$, 
and $\hat g_{(6)}$ stands for the determinant of the pull-back
of the metric in such a worldvolume.

As discussed in previous sections, we then think of the $N_f \to \infty$ branes as being
homogeneously
smeared along the four transverse directions parameterized by the coordinates
$\theta, \varphi$ and $\tilde \theta, \tilde \varphi$. 
The smearing erases
the dependence on the angular coordinates and makes it 
possible to consider an ansatz with functions only depending on $r$,
enormously simplifying  computations. One has:
\ba
-T_5 \sum^{N_f} 
\int_{{\cal M}_6} d^6x e^{\frac{\phi}{2}} \sqrt {-\hat g_{(6)}} &\to&
-\frac{T_5 N_f}{(4\pi)^2} \int d^{10}x
\sin\theta \sin \tilde \theta e^{\frac{\phi}{2}} \sqrt {-\hat g_{(6)}}\,\,,
\label{BIterm}\\
T_5 \sum^{N_f} \int_{{\cal M}_6} P[C_{6}] &\to& \frac{T_5 N_f}{(4\pi)^2}
\int Vol({\cal Y}_4) \wedge C_{(6)}\,\,,
\label{WZterm}
\ea
where we have defined $Vol({\cal Y}_4)=\sin \theta \sin \tilde \theta
d\theta \wedge d\varphi \wedge d\tilde\theta \wedge d\tilde\varphi$
and the new integrals span the full space-time.
We will need the expressions (with the choice explained 
above $\alpha'=g_s=1$).
\beq
T_5=\frac{1}{(2\pi)^5}\,\, ,\qquad\qquad
{2\kappa_{(10)}^2} ={ (2\pi)^7 }.
\label{T5value}
\eeq
From here, we will have a set of BPS equations describing the dynamics of 
this open-closed string system.
The same change of basis with the  purposes  described around 
eq. (\ref{changevariables}) can be performed - see 
\cite{HoyosBadajoz:2008fw} for details. The solution
in this case is dependent on the number of flavor branes $N_f$ and reads 
(reinstating momentarily the integration constant $\r_o$):
\be\label{tau}
\sinh\t=\frac{1}{\sinh(2(\r-\r_o))},
\ee
for the function $\tau$, while for $Q,\F$ we have,
\be\label{Q}
Q=\left(Q_o+\frac{2N_c-N_f}{2}\right)\cosh\t+\frac{2N_c-N_f}{2}\left(2\r\cosh\t-1\right),
\ee
\be\label{dilaton}
e^{4(\f-\f_o)}=\frac{\cosh^2(2\r_o)}{(P^2-Q^2)Y\sinh^2\t}.
\ee
In the case with flavors, like in the unflavored case previously discussed,  
both $Q$ and the dilaton are given {\em algebraically} in terms 
of the rest of the functions
parametrizing the backgrounds. 
As before,  $\r_o$, $Q_o$ and $\f_o$ are constants of 
integration and we have chosen the integration constant in (\ref{dilaton})
such that it admits a smooth 
limit as $\r_o\to -\infty$ (this limit gives $\t=\s=0$ and so corresponds 
to the type {\bf A} backgrounds).
The function $Y$ is determined in terms of $P$ as
\be\label{Y}
Y=\frac18(P'+N_f),
\ee
while the only remaining unknown, the function $P$, then satisfies the 
{\it new} decoupled second order master equation
\be\label{masterf}
P''+(P'+N_f)\left(\frac{P'+Q'+2N_f}{P-Q}+\frac{P'-Q'+2N_f}{P+Q}-4\coth(2\r-2\r_o)\right)=0.
\ee
One can redefine $P(\r)=N_c p(\r)$ and factor out $N_c $ from the master 
equation.
We will mention some solutions to eq. (\ref{masterf}), that explicitly 
include the quotient 
$x=\frac{N_f}{N_c}$, hence the solutions will  capture the nontrivial 
physics of the fields transforming in the fundamental representation of 
the gauge group.
\subsection{Study of solutions}
We now describe various solutions to the ``flavored'' master equation (\ref{masterf}).
Some solutions were found exactly, for the particular relation $N_f=2N_c$ while some other 
are known as asymptotic expansions, near the UV (large $\r$) and the IR (small $\r$). In 
these latter cases, a smooth numerical interpolation can be found.
\subsubsection{Exact solutions for $N_f=2N_c$} 
One can find some {\it exact} solutions for the 
case $N_f=2N_c$ or $x=2$. 
They were first discussed in the papers 
\cite{Casero:2006pt},\cite{HoyosBadajoz:2008fw}.

For $N_f=2N_c$ an exact type {\bf A} ($\r_o\to -\infty$) solution of  eq. (\ref{masterf}) is:
\be\label{self-dual-1}
P=N_c+\sqrt{N_c^2+Q_o^2}, \quad Q=Q_o\equiv 
4N_c\frac{(2-\x)}{\x(4-\x)},\quad 0<\x<4\,\,.
\ee
Another solution  with a qualitatively different UV behavior is: 
\be\label{self-dual-2}
P=\frac{9 N_c}{4}+c e^{4\r/3}, 
\quad c>0,\quad Q=\pm\frac{3N_c}{4}\,\,.
\ee 
One can check that in these solutions the radial coordinate moves all over 
the real axis and that for $\r\to -\infty$ the solutions take the same 
form, but as anticipated above, differ substantially in the far UV, for 
$\r\to \infty$. Also, for the case $N_f=2N_c$ the papers 
\cite{Caceres:2007mu} discuss  some extra solutions 
apart from the ones
mentioned, including, 
interestingly, the generalization to near-extremal solutions\footnote{
The metric for the simplest non-extremal solution can be written in
terms of a constant $\xi$ and a function ${\cal F}=1- (z_h/z)^4$ as:
\bear
ds^2&=&e^\frac{\phi_0}{2}\,z
\Big[-{\cal F}dt^2 + dx_1^2+ dx_2^2+ dx_3^2+ N_c \Big(
\frac{4}{z^2}{\cal F}^{-1}
dz^2 +\frac{1}{\xi} ( d\th^2+\sin^2\th d\varphi^2)+\rc
&&+\frac{1}{4-\xi}(d\tilde\th^2+\sin^2\tilde\th d\tilde\varphi^2)
+\frac14 (d\psi + \cos \theta d\varphi + \cos \tilde \th d\tilde\varphi)^2
\Big)\Big]\,\,.
\label{bhflav}
\eear
The solution also contains non-trivial RR $F_{(3)}$ and dilaton, see
\cite{Casero:2006pt} for details.
Different features of this black hole solution have been analysed in \cite{Bertoldi:2007sf},
\cite{talaveraetal}. An important remark is that the
theory is in a Hagedorn phase and, indeed, the temperature  coincides with the
Hagedorn temperature of Little String Theory. 
For this reason, this solution is a bit problematic for studying the effect of
quarks in a field theory plasma, 
unlike the finite temperature
solution of section \ref{D3D7plasma}. 
}.
\subsubsection{Asymptotic expansions of generic solutions}
Other solutions of interest have been discussed in the papers 
\cite{Casero:2006pt},
\cite{Casero:2007jj},\cite{HoyosBadajoz:2008fw}. We will summarize the 
results, but suggest to the interested reader to go over those papers for 
details of all the metric functions.

In the UV (for $\r\to\infty$), two possible asymptotics were found, that 
were called Class I and Class II in \cite{HoyosBadajoz:2008fw}. 
Table \ref{UV-classes} summarizes the situation.
\begin{table}
\begin{center}
\begin{tabular}{|c||l|l|}
\hline
$N_f$ & I & II \\
\hline\hline &&\\
$<2N_c$ & $P\sim Q\sim|2N_c-N_f|\r$ \mbox{}&  \\ 
	& $e^{2h}\sim\frac12\left(2N_c-N_f\right) \rho $			
&\\ 
	& $e^{2g}\sim N_c$			&\\ 
	& $Y\sim \frac{N_c}{4}$			&\\ 
	& $e^{4(\f-\f_o)}\sim \frac{e^{4 \left(\rho -\rho 
_o\right)}\sinh^2\left(2 \rho _o\right)}
{2N_c^2\left(2 N_c- N_f\right) \rho }$			&\\ 
	& $a\sim \frac{2}{N_c}\left(2N_c- N_f\right) e^{-2 \left(\rho 
-\rho _o\right)}\rho     $			&\\ 
	& &\\ \cline{1-2} && \\
$>2N_c$ & $P\sim -Q\sim|2N_c-N_f|\r$ &  $P\sim c_+e^{4\r/3}$  \\ 
	& $e^{2h}\sim \frac{1}{4} \left(N_f-N_c\right)$			& 
$e^{2h}\sim \frac14 c_+e^{4\r/3}$	\\ 
	& $e^{2g}\sim \frac{1}{2} \left(N_f-2 N_c\right) \rho$			
& $e^{2g}\sim c_+e^{4\r/3}$\\ 
	& $Y\sim \frac{1}{4} \left(N_f-N_c\right)$			& 
$Y\sim \frac16 c_+e^{4\r/3}$\\ 
	& $e^{4(\f-\f_o)}\sim \frac{e^{4 \left(\rho -\rho 
_o\right)}\sinh^2\left(2 \rho _o\right)}
{2 \left(N_c-N_f\right){}^2 \left(N_f-2 N_c\right) \rho }$			
& $e^{4(\f-\f_o)}\sim 1 $	\\ 
	& $a\sim  e^{-2 \left(\rho -\rho _o\right)}\rho $	& $a\sim2 
e^{-2(\r-\r_o)} $\\ 
	&&\\ \cline{1-2} && \\
$=2N_c$ & $P\sim N_c+\sqrt{N_c^2+Q_o^2}\sim \frac{8 N_c}{(4-\xi ) \xi }$ &    
\\ 
	& $e^{2h}\sim \frac{N_c}{\xi }$			&\\ 
	& $e^{2g}\sim \frac{4 N_c}{4-\xi }$			&\\ 
	& $Y\sim \frac{N_c}{4}$			&\\  
	& $e^{4(\f-\f_o)}\sim e^{4 \left(\rho -\rho 
_o\right)}\sinh^2\left(2 \rho _o\right)\frac{(4-\xi ) \xi   }
{16 N_c^3}$			&\\ 
	& $a\sim \frac{4}{\xi  }e^{-2 \left(\rho -\rho _o\right)}$			
&\\
\hline
\end{tabular}
\caption{The two classes of leading UV behaviors.}
\label{UV-classes}
\end{center}
\end{table}

In the IR ($\r \to 0$), three 
types of 
solutions were found, called Type I, II and III (there exist other, 
qualitatively different solutions reported in \cite{gmnp} ). The function 
$P(\r)$ in these cases is,
\bea
P=-N_f\r+P_o+\frac43c_+^3P_o^2\r^3-2c_+^3N_fP_o\r^4+\frac{4}{5}c_+^3\left(\frac{4}{3}P_o^2+N_f^2\right)\r^5+\co\left(\r^6\right),
\eea
for Type I.
For the Type II asymptotics, we assume that this behavior occurs when the 
IR
is located at $\r_{IR}>\r_o$. Without loss of generality we can choose 
$\r_{IR}=0$. With this choice we then
necessarily have $\r_o<0$. Expanding $Q$ in (\ref{Q}) around $\r=0$ we 
obtain
\be\label{Q-exp}
Q=b_0+b_1\r+\co(\r^2),
\ee
where
\bea
b_0&=&-\coth(2\r_o)\left(Q_o+\frac{2N_c-N_f}{2}\right)-\frac{2N_c-N_f}{2},\NO\\
b_1&=&-\frac{2}{\sinh^2(2\r_o)}\left(Q_o+\frac{2N_c-N_f}{2}\right)-(2N_c-N_f)\coth(2\r_o)\,\,.
\eea
Looking for IR solutions of (\ref{masterf}) we 
find  that we must require that $b_0>0$. The
corresponding asymptotic solution then takes the form
\bea\label{G-II-1}
P&=&Q+h_1\r^{1/2}-\frac{1}{6b_0}\left(h_1^2+12b_0(b_1+N_f)\right)\r\NO\\
&&+\frac{h_1}{72b_0^2}\left(5h_1^2+6(5b_1+2N_f)b_0-72b_0^2\coth(2\r_o)\right)\r^{3/2}+\co(\r^2),
\eea
where $h_1$ is an arbitrary constant. Note that this expansion for $P$ 
admits a smooth limit when $\r_o\to-\infty$ and so it 
is valid for both solutions of type \bA ($\r_o\to-\infty$). 

Finally, for Type III asymptotics we  consider 
$\r_{IR}>\r_o$ and we take $\r_{IR}=0$. In terms of the expansion 
(\ref{Q-exp}) this requires that $b_0=0$. We then find, 
\bea\label{III}
P=h_1\r^{1/3}-\frac{9N_f}{5}\r-\frac{2h_1}{3}\coth(2\r_o)\r^{4/3}-\frac{1}{175h_1}\left(50b_1^2-18N_f^2\right)\r^{5/3}
+\co(\r^2),
\eea
where $h_1\neq 0$ is an arbitrary constant.

To leading order the solutions for large $\r$ - UV solutions - are quoted in table 
\ref{UV-classes}. It is the presence of subleading terms that allow the smooth numerical 
interpolation with three possible IR behaviors discussed.

The physics of the dual field theory encoded in these solutions was 
discussed in detail in the papers \cite{Casero:2006pt},\cite{HoyosBadajoz:2008fw},\cite{Casero:2007jj} 
by computing various 
observables using the string solution of eq. (\ref{nonabmetric424}) 
evaluated on the solutions above\footnote{Finding a numerical 
interpolation between the IR solutions and the  solutions of Class 
I in the far UV is (numerically) delicate. One can see some plots in 
section V of the paper \cite{Nunez:2009da} .}. We move now to discuss 
general features of the dual field theory.

\subsection{The dual field theory}
The proposal here is the following: without the addition of the flavor branes, the field 
theory is 
known to be a twisted version of six-dimensional Yang-Mills, 
or as we discussed above, a four 
dimensional QFT with an infinite number of massive fields. 
See section \ref{qftwof}. 
To get an {\it intuitive} understanding of the modifications of the dynamics produced by the 
``quark'' fields (that feature below), we will consider that all the infinite massive fields 
are chiral multiplets 
and then argue that the dynamics is ruled by a lagrangian of the form:
\beq
L= Tr[-\frac{1}{4}F_{\mu\nu}^2 -i \bar{\lambda}\gamma^\mu D_\mu \lambda +
L(\Phi_k,W)]\sim \int d^2\theta W_\alpha W^\alpha +\sum_k \int d^4 
\theta\Phi^{\dagger}_k e^{V}\Phi_k +\int d^2\theta \mu_k |\Phi_k|^2+\cdots\,\,.
\eeq
When the 
flavor branes come into play, we are adding ``quark superfields'' that are 
realized as the 
open 
strings going from the non-compact flavor branes to the compact (or wrapped) color branes (as 
usual, the open strings that begin and end on a  
flavor brane decouple and do not contribute to the 
four dimensional dynamics). More concretely, we add the quark and anti-quark superfields 
($Q,\tilde{Q}$) and propose that we have a lagrangian 
for the massive fields interacting  with the quark-antiquark superfields $Q, \tilde{Q}$  
schematically of the form:\footnote{For more 
details, see the appendix A of the paper \cite{HoyosBadajoz:2008fw} }
\beq
L(\Phi_k,W, Q,\tilde{Q})= \sum_k \int d^4\theta \Phi_k^{\dagger} e^{V}\Phi_k + \kappa_k\int 
d^2\theta 
\tilde{Q}\Phi_k 
Q 
+\mu_k |\Phi_k|^2+\cdots\,\,,
\eeq
and canonical kinetic terms for $(Q,\tilde{Q})$. In this system, the $SU(N_f)_L\times 
SU(N_f)_R$ symmetry is explicitly broken to 
the diagonal $SU(N_f)_D$ by the presence of the coupling $\tilde{Q}\Phi_k Q$. In this 
respect, 
the theory is qualitatively different from ${\cal N}=1$ SQCD. 

One may be interested in the theory at low energies and hence integrate out the massive 
fields (either 
massive vectors or massive chirals) and after some algebra end with a theory of the form (again 
schematically):
\beq
L= \int d^4\theta\Big( Q^\dagger e^{V} Q + \tilde{Q}^\dagger e^{-V}\tilde{Q}\Big) + \int 
d^2\theta W_\alpha 
W^\alpha + \kappa \int d^2\theta (\tilde{Q}Q)^2\,\,,
\label{flavorslagzz}\eeq
where we have a (naively irrelevant) deformation of ${\cal N}=1$ SQCD. 

We emphasize that this is an {\it intuitive} way of understanding 
the field theory dual to the 
flavored system described above. 
As we will summarize below there are various observables that can be 
computed that match the predicted (or expected) result. So, the {\it precise} dual QFT should be 
something similar to what we described above, or at least with the same qualitative physics.
\subsection{Checks and predictions}
This subsection summarizes results developed in the papers
\cite{Casero:2006pt},
\cite{HoyosBadajoz:2008fw},
\cite{Casero:2007jj}. There is a point that should be emphasized here. All the solutions to the BPS 
equations or the master equation (\ref{masterf})
that have been found up to the time of writing this review, present a singularity in the IR (typically at 
$\r=0$). In spite of this being a ``good singularity'' according to some criteria developed in 
the literature \cite{Maldacena:2000mw}; 
 the presence of the singularity makes the interpretation of IR observables a bit 
unclear. In other words, though one gets the ``correct or expected'' result one should perhaps handle 
those particular computations with care.

Let us then concentrate on various quantities computed in the UV and then we will specify some that 
are mostly influenced by the IR of the geometry.
\subsubsection{Beta function and anomalies}
The gauge coupling and the theta angle of the dual QFT can be defined as explained in various places, 
see for example section 4.1 of the paper \cite{Casero:2007jj} or section 5 of the paper 
\cite{HoyosBadajoz:2008fw}. One gets, after some algebra,
that the gauge coupling is related to the functions of the background as:
\beq
\frac{8\pi^2}{g^2}= e^{-\tau}P\,\,.
\eeq
Choosing a particular radius-energy relation, that was discussed  in
\cite{Apreda:2001qb},
\cite{DiVecchia:2002ks},
one can compute the variation of the coupling with
 respect to energy. Using the solutions where the 
dilaton asymptotes to a linear function $(e^{4\phi}\sim\frac{e^{4\r}}{\r})$ and working to 
leading order in an 
expansion in inverse powers of the radial coordinate, we get
\beq
\beta_{\frac{8\pi^2}{g^2}}=\frac{3}{2}(2N_c-N_f)\,\,,
\eeq
that coincides with the result predicted by the NSVZ result, once we assign
anomalous dimensions to the quark superfields $\gamma_Q=\gamma_{\tilde{Q}}=-\frac{1}{2}$.

Similarly, one can define a geometrical quantity that can be associated with the quartic coupling. 
See section 4.2 of 
the paper \cite{Casero:2007jj}. The beta function can be computed using the anomalous dimensions 
discussed above and again get matching with the interesting fact that for $N_f>2N_c$ the quartic 
coupling is irrelevant, for $N_f<2N_c$ the coupling is relevant while for $N_f=2N_c$ the coupling is 
not running. See \cite{Strassler:2005qs} for a nice explanation of this fact.

One can also assign 
a value of the R-charge to the quark superfields to get the correct R-symmetry 
transformation properties of the quartic superpotential of 
eq. (\ref{flavorslagzz}), that is $R[Q]= R[\tilde{Q}]=\frac{1}{2}$. 
This predicts that the R-symmetry 
anomaly, the triangle with one R-current and two gauge currents is proportional to the 
quantity ($2N_c-N_f$) times the phase by which we are rotating the fermions. This is  the 
precise result that 
the string background gives. Indeed, if we compute the $\Theta$-angle as explained in section 4.1 of 
the paper \cite{Casero:2007jj} or in section 5 of the paper
\cite{HoyosBadajoz:2008fw}, we will get
\beq
\Theta=\frac{\psi-\psi_0}{2}(2N_c-N_f)\,\,,
\eeq
where we associated $\frac{\psi-\psi_0}{2}$ with the change in phase of the fermions in the 
quark multiplet 
and the gauge multiplet to get perfect matching.
In the same vein, it is possible to attempt a 't Hooft matching of anomalies, that is of triangles 
involving three global currents. The reader will find it quite instructive to go over section 
4.7 of the 
paper \cite{Casero:2007jj}. There, a detailed study of the matching of the 
correlator of three global 
currents -some of them corresponding to discrete symmetries, 
some of them being continuous symmetries- 
is presented. The treatment is performed in the case of the Type {\bf A} backgrounds, that 
are 
characterized by the fact that the functions $a=b=0$ in eq. (\ref{nonabmetric424}). 
This 
translates to the fact that the R-symmetry is broken to ${\mathbb Z}_{2N_c-N_f}$ without the further 
(spontaneous) 
breaking 
to ${\mathbb Z}_2$.
\subsubsection{Seiberg duality}
\label{sec:Seib}
It is known that Seiberg duality manifests beautifully in a QFT like the one 
of eq. (\ref{flavorslagzz}). This is explained in section 1.10 of \cite{Strassler:2005qs}. The 
backgrounds discussed here show this in a very nice way. Indeed, as discussed for example in the 
paper \cite{HoyosBadajoz:2008fw}, we can see that the master equation (and the whole system) is 
invariant under the change,
\beq
P\to P,\; Q\to -Q,\;\; \sigma\to - \sigma,\;\;\; N_c\to N_f-N_c,\;\; N_f\to N_f\,\,,
\label{changeszzz}
\eeq
while all other functions are invariant. 
Geometrically, this change amounts to swapping the two $S^2$ in the background,
namely those parameterized by $\theta,\varphi$ and $\tilde \theta, \tilde \varphi$ in
(\ref{nonabmetric424}).
This should be interpreted as follows: 
suppose that we are presented with 
a background, representing the dynamics of a field theory with $N_c$ colors and $N_f$ 
flavors. This 
implies that we have a particular solution to the master equation (\ref{masterf}) for the 
function $P$. 
With this solution, and applying the changes of eq.  (\ref{changeszzz}),  we can construct another 
solution, that will be related to the first one by a differomorphism and that will describe the 
physics of a field theory with $N_f-N_c$ colors and $N_f$ flavors. Various aspects of this 
interesting duality have been discussed in the papers \cite{Casero:2007jj},
\cite{HoyosBadajoz:2008fw}, \cite{Elander:2009bm}
and probably elsewhere.

The implementation of Seiberg duality in ${\cal N}=1$ subcritical string models
was discussed in 
\cite{Murthy:2006xt}, \cite{Ashok:2007sf}. One can think of the sphere exchange mentioned above as being the
geometrical version of the mechanism described in these papers. 
Interestingly, similar methods were used to propose a non-supersymmetric Seiberg duality in
\cite{Armoni:2008gg}.

\subsection{IR Physics: Domain Walls and some comments on Wilson/'t Hooft Loops}

One observable that can be computed and that strongly depends on the $\r\to 0$ region of the solution 
(the IR) is the tension of domain walls. Indeed, domain walls can be thought of as D5-branes that 
wrap a three cycle inside the internal six dimensional manifold and that extend on two of the 
Minkowski directions (and time, of course). We can compute the tension of a wall by considering a 
probe D5-brane that sits on the manifold 
$\Sigma_6=[t,x_1,x_2,\tilde{\theta},\tilde{\varphi},\psi]$, at 
constant $\r=0$, constant $\theta,\varphi$. The Born-Infeld action for this probe can be computed and 
one reads that the effective tension is given by (see for example section 5.6 of the paper 
\cite{Casero:2006pt}):
\beq
T_{DW}= 4 \pi^2 e^{2\phi+ 2g +k}\,\, T_{D5}\,\,,
\eeq
which,  when evaluated at $\r=0$,  gives a constant  proportional to $2N_c-N_f$. This is a good example 
for an observable, since even when a singularity is present (this typically reflects in some of the 
functions of the background being divergent), the combination above is finite. This is 
typical of ``good 
singularities''.
There are other observables that can be computed using D-branes or fundamental strings, examples of 
these are Wilson or 't Hooft loops. A similar 
conspiracy   of 
functions that avoids an infinite result occurs here. 
Nevertheless, one should be quite careful with these 
quantities as noted above. Indeed, it was found in section V of the paper \cite{Nunez:2009da}
that for the particular case of the backgrounds studied in this section, the Nambu-Goto action for 
the fundamental string might cease to be a good approximation as the string develops a cusp when 
approaching the singularity.

In other words, we believe that the solutions presented in this section  surely capture correctly 
many UV aspects of the field theory, together with some IR observables. Probably, we could think of 
the presence of the singularity in the same way as we think about the singularity in the 
Klebanov-Tseytlin background\footnote{In contrast to what happens for the solutions 
discussed here, the Klebanov-Tseytlin background presents a {\it bad} singularity and the 
IR physics computed with that solution is not trustable.}, that captures some of 
the physics, but some is lost and the singularity 
must be resolved. There are different ways of attempting a resolution of the singularity; for 
example, considering massive quarks. This is under present study.
\subsubsection{Wilson loops and first order phase transitions}
\label{CNPwilson}

One can consider the situation (of course, this is an idealized situation) in which all the flavors are 
massive and with a fixed sharp mass $M_0$ (corresponding to a given value of the 
radial coordinate, that we call $\rho=m_0$). The way to model this in a first approximation is 
to 
consider $N_f(\r)=N_f \Theta(\r-m_0)$, where $\Theta(\r)$ is a Heaviside 
step function. Once again, there is 
some 
dynamics that is being lost in doing this, for example, the matching of the derivatives of the 
solutions is not smooth at the point $\r=m_0$ and the curvature of the background is 
not well defined at that point. 
Nevertheless, it is possible (within this approximately correct way of proceeding) to find a solution 
that for energies below the scale set by $m_0$ corresponds to the theory {\it without} flavors, say 
those discussed around eq.  (\ref{solutionnear0xx}) and far in the UV corresponds to the flavored 
theory, as represented by solutions of the Class I in table \ref{UV-classes}. One can then compute 
the Wilson loop following the well-known prescription \cite{Maldacena:1998im}. This was done explicitly in 
\cite{Bigazzi:2008gd}. The qualitative result is the following: for a range of ratios between 
the 
mass of the quarks $M_0$ and the 
value of the gaugino condensate set by the function $a(\r)$ one observes that the 
relation between the quark-antiquark potential $V_{QQ}$ and their separation $L_{QQ} $ presents a 
first order 
transition\footnote{Notice that  we should talk of a 
``Quantum'' phase transition, as the system is at zero temperature.}.
 In other words, a point where $\frac{dV_{QQ}}{d L_{QQ}}$ is discontinuous. The 
same kind of behavior was observed in systems 
where a more careful study is possible. Indeed, in the backreacted
Klebanov-Witten (see section \ref{AdS5X5}) and Klebanov-Strassler 
(see section \ref{KS}) models,
it was possible to 
find the precise form for the function $p(\r)$ - that shows that the ``Heaviside 
approximation'' described above is not 
a bad one. 
The same qualitative behavior for the first order phase
 transition was found in 
the papers 
\cite{Bigazzi:2008zt}, \cite{Bigazzi:2008qq}, \cite{Ramallo:2008ew}.

This kind of first order transitions for the Wilson loop string configurations
are by no means particular of systems with dynamical fundamental fields. 
In fact, they were first found in a different context in \cite{sfetsos},
where a nice connection between these Wilson loops computations and the Van der Waals 
gas 
(paradigm of the first order transition) was put forward 
(further discussions can be found for instance in 
\cite{Bigazzi:2008qq},
\cite{Nunez:2009da}).
Different examples of such phase transitions in systems without flavors
have been worked out in 
\cite{vandertrans}.

The ``morale'' seems to be the following: when we have a physical system that has two {\it 
independent} 
scales (that is two scales that can be tuned independently, in the present example, the mass 
of the quarks $M_0$ and the gaugino condensate $\Lambda^3$) the first order phase transition 
for the 
quantity $V_{QQ}(L_{QQ})$ will be present. 
Of course, like in any other 
first order transition, it 
will happen that the discontinuity in the derivative will disappear  for some ratio between 
the scales 
mentioned above. 

\setcounter{equation}{0}

\section{A dual to a (2+1)-dimensional ${\cal N}=1$ SQCD-like  model}
\label{sec:2+1}
In this section we will study gravity duals to minimal supersymmetric theories in 2+1 dimensions. These backgrounds can be obtained by wrapping D5-branes along three-cycles of manifolds with $G_2$ holonomy \cite{Chamseddine:2001hk,Maldacena:2001pb,Schvellinger:2001ib}. The corresponding field theory dual is a (2+1)-dimensional ${\cal N}=1$ supersymmetric $U(N_c)$ Yang-Mills theory with a level $k$ Chern-Simons interaction. Such a theory coupled to an adjoint massive scalar field should arise on the domain walls separating the different vacua of pure ${\cal N}=1$ super-Yang-Mills in 3+1 dimensions.  The corresponding unflavored background was studied in ref. \cite{Maldacena:2001pb}, where it was argued to be dual to a $U(N_c)$ gauge theory with Chern-Simons level $k=N_c/2$. In what follows we will review a generalization of these results, following closely ref. \cite{Canoura:2008at}. We will present the deformation of the background of 
\cite{Chamseddine:2001hk,Maldacena:2001pb,Schvellinger:2001ib} induced by a smeared distribution of massless flavors. In order to formulate these generalized backgrounds,  let $\sigma^i$ and $\omega^i$ $(i=1,2,3)$ be two sets of SU(2) left-invariant one forms, obeying:
\beq
d\sigma^i=-{1\over 2}\,\epsilon_{ijk}\,\sigma^j\wedge \sigma^k\,\,,
\qquad\qquad
d\omega^i=-{1\over 2}\,\epsilon_{ijk}\,\omega^j\wedge \omega^k\,\,.
\label{sigma-w}
\eeq
The forms $\sigma^i$ and $\omega^i$ parameterize two three-spheres. In the geometries we  will be dealing with,  these spheres are fibered by a one-form $A^i$. The corresponding  ten-dimensional metric of the type IIB theory in the Einstein frame is given by:
\beq
ds^2\,=\,e^{{\phi\over 2}}\left[ dx_{1,2}^2+dr^2+\frac{e^{2h}}{4}(\sigma^i)^2+\frac{e^{2g}}{4}(\omega^i-A^i)^2\right]\,\,,
\label{2+1ansatz}
\eeq
where  $\phi(r)$ is the dilaton of type IIB supergravity and $g$ and $h$ are functions of the radial variable  $r$.  In addition, the one-form $A^i$ will be taken as:
\beq
A^i\,=\,{1+w(r)\over 2}\,\,\sigma^i\,\,,
\label{Ai-w}
\eeq
with $w(r)$ being a new function of $r$.  For convenience in this section  we will take $g_s=\alpha'=1$, as we did in section \ref{sec:D5D5}.  The backgrounds considered here are also endowed with a RR three-form $F_{3}$. We will represent $F_{3}$ as the sum of two contributions:
\beq
F_3\,=\,{\cal F}_3\,+\,f_3\,\,,
\label{F3f3}
\eeq
where $d{\cal F}_3=0$ and $f_3$ is the part of the RR three-form which is responsible for the violation of the Bianchi identity ($df_3\not=0$) and which is sourced by the flavor D5-branes. Let us first parametrize the component ${\cal F}_3$ as:
\beq
\frac{{\cal F}_3}{N_c}=-\frac{1}{4}(\omega^1-B^1)\wedge(\omega^2-B^2)\wedge(\omega^3-B^3)+\frac{1}{4}F^i\wedge(\omega^i-B^i)+H\,\,,
\label{F3ansatz}
\eeq
where $B^i$ is a new one-form and $F^i$ are the components of its field strength, given by:
\beq
F^i\,=\,dB^i\,+\,{1\over 2}\,\epsilon_{ijk}\,B^j\wedge B^k\,\,.
\label{Fi}
\eeq
In (\ref{F3ansatz}) $H$ is a three-form that is determined by imposing the Bianchi identity for ${\cal F}_3$, namely:
\beq
d{\cal F}_3\,=\,0\,\,.
\label{Bianchi-id}
\eeq
By using (\ref{sigma-w}) one can easily  check  from the explicit expression written in 
(\ref{F3ansatz}) that, in order to fulfill (\ref{Bianchi-id}), the three-form $H$ must satisfy the equation:
\beq
dH\,=\,{1\over 4}\,F^i\wedge F^i\,\,.
\label{dH}
\eeq
In what follows we shall adopt the following ansatz for $B^i$:
\beq
B^i\,=\,{1+\gamma(r)\over 2}\,\,\sigma^i\,\,,
\label{Bi}
\eeq
where $\gamma(r)$ is a new function.  After plugging the ansatz of $B^i$ written in (\ref{Bi}) into (\ref{Fi}), one gets the expression for $F^i$ in terms of $\gamma(r)$:
\beq
F^i\,=\,{\gamma'\over 2}\,\,dr\wedge \sigma^i\,+\,
{\gamma^2-1\over 8}\,\,\epsilon_{ijk}\,\sigma^j\wedge \sigma^k\,\,,
\label{Fi-gamma}
\eeq
where the prime denotes the derivative with respect to the radial variable $r$. Using this result for $F^i$ in (\ref{dH}) one can easily determine the three-form $H$ in terms of $\gamma$. Let us parameterize $H$  as:
\beq
H\,=\,{1\over 32}\,\,{1\over 3!}\,\,{\cal H}(r)\,\,\epsilon_{ijk}\,\,
\sigma^i\wedge \sigma^j\wedge\sigma^k\,\,.
\eeq 
Then, by solving (\ref{dH}) for $H$,  one can verify that 
${\cal H}(r)$ is the following function of the radial variable:
\beq
{\cal H}\,=\,2\gamma^3-6\gamma\,+\,8\kappa\,\,,
\label{calH}
\eeq
with $\kappa$ being an integration constant.

Let us now consider the contribution $f_3$ to the RR three-form $F_3$. As explained above, this contribution violates the Bianchi identity and is non-zero when flavor branes are present. Indeed, let us write the WZ term of the action of a system of flavor D5-branes as:
\beq
S_{flavor}^{WZ}\,=\,T_5\,\int_{{\cal M}_{10}}\,\Omega\wedge C_6\,\,,
\label{C6-Omega}
\eeq
with $\Omega$ being a four-form with components along the space transverse to the worldvolume of the branes. Then, the coupling to the RR potential $C_6$ written in (\ref{C6-Omega}) gives rise to the following modified Bianchi identity:
\beq
dF_3\,=\,df_3\,=\,{4\pi^2\Omega}\,\,.
\label{Bianchi2+1}
\eeq
To write an specific ansatz for $\Omega$ and $f_3$  we have to select some family of supersymmetric embeddings for the flavor branes. As explained above, this  can be done by using kappa symmetry. In the simplest case one looks for massless embeddings, which extend along the full range of the radial coordinate $r$. Those are the configurations considered in \cite{Canoura:2008at}, in which the D5-brane is extended along the three Minkowski directions $x^{\mu}$, as well as along a three-dimensional cylinder spanned by $r$ and two other angular directions. Actually, it was shown in \cite{Canoura:2008at} that these two angular directions could be the ones corresponding to $\sigma^{3}$ and $\omega^3$. The corresponding transverse volume for this configuration is just:
\beq
{\rm Vol}(\,{\cal Y}_4^{1,2}\,)\,=\,
\sigma^1\wedge \sigma^2\wedge \omega^1\wedge \omega^2\,\,.
\label{Vol12}
\eeq
However, there is nothing special in our background about these directions. Indeed, both in the metric and in the RR three-form ${\cal F}_3$, we are adopting a round ansatz which does not distinguish among the directions of the two three-spheres. Thus  we could as well consider supersymmetric cylinder embeddings that span the $1, \hat 1$ or $2, \hat 2$ directions. The volume forms of the spaces transverse to these embeddings are clearly:
\beq
{\rm Vol}(\,{\cal Y}_4^{2,3}\,)\,=\,\sigma^2\wedge \sigma^3\wedge \omega^2\wedge \omega^3\,\,,
\qquad\qquad
{\rm Vol}(\,{\cal Y}_4^{1,3}\,)\,=\,\sigma^1\wedge \sigma^3\wedge \omega^1
\wedge \omega^3\,\,.
\label{Vol23-Vol13}
\eeq
To construct a backreacted supergravity solution with the same type of ansatz as in (\ref{2+1ansatz}) we should consider a brane configuration that combines these three possible  types of  embeddings in an isotropic way.  The corresponding transverse volume form ${\rm Vol}(\,{\cal Y}_4)$  of this three-branch brane system would be just the sum of the three four-forms written in eqs. (\ref{Vol12}) and (\ref{Vol23-Vol13}). The corresponding smearing form $\Omega$ is obtained by multiplying by the suitable normalization factor, namely :
\beq
\Omega\,=\,-{N_f\over 16\pi^2}\,\,{\rm Vol}(\,{\cal Y}_4\,)\,=\,-{N_f\over 64\pi^2}\,
\epsilon_{ijk}\,\epsilon_{ilm}\,
\sigma^j\wedge \sigma^k\wedge \omega^l\wedge \omega^m\,\,,
\label{Omega2+1}
\eeq
where the minus sign has its origin in the different orientation (required in the kappa symmetry analysis of \cite{Canoura:2008at}) of the  D5-brane worldvolume with respect to the ten-dimensional space. It is now straightforward to use the $\Omega$ written in (\ref{Omega2+1})  and get an expression of $f_3$ whose modified Bianchi identity is the one of  (\ref{Bianchi2+1}). One has:
\begin{equation}
f_3=\frac{N_f}{8}\epsilon_{ijk}(\omega^i-\frac{\sigma^i}{2})\wedge\sigma^j\wedge\sigma^k\,.
\label{f3}
\end{equation}
Eq. (\ref{f3}) completes our ansatz for the general flavored case. Using these expressions of the metric and RR three-form in the supersymmetry variations of the dilatino and gravitino of type IIB supergravity, after imposing that the background preserves two supersymmetries,  we arrive at a system of first-order BPS equations. These equations, which  are rather involved, have been derived and analyzed in detail in \cite{Canoura:2008at}. They admit several consistent  truncations which lead to simpler solutions. One can, for example,  first consider the unflavored case $N_f=0$. If, in addition,  we  require that the function $g$ is constant and that the fibering functions $w$ and $\gamma$ are equal, our ansatz reduces to the one considered in  \cite{Chamseddine:2001hk,Maldacena:2001pb,Schvellinger:2001ib}. Actually, in this case the BPS equations fix the value of $g$ to be $e^{2g}=N_c$ and, in order to have a regular solution one should take  the constant $\kappa$ of (\ref{calH}) to be equal to $1/2$ ($\kappa$ is related to the Chern-Simons level $k$ of the dual field theory). Other unflavored solutions exist and have been studied in detail in ref. \cite{Canoura:2008at}. Here we will concentrate on reviewing the case in which $N_f\not=0$, starting from a particular truncation of the BPS system which is very interesting and serves to classify the different more involved solutions in the UV.

\subsection{The truncated system}
\label{Mnas-truncated}

In this section we will analyze the truncation of the general system of BPS equations that corresponds to taking $w=\gamma= \kappa=0$. In this case the BPS equations of 
\cite{Canoura:2008at} for the remaining functions $h$ and $g$ of the metric  and for  the dilaton $\phi$ consistently reduce to the following simple system of differential equations:
\bear
&&\phi'\,=\,N_c\,e^{-3g}\,\,-\,{3\over 4}\,(\,N_c-4N_f\,)\,e^{-g-2h}\,\,,\rc\rc
&&h'\,=\,{1\over 2}\,e^{g-2h}\,+\,{N_c-4N_f\over 2}\,\,e^{-g-2h}\,\,,\rc\rc
&&g'\,=\,e^{-g}\,-\,{1\over 4}\,e^{g-2h}\,-\,N_c\,e^{-3g}\,+\,{N_c-4N_f \over 4}\,
e^{-g-2h}\,\,.
\label{ab-system-flavor}
\eear
By inspecting the system (\ref{ab-system-flavor}) one readily realizes that there is a special solution for which the metric functions $h$ and $g$ are constant. Actually this solution only exists when $N_c<2N_f$ and the  corresponding expressions for $g$ and $h$ are the following:
\beq
e^{2g}\,=\,4N_f-N_c\,\,,
\qquad\qquad
e^{2h}\,=\,{1\over 4}\,\,{(4N_f-N_c)^2\over 2N_f-N_c}\,\,,
\qquad\qquad (\,N_c\, <\,2N_f\,)\,\,,
\eeq
while the dilaton grows linearly with the holographic coordinate $r$, namely:
\beq
\phi\,=\,{2(3N_f-N_c)\over \big[\,4N_f-N_c]^{{3\over 2}}}\,\,r\,\,+\,\,\phi_0\,\,.
\eeq

Let us next consider solutions for which the function $h$ is not constant. In this case we can use $\rho=e^{2h}$ as a radial  variable and one can define a new function $F(\rho)$ as $F(\rho)=e^{2g}$. It follows from (\ref{ab-system-flavor}) that the BPS equation for $F(\rho)$ is now:
\beq
{dF\over d\rho}\,=\,{(F-N_c)\,\Big(\,2-{F\over 2\rho}\,\Big)\,-\,{2N_f\over \rho}\,\,F
\over F+N_c-4N_f}\,\,,
\label{F-ab-flavor}
\eeq
while the equation for the dilaton  as a function of $\rho$ can be written as:
\beq
{d\phi\over d\rho}\,=\,{N_c\over F(F+N_c-4N_f)}\,\Big[\,1\,-\,{3\over 4\rho}\,
\Big(\,1\,-\,{4N_f\over N_c}\,\Big)\,F\,\Big]\,\,.
\label{dilaton-ab-flavor}
\eeq
Moreover,  from the second equation in (\ref{ab-system-flavor}) we can obtain the relation between the two radial variables $r$ and $\rho$, namely:
\beq
{dr\over d\rho}\,=\,{\sqrt{F(\rho)}\over F(\rho)+N_c-4N_f}\,\,.
\label{r-rho-flavored-ab}
\eeq
Notice that the sign of the right-hand side of (\ref{r-rho-flavored-ab}) could be negative when $N_f\not=0$. This means that we have to be careful in identifying the UV and IR domains in terms of the new radial variable $\rho$.  We can use the result of integrating eqs. (\ref{F-ab-flavor})-(\ref{r-rho-flavored-ab}) to obtain the metric  in terms of the new variable $\rho$, which takes the form:
\beq
ds^2\,=\,e^{{\phi\over 2}}\,\Bigg[\,dx^2_{1,2}\,+\,
\Big(\,{dr\over d\rho}\,\Big)^2\,(d\rho)^2\,+\,
{\rho\over 4}\,(\sigma^i)^2\,+\,{F\over 4}\,\Big(\,\omega^i\,-\,A^i\,\Big)^2
\,\Bigg]\,\,.
\label{metric-in-rho}
\eeq

Let us now study the different solutions of eqs. (\ref{F-ab-flavor})-(\ref{dilaton-ab-flavor}).

\subsubsection{Linear dilaton backgrounds}
\label{FMnas-abe-subsection}

When $N_f=0$,  eq. (\ref{F-ab-flavor}) can be simply solved by taking $F=N_c$. However, 
it is clear from (\ref{F-ab-flavor}) that in the flavored case $F=N_c$ is no longer a solution of the equations. Nevertheless, there are solutions for which this constant value of $F$ is reached asymptotically when $\rho\to\infty$. Indeed, one can check this fact by solving (\ref{F-ab-flavor})  as an expansion in powers  of $1/\rho$. One gets:
\beq
F\,=\,N_c\,+\, N_c\, N_f\,{1\over \rho}\,-\,
{3\over 4}\, N_c\, N_f\,(N_c-4N_f)\,{1\over \rho^2}\,+\,\cdots\,\,,
\qquad\qquad (\rho\to\infty)\,\,.
\label{UVF}
\eeq
By plugging the expansion (\ref{UVF}) into (\ref{dilaton-ab-flavor}) one can prove that, when $N_c\not=2N_f$,  these solutions have a  dilaton  that depends  linearly on $\rho$  in the UV and, actually,  one can verify that:
\beq
{d\phi\over d\rho}\,=\,{1\over 2(N_c-2N_f)}\,-\,
{3N_c^2\,-\,12 N_c N_f\,+\,16N_f^2\over 8(N_c-2N_f)^2}\,\,\,{1\over \rho}
\,+\,\cdots\,\,,
\qquad\qquad (\rho\to\infty)\,\,.
\label{UVdilaton}
\eeq
Notice the different large $\rho$  behavior of the dilaton in the two cases $N_c>2N_f$ and  $N_c<2N_f$. Indeed, when $N_c>2N_f$ the dilaton grows linearly with the holographic coordinate $\rho$ (the behavior expected for a confining theory in the UV), while for $N_c<2N_f$ the field $\phi$ decreases linearly with $\rho$. This seems to suggest that the sign of the beta function of the dual gauge theory depends on $N_c$ and $N_f$ through the combination $N_c-2N_f$. Actually one can verify by means of a probe calculation in the complete system that the beta function is positive for $N_c>2N_f$ and changes its sign when $N_c<2N_f$ \cite{Canoura:2008at}. 

Eq. (\ref{F-ab-flavor}) can be solved numerically by imposing the behavior (\ref{UVF}) for large $\rho$. Once $F(\rho)$ is known one can obtain the dilaton $\phi(\rho)$ by direct integration of the right-hand side of (\ref{dilaton-ab-flavor}). The result of this numerical calculation was analyzed in detail in \cite{Canoura:2008at}. Let us only mention here that, in the most interesting case $N_c>2N_f$, the function $F$ diverges for $\rho\to 0$, while the dilaton $\phi$ remains finite for small $\rho$. This bad IR behavior of $F$ is cured in the untruncated solution with the same leading UV form of $F$ and $\phi$  but with $w, \gamma\not=0$ (see below). 

\subsubsection{Flavored $G_2$ cone}
\label{G2cone}
Let us now consider the solution of the equations (\ref{F-ab-flavor}) and (\ref{dilaton-ab-flavor}) that leads to a metric which is asymptotically a $G_2$-cone with constant dilaton  in the UV.  It can be checked that there exists a solution of (\ref{F-ab-flavor}) which can be expanded for large values of $\rho$ as:
\beq
F\,=\,{4\over 3}\,\,\rho\,+\,4(N_f-N_c)\,+\,
{15 N_c^2-39N_c N_f\,+\,24 N_f^2\over \rho}\,+\,\cdots\,\,.
\label{G2-F-flavored}
\eeq
The corresponding expansion for $\phi(\rho)$ is:
\beq
\phi\,=\,\phi_{*}\,-\,{9N_f\over 4}\,{1\over \rho}\,-\,{27\over 32}\,N_c\,(N_c+2N_f)\,
{1\over \rho^2}\,+\,\cdots\,\,,
\label{G2-int-phi-flavored}
\eeq
where $\phi_{*}$ is the constant limiting value of $\phi$ in the UV. In order to explore the asymptotic form of the metric for large $\rho$, it is convenient to perform a change in the radial variable, namely:
\beq
\rho\,=\,{1\over 3}\,\tau^2\,\,,
\eeq
in terms of which  the  metric   asymptotically becomes the one corresponding to the direct product of (2+1)-dimensional Minkowski space and a seven-dimensional cone with $G_2$ holonomy, namely:
\beq
ds^2\,\approx\,e^{{\phi_{*}\over 2}}\,\Big[\,dx^2_{1,2}\,+\,
(d\tau)^2
\,+\,
{\tau^2\over 12}\,(\sigma^i)^2\,+\,{\tau^2\over 9}\,\,
\Big(\,\omega^i\,-\,{\sigma^i\over 2}\,\Big)^2
\,\Big]\,\,.
\label{UV-asymp-metric}
\eeq
To find the solution in the whole range of the radial coordinate one can numerically integrate the system (\ref{F-ab-flavor})-(\ref{dilaton-ab-flavor}) by imposing the asymptotic behavior (\ref{G2-F-flavored}) to the function $F(\rho)$.  For $N_c\ge 2 N_f$ one can show that $F(\rho)$ is well-defined for $\rho>0$, while it diverges for $\rho\to 0$
(see \cite{Canoura:2008at} for further details). Notice that, at least in the unflavored case $N_f=0$, it is natural to regard these solutions with finite dilaton in the UV as corresponding to D5-branes wrapped on a three-cycle of a $G_2$ cone, in which the near horizon limit has not been taken and, thus, as we move towards the large $\rho$ region the effect of the branes on the metric becomes asymptotically negligible and we recover the  geometry of the $G_2$ cone where the branes are wrapped.

\subsection{The complete system}
Let us now consider the solutions of the BPS equations for our general ansatz. These complete BPS equations have been derived in the appendix A of \cite{Canoura:2008at}. Here we will restrict ourselves  from now on to the cases  with $N_c> 2 N_f$, which are the ones that lead to more sensible solutions. As in the truncated case of section \ref{Mnas-truncated}, we will use $\rho=e^{2h}$ as radial variable and $F=e^{2h}$ as a function of $\rho$. In  order to solve the general BPS equations we must impose initial conditions to the functions $w(\rho)$ and $\gamma(\rho)$ introduced in (\ref{Ai-w}) and (\ref{Bi}), and we must fix the value of the constant $\kappa$  of (\ref{calH}). These initial conditions are determined by imposing some regularity requirements at $\rho=0$ that we now review (see \cite{Canoura:2008at} for additional details). First of all, we will demand that the function $F$ approaches a constant finite value when $\rho\to 0$ (\ie\ $F\sim F_0$ for $\rho\to 0$). In order  to fix the value of the function $w(\rho)$ at $\rho=0$ let us recall (see (\ref{Ai-w})) that $w$ parameterizes the one-form $A^i$ which, in turn, determines the mixing of the two three-spheres in the ten-dimensional fibered geometry. The curvature of the gauge connection $A^i$ (defined as in (\ref{Fi}) with $B^i\to A^i$) determines the non-triviality of this mixing. When this curvature vanishes  one can choose a new set of three one-forms in which  the two three-spheres are disentangled in a manifest way and one can  factorize the directions parallel and orthogonal to the color brane worldvolume in a well-defined way.  From the wrapped brane origin of our solutions, one naturally expects such an un-mixing of the two $S^3$'s  to occur in the IR limit $\rho=0$ of the metric. Moreover, by  a direct calculation using (\ref{sigma-w}) it is easy to verify that for $w=1$ the curvature of the  one-form $A^i$ vanishes . Thus, it follows that the natural initial condition for $w(\rho)$ is:
\beq
w(\rho=0)=1\,\,.
\label{initialw}
\eeq
Actually, the three-cycle that the color branes wrap can be identified with the one that shrinks when $\rho\to 0$, which is the one given by:
 \beq
 \Sigma\equiv\{\omega^i=\sigma^i\}.
 \label{mixc}
 \eeq
In order to have a non-singular flux at the origin, the RR three-form $F_3$ should vanish on $\Sigma$ when $\rho\to 0$. It is easy to check that this occurs if the constant $\kappa$ takes the value:
\beq
\kappa\,=\,{1\over 2}\,-\,{3N_f\over 2 N_c}\,\,.
\label{kappa-flavors}
\eeq
Actually, (\ref{kappa-flavors}) is also a necessary  condition to have a finite dilaton at $\rho=0$. Indeed, it was shown in \cite{Canoura:2008at} that, in addition to (\ref{kappa-flavors}),  the dilaton remains finite in the IR if the function $\gamma(\rho)$  takes the following value for $\rho=0$:
\beq
\gamma(\rho=0)\,=\,1\,-\,{2N_f\over N_c}\,\,.
\label{gamma-initial-flavored}
\eeq
Eqs. (\ref{initialw}) and (\ref{gamma-initial-flavored}) provide the initial conditions for the functions $w$ and $\gamma$ we were looking for.

\subsubsection{Asymptotic linear dilaton}
\label{FlaMnas-subsection}

As explained above,
we are interested in solutions of the BPS equations such that asymptotically $F$ is constant. Actually, by solving the BPS system in powers of $1/\rho$,
one can check that there are  solutions in which $F$ has the following  asymptotic 
behavior:
\beq
F\,=\,N_c\,+\,{a_1\over \rho}\,+\,{a_2\over \rho^2}\,+\,{a_3\over \rho^3}\,+\,\cdots\,\,,
\label{F-inseries-untruncated}
\eeq
where the coefficients $a_1$, $a_2$ and $a_3$ are given by:
\bear
&&a_1\,=\,N_f\, N_c\,\,,\rc\rc
&&a_2\,=\,-{3\over 4}\, N_c N_f(\,N_c\,-\,4N_f\,)\,\,,\rc\rc
&&a_3\,=\,{N_f\,N_c\over 16}\,
\Big[\,21N_c^2\,-\,148\,N_f\, N_c\,+\,240\,N_f^2\,\Big]\,\,.
\label{a-coefficients-untruncated}
\eear
Notice that the first two terms in (\ref{F-inseries-untruncated}) and 
(\ref{a-coefficients-untruncated}) coincide with the one written in (\ref{UVF}) for the truncated system.  Similarly, the functions $w$ and $\gamma$ can be represented as:
\bear
&&w\,=\,{b_1\over \rho}\,+\,{b_2\over \rho^2}\,+\,{b_3\over \rho^3}\,+\,\cdots\,\,,\rc\rc
&&\gamma\,=\,{c_1\over \rho}\,+\,{c_2\over \rho^2}\,+\,{c_3\over \rho^3}\,+\,\cdots\,\,,
\label{wgamma-inseries-untruncated}
\eear
where the coefficients $b_i$ and $c_i$ are the following:
\bear
&&b_1\,=\,c_1\,=\,{1\over 2}\,\,(N_c\,-3N_f)\,\,,\rc\rc
&&b_2\,=\,c_2\,=\,{5\over 8}\,\,(N_c\,-3N_f)\,(N_c-2N_f)\,\,,\rc\rc
&&b_3\,=\,{1\over 32}\,\,(N_c\,-3N_f)\,
\Big[\,49 N_c^2\,-\,184\,  N_c\,N_f\,+\,204\,N_f^2\,\Big]\,\,,\rc\rc
&&c_3\,=\,{1\over 32}\,\,(N_c\,-3N_f)\,
\Big[\,49 N_c^2\,-\,208\,N_f N_c\,+\,252\,N_f^2\,\Big]\,\,.
\label{bc-coefficients-untruncated}
\eear
Moreover, for $N_c>2N_f$ the dilaton grows linearly with $\rho$ as in (\ref{UVdilaton}), \ie\
$\phi\sim \rho/[2(N_c-2N_f)]$ for large $\rho$.  

The solution for the full range of the holographic coordinate can be found by numerical integration of the BPS system with the IR regularity conditions (\ref{initialw}), (\ref{kappa-flavors}) and (\ref{gamma-initial-flavored}) and with  $F(\rho=0)=F_0$ finite. One has to perform an interpolation between the $\rho\to 0$ and $\rho\to\infty$ behaviors by means of a shooting technique in which the only free parameter $F_0$ is varied until a solution with $F(\rho)\approx N_c$ for large $\rho$ is obtained (which only occurs when $F_0$ is fine tuned to a very precise value). 

After obtaining this solution of the equations of motion of the gravity plus brane system, we can  see if it incorporates some of the features that the supergravity dual of 2+1 dimensional gauge theory plus flavors should exhibit. In particular, we can  study the evolution of the gauge  coupling constant with the holographic coordinate. In order to do that, let us consider a D5-brane  probe extended along the three Minkowski directions and wrapping the internal three-cycle  $\Sigma$ defined in (\ref{mixc}) at a fixed value of the holographic coordinate $\rho$. By looking at the ${\cal F}^2$ terms in the DBI action of this probe, we get the value of the Yang-Mills coupling constant of the dual (2+1)-dimensional gauge theory, namely:
\beq
{1\over g^2_{YM}}\,\sim\,e^{-{3\over 4}\,\phi}\,\,
\int_{\Sigma}\,\,\sqrt{- \det \big(\,\hat G_3\,\big)}\,\, \,d^3\,\xi\,\,\sim\,
\Big[\,\rho+\,{F\over 4}\,(1-w)^2\,\Big]^{{3\over 2}}\,\,,
\label{gYM}
\eeq
where $\hat G_3$ is  the induced metric on the three-cycle $\Sigma$  and we have  neglected   all constant  numerical factors.   Due to our initial condition (\ref{initialw}), the right-hand side of (\ref{gYM}) vanishes for $\rho=0$, which corresponds to having $g^2_{YM}\to\infty$ in the IR, as expected in a confining theory. Moreover, $1/g^2_{YM}$ grows as we move towards the UV region $\rho\to\infty$, in agreement with the expected property of asymptotic freedom. Other gauge theory observables for these backgrounds, such as the Wilson loops, can be also analyzed (see ref.  \cite{Canoura:2008at}). Notice that, despite the regularity conditions we have imposed, in the flavored case $N_f\not=0$ the explicit calculation of the scalar curvature  for the linear dilaton solutions shows that the metric is singular at the origin of the radial coordinate. Notice that, as argued for other backgrounds,  it  is physically reasonable to expect that massless flavors drastically alter  the backreacted geometry in the deep IR.  However, as our initial conditions are such that the dilaton is finite at the origin, the value of the $g_{tt}$ component of the metric is also bounded and then, according to the criterium of  \cite{Maldacena:2000mw}, the singularity is ``good" and the background can be used to extract non-perturbative information of the dual gauge theory.

\subsubsection{Asymptotic $G_2$ cones}
\label{G2cones-untruncated-subsection}

		When $F(\rho=0)$ takes values in a certain range, the solutions of the BPS equations lead to  the metric (\ref{UV-asymp-metric})  at the UV,  which is the direct product of 2+1 dimensional Minkowski space and a $G_2$ cone. The solutions in this case are very similar in the UV to the ones discussed in subsection \ref{G2cone} (with better IR behavior) and we will not discuss them further here. Let us only mention that the asymptotic values of $F$, $w$ and $\gamma$  for $\rho\to\infty$ can be determined analytically and are given by:
\bear
&&F\,\approx\,{4\over 3}\,\rho\,+\,4\,(\,N_f-N_c)\,+\,\cdots\,\,,\rc\rc
&&w\,\approx\,{3(N_c-3N_f)\over 2\rho}\,+\,\cdots\,\,,
\qquad\qquad (\rho\to\infty)\,\,,
\rc\rc
&&\gamma\,\approx\,{1\over 3}\,-\,{N_f\over N_c}\,+\,\cdots\,\,\,. 
\eear

\setcounter{equation}{0}

\section{Flavors in the Klebanov-Strassler Model}
\label{KS}

The so-called Klebanov-Strassler (KS) solution \cite{Klebanov:2000hb} is dual
to a cascading, confining theory,  and has been a popular and successful laboratory in which
to study numerous issues related to gauge-gravity duality and to cosmology.
The gauge theory lives on a stack of regular and fractional D3-branes  at the tip of the
deformed conifold, as we now briefly review.

The deformed conifold is a regular, six dimensional, non-compact manifold defined by the equation $z_1\,z_2-z_3\,z_4 =\hat\mu^2$ in $\IC^4$. When the complex deformation parameter $\hat\mu$ is turned off, it reduces to the singular conifold, which is invariant under complex rescaling of the $z_i$.
The base of the conifold
 has $SU(2)\times SU(2)\times U(1)$ isometry and $S^2\times S^3$ topology. The deformation parameter breaks the scale invariance, produces a blown-up $S^3$ at the apex of the conifold and breaks the $U(1)$ isometry to  ${\mathbb Z}_2$. 

The low energy dynamics of $N$ regular and $M$ fractional D3-branes on the deformed conifold is described by a cascading ${\cal N}=1$ 4d gauge theory with gauge group $SU(N+M)\times SU(N)$ and bifundamental matter fields $A, B$ transforming as $SU(2)\times SU(2)$ doublets and interacting with a quartic superpotential $W_{KW}=\epsilon^{ij}\epsilon^{kl}A_iB_kA_jB_l$. 
The dual to this theory is the
KS solution \cite{Klebanov:2000hb}, that 
 is relevant for the $N=n\,M$ case, where $n$ is an integer. The related theory develops a Seiberg duality cascade which stops after $n-1$ steps when the gauge group is reduced to $SU(2M)\times SU(M)$. The regular KS solution precisely accounts for the physics of an $A\leftrightarrow B$-symmetric point in the baryonic branch of the latter theory, which exhibits confinement and $U(1)_R\rightarrow {\mathbb Z}_{2N}\rightarrow {\mathbb Z}_2$ 
 where the second
 breaking is
  due to the formation of a gluino condensate $\langle\lambda\lambda\rangle\sim\Lambda^3_{IR}$. The complex parameter $\epsilon$ is the geometric counterpart of this condensate.

In this section, we will discuss how the solution is modified when a smeared distribution
of D7-branes is introduced. In the dual theory, they correspond to fundamental fields, but
the precise way in which they couple to the rest of fields depends on the D7-brane embeddings,
as we will discuss below. 
In what follows, we only discuss cases in which the flavor D7-branes do not break any supersymmetry,
such that the four-dimensional ${\cal N}=1$ of the KS solution is preserved.
The material we summarize in this section was developed in \cite{Benini:2007gx},
\cite{Benini:2007kg}, \cite{Bigazzi:2008qq}.

\subsection{Backreaction with non-chiral flavors}
\label{nonchiralKS}

\subsubsection{Brane embeddings}
Let us start by choosing an appropriate family of supersymmetric D7-brane embeddings.
A particularly interesting
example is given by D7-branes wrapping the holomorphic 4-cycle defined by an equation of the form \cite{Kuperstein:2004hy}:
\be
z_1-z_2=\mu\,,
\label{kuperemb}
\ee
where $\mu$ is a constant.
It was shown in \cite{Kuperstein:2004hy} that this embedding is $\kappa$-symmetric and hence preserves the four supercharges of the deformed conifold theory. 

\begin{figure}[ht]
\begin{center}
\includegraphics[width=0.5\textwidth]{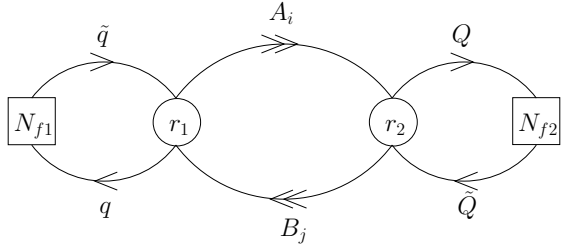}
\end{center}
\caption{The quiver diagram of the gauge theory. Circles are gauge groups, squares are flavor groups, and arrows are bifundamental chiral superfields. $N_{f1}$ and $N_{f2}$ sum up to $N_f$. } 
\label{quiverN}
\end{figure}

A D7-brane wrapping the 4-cycle defined above is conjectured to add a massless (if $\mu=0$) or massive (anti) fundamental flavor to a node of the KS model. The resulting gauge theory is said to be
``non-chiral'' because the flavor mass terms do not break the classical flavor symmetry of the massless theory. The related perturbative superpotential is, just as in the singular conifold case \cite{Ouyang:2003df}, which
we wrote in (\ref{KWsuperpquark}).
 The complex mass parameter $m$ in $W$ is mapped to the geometrical parameter $\mu$. The different fields are summarized in the quiver diagram of figure \ref{quiverN}.

By acting on this fiducial embedding (\ref{kuperemb}) with the generators of the broken symmetries,
we can build the family of embeddings over which we want to smear. 
This is the obvious generalization to the deformed conifold case
of the discussion in section \ref{AdS5X5} and, in fact, a generic non-chiral embedding
is still given by (\ref{genembeddingks}).

\subsubsection{The ansatz}

We now write the ansatz for the metric and forms. It is similar to the ansatz for the
KS solution, but, due to the presence of D7-branes, the RR one-form $F_{(1)}$ is non-trivial
and the dilaton runs. 
 It is useful to introduce the $g_i$ one-forms 
used in \cite{Klebanov:2000hb}:
\bear
g^1 &=& \frac{-\sin\theta_1\,d\varphi_1 -\cos\psi\sin\theta_2\,d\varphi_2 +\sin\psi\,d\theta_2}{\sqrt{2}}\,,\quad
g^2 = \frac{d\theta_1-\sin\psi\sin\theta_2\,d\varphi_2 -\cos\psi\,d\theta_2}{\sqrt{2}}\,,\rc
g^3 &=& \frac{-\sin\theta_1\,d\varphi_1 +\cos\psi\sin\theta_2\,d\varphi_2 -\sin\psi\,d\theta_2}{\sqrt{2}}\,,\quad
g^4 = \frac{d\theta_1+\sin\psi\sin\theta_2\,d\varphi_2 +\cos\psi\,d\theta_2}{\sqrt{2}}\,,\rc
g^5&=& d\psi +\cos\theta_1\,d\varphi_1 + \cos\theta_2\,d\varphi_2\,.
\label{gis}
\eear
The Einstein frame metric ansatz is\footnote{A more generic form of the ansatz was
used in \cite{Benini:2007gx},
\cite{Bigazzi:2008qq}. By requiring supersymmetry and performing some algebra, one
ends up with (\ref{metric}). We will skip those intermediate steps here for the sake of briefness.}:
\bear
ds^2 &=& h^{-1/2}(\tau)\,dx^2_{1,3} + h^{1/2}(\tau)
{1\over 2}\,\,\hat\mu^{{4\over 3}}\,\,e^{-{\phi(\tau)\over 3}}\,{\cal K}(\tau)\,\,
\Bigg[\,{1\over 
3{\cal K}^3(\tau)}\,\,\big(\,d\tau^2\,+\,(g^5)^2\,\big)\,+\rc\,
 &+& \cosh^2\Big({\tau\over 2}\Big)\,\Big(\,(g^3)^2\,+\, 
(g^4)^2\,\Big)\,
  +\,\sinh^2\Big({\tau\over 2}\Big)\,
\Big(\,(g^1)^2\,+\, (g^2)^2\,\Big)\,
\,\Bigg]\,\,,
\label{metric}
\eear
where $\hat \mu$ is the complex deformation parameter of the conifold, $dx^2_{1,3}$ denotes the four-dimensional Minkowski metric  and
${\cal K}(\tau)$, $h(\tau)$ and the dilaton $\phi$  are unknown
 functions of the radial variable to be determined.\footnote{
The relation of the $z_i$ complex variables as used above to the $\tau, \theta_1,\varphi_1,
\theta_2,\varphi_2,\psi$ coordinates can be found, for instance in \cite{Bigazzi:2008qq}.
 The embedding equation (\ref{genembeddingks})
 expressed in terms of the ``deformed conifold $\tau$ variable'' looks the same in terms of the ``backreacted ansatz $\tau$ variable''. See \cite{Bigazzi:2008qq} for details. }

For the forms we will adopt the following ansatz:
\bear
F_5&=&d h^{-1}(\tau)\wedge dx^0\wedge\cdots\wedge dx^3\,-\,\dot h \frac{\hat\mu^{\frac83}}{16}
e^{-\frac{2\phi}{3}}\sinh^2 \tau\, {\cal K}^4 g_1\wedge g_2\wedge g_3\wedge g_4\wedge g_5\,,\rc
B_2 &=& g_s\alpha'\frac{M}{2} \Bigl[ f\, g^1 \wedge g^2\,+\,k\, g^3 \wedge 
g^4 \Bigr]\,,\rc
H_3&=&  g_s\alpha'\frac{M}{2} \, \Bigl[ d\tau \wedge (\dot f \,g^1 \wedge g^2\,+\,
\dot k\,g^3 \wedge g^4)\,+\,{1 \over 2}(k-f)\, g^5 \wedge (g^1
\wedge g^3\,+\,g^2 \wedge g^4) \Bigr]\,,\rc
F_1&=&g_s{N_f\,p(\tau)\over 4\pi}\,\,g^5\,,\rc
F_3&=& g_s\alpha'\frac{M}{2} \Big\{ g^5\wedge \Big[ \big( F+\frac{g_sN_f\,p(\tau)}{4\pi}f\big)g^1\wedge g^2 + \big(1- F+\frac{g_sN_f\,p(\tau)}{4\pi}k\big)g^3\wedge g^4 \Big] +\rc
&&+\dot F d\tau \wedge \big(g^1\wedge g^3 + g^2\wedge g^4   \big)\Big\} \,,
 \label{theansatz}
\eear
where  $f=f(\tau)$, $k=k(\tau)$, $F=F(\tau)$ are functions of the radial coordinate (and
where the dot denotes derivatives with respect to $\tau$). We have implemented the self-duality
condition for $F_5$.

Notice that, consistently, $dF_1=-g_s \Omega$, where $\Omega$ is the 
symmetry preserving
D7-brane density distribution form analogous to (\ref{gsomega}):
\begin{equation}
\Omega =\frac{N_f}{4\pi}\Big(p(\tau)(\sin\theta_1 d\theta_1 \wedge d\varphi_1
+ \sin\theta_2 d\theta_2 \wedge d\varphi_2)
-\dot p(\tau) d\tau\wedge g_5\Big)\,.
\label{massive_2form}
\end{equation}
When quarks are massless \cite{Benini:2007gx}, one just has $p(\tau)=1$, whereas $p(\tau)$
 becomes non-trivial when quarks are massive. 
In what follows, we will keep $p(\tau)$ generic. We refer the reader to \cite{Bigazzi:2008qq} for the computation of $p(\tau)$  from the massive non-chiral brane embeddings 
(\ref{genembeddingks}).
The source contributions to the modified Bianchi identities for $F_3$ and $F_5$
\bear
&&dF_3\,=\,H_3\wedge F_1\,-\,g_s\Omega\wedge B_2\,\,,\rc
&&dF_5\,=\,H_3\wedge F_3\,-\,{1\over 2}\,g_s\Omega\wedge
 B_2\wedge  B_2\,,
 \label{newBianchi}
\eear
follow from the WZ term of the smeared D7-brane action \cite{Bigazzi:2008qq}.
Given (\ref{theansatz}) and (\ref{massive_2form}), the equations (\ref{newBianchi}) are
satisfied provided:
\be
\dot h \frac{\hat \mu^{\frac83}}{16}
e^{-\frac{2\phi}{3}}\sinh^2 \tau\, {\cal K}^4 = \text{const}
- \frac14 (g_s \alpha' M)^2 \left[ f - (f-k)F + \frac{g_s N_f }{4\pi}p(\tau) f\,k\right]\,\,.
\label{hdot}
\ee

\subsubsection{The BPS equations}\label{kssolution}

By requiring the vanishing of the bulk fermionic supersymmetry variations, one
finds a set of first order BPS equations. The computation is lengthy but straightforward
and was carried out in \cite{Benini:2007gx}
 (since in that paper $p(\tau)=1$, the substitution $N_f \to N_f\, p(\tau)$
 has to be implemented in the equations of \cite{Benini:2007gx}). 
In the present notation, the differential equations are:
\bear
\dot\phi &=& \frac{g_s N_f p(\tau)}{4\pi}e^{\phi}\,\,,\rc
\dot k &=& e^{\phi}\left(F+\frac{g_s N_f p(\tau)}{4\pi}\,f\right)\coth^2\frac{\tau}{2}\,,\nonumber \\
\dot f &=& e^{\phi}\left(1-F+\frac{g_s N_f p(\tau)}{4\pi}\,k\right)\tanh^2\frac{\tau}{2}\,,\nonumber \\
\dot F &=& \frac{1}{2}e^{-\phi}(k-f)\,,\rc
\frac{\dot {\cal K}}{\cal K} &=& \frac{2}{3{\cal K}^3 \sinh \tau } + \frac{\dot \phi}{3} -\coth \tau\,\,,
\label{BPSKS}
\eear
supplemented  by the algebraic constraint
\be
e^{-\phi}(k-f)=\tanh\frac{\tau}{2}- 2F\,\coth\tau + \frac{g_s N_f p(\tau)}{4\pi}\left[k\,\tanh\frac{\tau}{2} -f\,\coth\frac{\tau}{2}\right]\,.
\label{fluxconstr}
\ee
Quite remarkably, the equations (\ref{BPSKS})-(\ref{fluxconstr})
can be (almost) explicitly integrated. In the following, 
we will use notations similar to those employed in section \ref{AdS5X5}.
We introduce an arbitrary value of the radial coordinate $\tau_*$ at which the
dilaton is $\phi_*$. Then, we can write the dilaton as:
\be
e^{\phi-\phi_*}=\frac{1}{1+\epsilon_* \int_{\tau}^{\tau_*} p(\xi) d\xi}\,\,,
\label{dilKS}
\ee
where we have introduced the deformation parameter which 
weighs the flavor loops as:
\be
\epsilon_* = \frac{N_f}{16\pi^2 M}\lambda_* \qquad {\rm with}\qquad
\lambda_* \equiv 4\pi g_s M\, e^{\phi_*}\,\,.
\label{KSdefs}
\ee
Let us also introduce a function:
\be
\eta(\tau) = \epsilon_* e^{\phi-\phi_*} \int_0^\tau (\sinh 2 \xi -2\xi) p(\xi) d\xi\,\,.
\ee
Then, we can integrate for the rest of the functions of the ansatz:
\bear
{\cal K}\,&=&\,{\big[\sinh 2\tau-2\tau\,+\,\eta(\tau)\big]^{{1\over 3}}\over
2^{{1\over 3}}\,\sinh\tau}\,\,,\qquad\quad F= \frac{\sinh \tau - \tau}{2 \sinh \tau}\,, \rc
f&=& e^{\phi}\,\frac{\tau \coth \tau -1}{2 \sinh \tau} (\cosh \tau -1)\,\,,\qquad
k= e^{\phi}\,\frac{\tau \coth \tau -1}{2 \sinh \tau} (\cosh \tau +1)\,\,.
\label{fkFexplicit}
\eear
Finally, the function $h$ can be obtained by integrating (\ref{hdot}).
The KS solution without flavors \cite{Klebanov:2000hb} is obtained by taking $\epsilon_*=0$, such that
the dilaton is constant and $\eta(\tau)=0$. For $p(\tau)=1$, we find the solution
backreacted with massless flavors \cite{Benini:2007gx}. 
In this case, the integrals for the dilaton and $\eta(\tau)$ can be explicitly performed:
\bear
\eta(\tau) &=& \epsilon_* e^{\phi-\phi_*} (\sinh^2 \tau - \tau^2) \,\,,\rc
e^{\phi-\phi_*} &=& \frac{1}{1+\epsilon_* ( \tau_* - \tau)}\,\,. \qquad\qquad\qquad
(\text{for}\, p(\tau)=1)
\eear
In this massless case, the solution has a curvature singularity in the IR $\tau=0$.
Some cases where $p(\tau)$ is non-trivial
were discussed in \cite{Bigazzi:2008qq}.

\subsubsection{Some physical features}

The solution presented in the preceding  sections has been used to extract some
of the physics encoded in the unquenched background. In \cite{Benini:2007gx},
the running of the couplings and anomalies were discussed.  As anticipated above, in 
\cite{Bigazzi:2008qq}, the solution with massive flavors was found. 
Quark masses erase the IR singularity in the same way as explained in section
\ref{heuristiczzz} or in section \ref{massiveKW}.
Quark-antiquark potentials, screening lengths and associated quantum phase transitions
were discussed in the same paper. Finally, in \cite{unquenchedmesons}, it was computed
how the screening effects due to unquenched fundamental matter affect the mass spectra of
the KS model, with results similar to section \ref{sec:screening}. 
Due to space constraints, we cannot go explicitly through all of these features and we refer
the interested reader to the original papers. Here, we will just briefly discuss how the
solution captures the phenomenon of a duality wall \cite{Benini:2007gx} and how gauge
groups ranks
change upon Seiberg duality.

We will make use of the following holographic formulae, 
which can be derived in the $\mathcal{N}=2$ orbifold case by looking at the lagrangian of the low energy field theory living on probe (fractional) D3-branes:%
\begin{equation}
 \label{holographic relations}
\frac{4\pi^2}{g_{YM}(l)^2} + \frac{4\pi^2}{g_{YM}(s)^2} = \frac{\pi\, e^{-\phi}}{g_s} \,\,,\qquad
\frac{4\pi^2}{g_{YM}(l)^2} - \frac{4\pi^2}{g_{YM}(s)^2} = 
\frac{2 \pi\,e^{-\phi}}{g_s} \Bigl[\frac{1}{4\pi^2 \alpha'} \int_{S^2} B_2 - \frac12 \; 
 \Bigr]\,\,.
\end{equation}
The labels  $(l),(s)$  in (\ref{holographic relations}) refer to the gauge group with the larger or smaller rank. Strictly speaking, these formulae need to be corrected for small values of the gauge couplings and are only valid in the large 't Hooft coupling regime (see \cite{Benvenuti:2005wi,Strassler:2005qs,Benini:2006hh}), which is the case under consideration. Moreover, they are also expected to be precise just in the UV region, where the cascade takes place
and the region on which we will focus below.
 The expressions (\ref{holographic relations})
   give positive squared couplings only if the expression inside the
 square bracket is in the range $[-\frac12,\frac12]$. Define:
\be
b_0(\tau) \equiv \frac{1}{4\pi^2\alpha'}\int_{S^2} B_2 = \frac{g_s M}{\pi}f
= \frac{\lambda_*}{8\pi^2} \frac{\tau -1}{1-\epsilon_*(\tau-
\tau_*)}\,,\qquad
\tilde b_0 \equiv b_0 - [b_0] \in [0,1] \,,
\label{b0val}
\ee
where $[b_0]$ denotes the integer part of $b_0$.
In order to get the explicit expression for $b_0$ we have integrated over the $S^2$ parameterized by
$\theta_1=\theta_2$, $\varphi_1 = 2\pi - \varphi_2$, $\psi = const$ \cite{Benini:2007gx}, considered the UV limit of (\ref{dilKS}), (\ref{fkFexplicit}) such that
$p(\tau)\approx 1$ and $f\approx k \approx e^\phi (\tau -1)/2$ and inserted the
definitions (\ref{KSdefs}).
 Now we see that what we
 have to insert in (\ref{holographic relations}) is indeed $\tilde b_0$.
  This is the physical content of the cascade: at a given energy scale we must perform a large gauge transformation on $B_2$ in supergravity to shift $\int B_2$ by a multiple of $4\pi^2\alpha'$
    to get a field theory description with positive squared couplings.

Let us restrict our attention to an energy range, between two subsequent Seiberg dualities, 
where a field theory description in terms of specific ranks holds.
When flowing towards the IR, $\tilde b_0$ decreases from 1 to 0. From 
(\ref{holographic relations}) and inserting the solution, we can find an expression
for each of the gauge couplings:
\be
\frac{1}{\lambda_l}=\frac{1}{\lambda_*}(1-\epsilon_* (\tau-\tau_*))\, \tilde b_0\,\,,\qquad
\frac{1}{\lambda_s}=\frac{1}{\lambda_*}(1-\epsilon_* (\tau-\tau_*))\, (1-\tilde b_0)\,\,.
\ee
 In this energy range, the coupling $\lambda_l$ 
 starts different from zero and flows to $\infty$ at the end of this range, 
 where a Seiberg duality on its gauge group is needed. The coupling $\lambda_s$ of 
 the gauge group with smaller rank is the one which starts very large (actually 
divergent) after the previous Seiberg duality on its gauge group, and then flows toward weak coupling. 

\begin{figure}[ht]
\begin{center}
\includegraphics[width=0.45\textwidth]{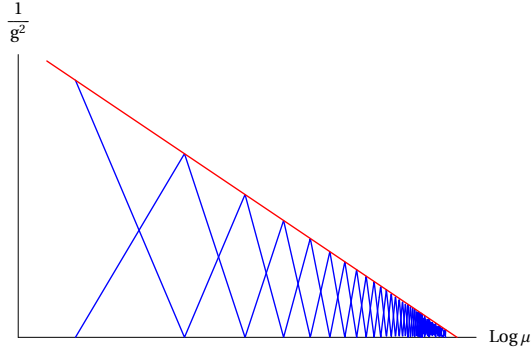} 
\end{center}
\caption[wall]{Qualitative plot of the running gauge couplings as functions of the logarithm of the energy scale in the cascading gauge theory. The blue lines are the inverse squared gauge couplings, while the red line is their sum. \label{wall}}
\end{figure}

The qualitative picture of the RG flow in the UV can be extracted from our supergravity solution  even without discussing
 the precise radius-energy relation, simply recalling that the radius must be a monotonic function of the energy scale.
First, notice that at a finite $\tau$
(and therefore at a finite energy scale $E_{UV}$), the dilaton diverges making both gauge couplings diverge.
This happens at:
\be
\tau_{dw}= \tau_* + \frac{1}{\epsilon_*}\,\,.
\ee
From (\ref{b0val}), we see that the derivative $\frac{db_0}{d\tau}$ grows unbounded near 
$\tau_{dw}$, meaning that the interval (in $\tau$) between Seiberg dualities becomes shorter and
shorter.
 The Seiberg dualities pile up the more we approach the UV cut-off $E_{UV}$.
The picture which stems from the flavored Klebanov-Tseytlin/Strassler solution is that 
$\tau_{dw}$ is a so-called ``Duality Wall", namely an accumulation point of energy scales at which a Seiberg duality is required in order to have a weakly coupled description of the gauge theory \cite{Strassler:1996ua}. Above the duality wall, Seiberg duality does not proceed and a weakly coupled dual description of the field theory is not known. 
See Figure \ref{wall}. Nevertheless, in full analogy with the discussion of section \ref{AdS5X5}, the
derivative of the holographic $a$-function changes sign at a finite distance in $\tau$ below $\tau_{dw}$
so one should not trust the solution all the way up to the singular point $\tau_{dw}$.

Duality walls were studied in the context of quiver gauge theories first by Fiol \cite{Fiol:2002ah} and later in a series of papers by Hanany and collaborators \cite{Hanany}. 
To our knowledge, the solution above
 is the only explicit realization of this exotic ultraviolet phenomenon on the supergravity side of the gauge/gravity correspondence.

To end this section, we discuss how the effective number of regular and fractional
D3-branes change when undergoing a step of the cascade of  Seiberg dualities. 
We will not compute the explicit shift in $\tau$ but rather the shift in the function $f$ ($\approx k$).
From (\ref{b0val}), we have:
\begin{equation}
\begin{split}
b_0 (\tau) \to b_0(\tau') = b_0 (\tau) - 1 
\end{split}
\quad 
\Longrightarrow \quad 
\begin{split}
f(\tau) &\to f(\tau') = f(\tau)  - \frac{\pi}{g_s M} \,\,.
\end{split} \label{shift f k}
\end{equation} 
On the other hand, we compute the effective number of branes at a given
energy scale by integrating the
appropriate RR-forms:
\bear 
N_{eff} (\tau) &\equiv&  \frac{1}{(2\pi)^4 g_s \alpha'^2} \int_{{\cal M}_5} F_5 \,=N_0 +
  \,{g_s M^2\over \pi}\,\Big[\,
f +{g_s N_f\over 4\pi}f^2 \Big]\,\,,\rc
M_{eff}(\tau) &\equiv& \frac{1}{4\pi^2g_s \alpha'} \int_{S^3} F_3 = M \Bigl[ 1 + \frac{g_s N_f}{2\pi} f
 \Bigr]\;.
 \label{neffmeff}
\eear
In these expressions we have substituted (\ref{theansatz}), (\ref{hdot}) and
already taken the UV limit $f = k$ and $p(\tau)=1$.
The $S^3$ for the second integral is the one parameterized by
$\theta_2=\text{constant}$, $\varphi_2=\text{constant}$. Notice that $N_{eff}$ and $M_{eff}$ are not quantized. This is because they are {\it Maxwell} charges, as opposed to
{\it Page} charges. See \cite{Benini:2007gx} for thorough explanations.

We can compute how $N_{eff}$, $M_{eff}$ vary in a Seiberg duality step
(\ref{shift f k}).
A bit of algebra shows that:
\bear
M_{eff} (\tau) &\to& M_{eff} (\tau') = M_{eff} (\tau) - \frac{N_f}{2} \label{scaling sugra Meff}\,\,,
\nonumber\\
N_{eff} (\tau) &\to& N_{eff} (\tau') = N_{eff} (\tau) - M_{eff} (\tau) + \frac{N_f}{4} \,\,,
\nonumber
\label{scaling sugra 2}
\eear
whereas $N_f$ remains unchanged. A careful analysis in 
\cite{Benini:2007gx} showed that this is in full agreement with field theory expectations.

\subsection{Backreaction with chiral flavors}

\begin{figure}[ht]
\begin{center}
\includegraphics[width=0.3\textwidth]{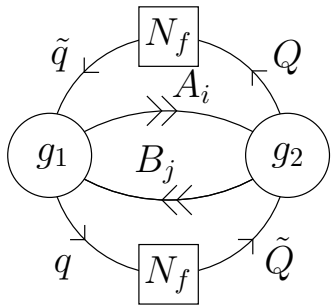}
\caption{Quiver diagram of the KS theory with chiral flavors. }
\end{center}
\label{quiverBenini}
\end{figure}

In a remarkable paper \cite{Benini:2007kg}, Benini discussed the solution dual to
having smeared chiral flavors on the conifold. In the probe approximation, the D7-brane embeddings
that correspond to chiral flavors were discussed in \cite{Ouyang:2003df}, \cite{Levi:2005hh}.
How these flavors transform under the gauge groups is shown in the quiver diagram 
\ref{quiverBenini}. In this case, 
the quiver theory is not self-similar under the duality cascade; in each step 
of the cascade a meson field is generated. Its couplings to the 
rest of the fields are, however, irrelevant \cite{Benini:2007kg}.

The backreacted solution of \cite{Benini:2007kg} uses the singular conifold and therefore it
can be considered as the deformation of the Klebanov-Tseytlin solution  \cite{Klebanov:2000nc}
due to smeared chiral flavors. 
From the gravity point of view, the extra complication with respect to section \ref{nonchiralKS}
is that the worldvolume gauge field on the D7s has to be turned on.
In fact, this is crucial when matching the shifts in the ranks of the gauge groups upon Seiberg
dualities to the supergravity background (there are subtle differences with respect to the
non-chiral case). We will not report further on this solution here, but
refer the reader to \cite{Benini:2007kg}.

\setcounter{equation}{0}

\section{Models with cohomogeneity 2}\label{models2}

In this section we present some situations in which, even smearing the flavor
branes, the system cannot be reduced to a one-dimensional problem. In fact,
the different fields will depend on two different radial coordinates and,
accordingly, one has to solve partial differential equations rather than
ordinary differential equations.

\begin{figure}[ht]
\begin{center}
\includegraphics[width=0.45\textwidth]{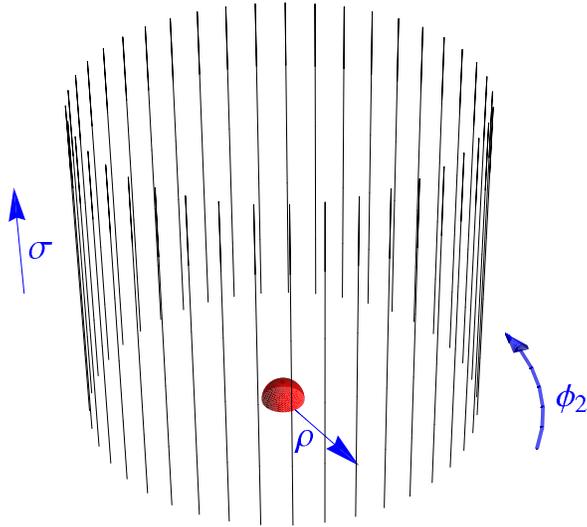}
\end{center}
\caption{A qualitative plot of the situation with
cohomogeneity 2 models. The red dot in the center represents the color branes and each
vertical line is a flavor brane. Taking a lot of them smeared along the 
$\phi_2$ angle, the rotational symmetry associated to this angle is
effectively recovered. Any function of the ansatz depends on the radial coordinates
$\rho$ and $\sigma$.}
\label{smearingfig2} 
\end{figure}

In order to provide a heuristic picture, the situation is depicted in
figure \ref{smearingfig2}. Concretely, we will refer here to the model of section 
\ref{secN2}, but the situation  is
very similar
for all the cases discussed in this section.

In figure \ref{smearingfig2}, the color branes are placed at the tip of a Calabi-Yau
($\sigma$ is a radial coordinate along the CY, the rest of the  directions of the CY
are omitted from the plot).
The $\rho-\phi_2$ plane is transverse both to the color branes and to the CY. Each flavor
brane lies at a point in this plane and is extended along $\sigma$.
Distributing the flavor branes along $\phi_2$, it is possible to recover 
(in the smeared limit)
the associated $U(1)$ isometry. On the contrary, as is apparent from 
figure \ref{smearingfig2}, there is no way in which one can place the flavor brane to
recover the full radial symmetry. Hence, the solution associated to this brane configuration
must be cohomogeneity two, meaning that all functions of the eventual ansatz
will depend on $\rho$ and $\sigma$.

As a matter of fact, if one wishes to construct a deformation
of $AdS_5\times S^5$ with smeared flavor such that the supersymmetry 
preserved is ${\cal N}=2$ (rather than ${\cal N}=1$ as 
 in section \ref{AdS5X5}), the solution would have cohomogeneity two
and, presumably, would share some similarities with the examples presented in
this section.
This is an interesting open problem for the future.

\subsection{A dual to (3+1)-dimensional ${\cal N}=2$ SQCD-like theory}
\label{secN2}

In this section, we study the dual solution to the brane intersection summarized in
table \ref{table: N2}. The gauge theory lives on $N_c$
D5-branes wrapping a two-sphere with the appropriate twisting
to preserve eight supercharges, {\it i.e.} ${\cal N}=2$ in the
effective four-dimensional low energy theory. Geometrically, it corresponds
to wrapping the branes along a compact SLag two-cycle inside a 
non-compact Calabi-Yau two-fold.
This leaves two flat transverse dimensions which are identified with
the moduli space corresponding to giving vevs to the complex scalar
inside the ${\cal N}=2$ vector multiplet.
The $N_f$ flavor D5-branes do not further break supersymmetry and 
provide
fundamental hypermultiplets in order to
build 
${\cal N}=2$ SQCD. 
They are extended in the non-compact $\s$ direction and, thus, their volume
is infinite, making exactly zero the
effective four-dimensional gauge coupling living on them.
They would provide a global symmetry group $U(N_f)$ if they were placed on top of each other,
but due to the smearing, only $U(1)^{N_f}$ is left.
The dual solution without flavors was found in \cite{Gauntlett:2001ps}, and the flavored
case was discussed in \cite{Paredes:2006wb}.

\begin{table}[ht!]
\begin{center}
\begin{tabular}{|c|c|c|c|c|c|c|c|}
\multicolumn{1}{c}{ }
&
\multicolumn{1}{c}{ }
&
\multicolumn{4}{c}{$\overbrace{\phantom{\qquad\ \qquad
\qquad
\ \ \,}}^{CY_2}$}
&
\multicolumn{2}{c}{$\overbrace{\phantom{\ \quad
\ \ \,}}^{\mathbb{R}_2}$}
\\
\hline
\multicolumn{1}{|c|}{ }
&
\multicolumn{1}{|c|}{$x_{1,3}$}
&\multicolumn{1}{|c|}{$\sigma$}
&\multicolumn{1}{|c|}{$\phi_1$}
&\multicolumn{1}{|c|}{$\tt$}
&\multicolumn{1}{|c|}{$\tvf$}
&\multicolumn{1}{|c|}{$\r$}
&\multicolumn{1}{|c|}{$\phi_2$}
\\
\hline
$N_c$ D5 &$-$&$\cdot$&$\cdot$&$\bigcirc$&$\bigcirc$&$\cdot$&$ \sim $\\
\hline
$N_f$ D5 &$-$&$-$&$\bigcirc$&$ \sim $&$ \sim $&$\cdot$&$ \sim $\\
\hline
\end{tabular}
\caption{A scheme of the setup:
for the brane configuration,
a line $-$ means that the brane spans a non-compact dimension, a point
$\cdot$ that it is point-like in that direction, a circle
$\bigcirc$ that it wraps a compact cycle and $\sim$ indicates smearing in the
direction.
 Above, it is shown which directions spanned the Calabi-Yau
and which the transverse plane before backreaction.
\label{table: N2}}
\end{center}
\end{table}
We start by writing an ansatz for the metric consistent with the symmetries of the problem.
In Einstein frame:
\bear
ds_{10}^2 &=& g_s N_c \a' e^\frac{\Phi}{2} \left[ 
\frac{1}{g_s N_c \a'} dx_{1,3}^2 +  z (d\tt^2 +
\sin^2 \tt d\tvf^2) + \right.\rc
&&
\left.
+ e^{-2\Phi} (d\rho^2 + \rho^2 d\phi_2^2)
+ \frac{e^{-2\Phi}}{z} \left( d\s^2 + \s^2
(d\phi_1 + \cos \tt d\tvf)^2 
\right) \right]\,\,,
\label{metricN2}
\eear
where $z$ and $\Phi$ depend on both radial coordinates $\rho$, $\sigma$.
The Calabi-Yau twofold directions are $0 \leq \tt \leq \pi$, $0\leq \tvf < 2\pi$, $0\leq\s<\infty$,
$0\leq \phi_1 < 2\pi$
(of course, in this solution with fluxes there is not a 
Calabi-Yau any more, but it can be
thought of as a deformation of the Calabi-Yau that was present 
before backreaction).
The coordinates
$0\leq \rho < \infty$, $0 \leq \phi_2 < 2 \pi$ 
span the transverse two-dimensional plane, so they should be identified with the moduli space,
and therefore rotations in $\phi_2$ are related to the $U(1)_R$ symmetry of the field theory.
Out of the $SU(2)_R$ symmetry, only its diagonal $U(1)_J$ is manifest in the geometry, as
rotations in $\phi_1$. The extra $SO(3)$ isometry which acts on $\tt,\tvf,\phi_1$ 
 does not play
a role in the low energy ${\cal N}=2$ SQCD theory \cite{Gauntlett:2001ps}.

As anticipated in table \ref{table: N2}, we want to consider a set of $N_f$ D5-branes extended
in $x_{1,3}$, $\sigma$ and wrapped in $\phi_1$. They lie at fixed $\rho=\rho_Q$, where $\rho_Q$ is
proportional to the modulus of the mass of the fundamental hypermultiplets. These D5-branes are
homogeneously smeared over the $S^2$ parameterized by $\tilde \theta$, $\tilde \varphi$ and on the
angle $\phi_2$, which corresponds to the phase of the mass of the hypers.
This distribution is described by the four-form: 
\beq
\Omega=\frac{N_f}{8\pi^2} 
 \delta(\rho-\rho_Q)\ \sin \tt d\r \wedge d\phi_2 
\wedge d\tt \wedge d\tvf \,\,,
\label{y4}
\eeq
such that the source-modified Bianchi identity for $F_{(3)}$ reads:
\beq
dF_{(3)}= 2\kappa_{(10)}^2 T_5 \Omega=
 g_s \a' \ \frac{N_f}{2} 
 \delta(\rho-\rho_Q)\ \sin \tt d\r \wedge d\phi_2 
\wedge d\tt \wedge d\tvf \,\,.
\label{newdF}
\eeq
We can write an ansatz for $F_{(3)}$ consistent with this expression:
\bear
F_{(3)}&=& N_c g_s \a' \left[ -g' d\phi_2 \wedge d\r \wedge
(d\phi_1 + \cos \tt d\tvf) - \dot g d\phi_2 \wedge d\s \wedge
(d\phi_1 + \cos \tt d\tvf)+\right. \rc
&& \left. + (g + \frac{N_f}{2N_c}\Theta(\rho-\rho_Q))
 \sin \tt d\phi_2\wedge d\tt \wedge 
d\tvf \right]\,.
\label{F3N2}
\eear 
where $\Theta$ is the Heaviside step function\footnote{Notice that,
as opposed to section \ref{CNPwilson} where a Heaviside function was introduced as an approximation to
the effect of the massive flavors, 
the $\Theta$ here is exactly what comes from the family of D-brane embeddings considered, since
they all lie at fixed $\rho=\rho_Q$.}, 
$g$ a new function of $\rho$ and $\sigma$ that
needs to be determined, and we have introduced the following notation for the partial
derivatives:
\beq
' \equiv \partial_\rho \,,\qquad \dot{}  \equiv \partial_\s
\label{prime&dot}
\eeq
The next step is to insert the ansatz (\ref{metricN2}), (\ref{F3N2}) into the type 
IIB supersymmetry transformations $\delta \psi_\mu = \delta \lambda =0$,
as outlined in section \ref{sec: kappa}. 
This procedure was carefully performed in 
\cite{Paredes:2006wb},
whereas here we just quote the resulting system of first order equations:
\bear
&&g + \frac{N_f}{2N_c}\Theta(\rho-\rho_Q) = - \r z' \,,\qquad\quad\qquad\qquad
g' = -2 e^{-2\Phi} \r\s \dot \Phi
 \,,\qquad\qquad\rc
 &&e^{2\Phi}= \frac{\s}{z \dot z}
 \,, \qquad\quad\quad\qquad
\dot g = - z^{-2} e^{-2\Phi} \s (g+\frac{N_f}{2N_c}\Theta(\rho-\rho_Q)) +
 2 z^{-1} \r\s e^{-2\Phi} \Phi'\,.\qquad
\label{neweq4}
\eear
It is easy to check that the last equation is not independent of the previous ones and
that equations (\ref{neweq4}) ensure the equation of
motion for the 3-form
$d(e^\Phi\ {}^*F_{(3)}) = 0$. This system of equations can be recast as a single, non-linear,
second order PDE for $z(\rho,\sigma)$:
\beq
\s \frac{N_f}{2N_c}\delta(\rho-\rho_Q) +
\r z (\dot z - \s \ddot z)=\s (\r \dot z^2 + z' + \r z'')\,\,.
\label{neweqinz}
\eeq
Once $z(\rho,\sigma)$ is computed, $g$ and $\Phi$ are read from (\ref{neweq4}).
In general, the equation (\ref{neweqinz}) cannot be solved explicitly.
In the unflavored case $N_f=0$, there is in fact an exact solution
\cite{Gauntlett:2001ps} (see \cite{Paredes:2006wb} and the first paper
of \cite{DiVecchia:2002ks} for the adaptation of the solution \cite{Gauntlett:2001ps} to 
the present coordinate system). 

Equation (\ref{neweqinz}), however, can be studied numerically \cite{Paredes:2006wb}.
We will not pursue that here, but we will verify using (\ref{neweq4}) that the
expected beta-function for the gauge coupling stems from the differential equations.
In order to read the effective four-dimensional gauge coupling from the geometry, we consider 
a ``color" D5-brane probing the Coulomb branch of the theory, namely, a D5 wrapping
the $S^2$ parameterized by $\tilde \theta$, $\tilde \varphi$, sitting at $\sigma=0$
\cite{DiVecchia:2002ks}. After integrating the volume of the $S^2$, we find:
\be
\frac{1}{g_{YM}^2}=\frac{N_c}{4\pi^2}(z|_{\sigma=0})\,\,.
\ee
Thus, in order to understand the running of the coupling it is not necessary to know the
geometry everywhere, but just at $\sigma=0$. From the second equation of (\ref{neweq4}),
we see that $g$ is a constant at $\sigma=0$, which then results in the fact that the first 
equation of (\ref{neweq4}) can be trivially integrated. 
But before doing that, let us find out which is the value of $g|_{\sigma=0}$.
With that purpose, let us consider the normalization condition:
\beq
\frac{1}{2\kappa_{(10)}^2}\int F_{(3)} = N_c T_5\,\,,
\label{quantt}
\eeq
where we have to integrate along $\phi_1$, $\phi_2$ and an angle built in the
``plane" of the two radial directions $\rho,\sigma$
(heuristically, think of introducing some polar coordinates
$r,\theta$ such that $\rho=r \sin\theta$ and $\sigma=r \cos\theta$. Then
we want to integrate in $\theta$ from 0 to $\frac{\pi}{2}$ at fixed and large
$r$). Inserting (\ref{F3N2}),
we find
\be
g|_{(\s=\infty ,\r=0 )} - g|_{(\s=0,\r=\infty )}=1\,\,.
\ee
But from the first equation in (\ref{neweq4}) we read that 
$g|_{\r=0 }=0$ and, thus, $g|_{\s=0}=-1$.
We are now ready to integrate the first equation in (\ref{neweq4})
at $\sigma=0$:
\be
\frac{N_c}{4\pi^2}(z|_{\sigma=0})=\frac{1}{g_{YM}^2}=\frac{1}{4\pi^2}\Big[
\Big(N_c - \frac{N_f}{2} \Theta(\r-\r_Q)\Big) \log \r
+ \frac{N_f}{2} \Theta(\r-\r_Q) \log \r_Q+const\Big]\,\,,
\label{flavgym}
\ee
where the next to last term comes from requiring continuity of the metric
at $\rho=\rho_Q$.
Making use of the radius-energy relation
$\rho=\frac{\mu}{\Lambda}$ found in \cite{DiVecchia:2002ks},
we get:
\be
\b (g_{YM})(\m) = -\frac{g_{YM}^3}{8\pi^2} \Big(N_c - \frac{N_f(\m)}{2}\Big)\,\,,
\label{bgym}
\ee
where $N_f (\m)$ is defined as the number of flavors for which the
modulus of their masses is
smaller than the scale. Matter fields with bigger mass are holomorphically decoupled
at lower scales, as expected. 
The expression (\ref{bgym}) fits field theory expectations and is a non-trivial check
of the described unquenched set-up.
For further discussion of this model, see \cite{Paredes:2006wb}.

\subsection{Flavors in lower-dimensional SQCD Models}

 The approach described in subsection \ref{secN2} can be also applied to construct supergravity duals of SQCD-like models in two and three dimensions by considering lower dimensional branes wrapping different cycles of Calabi-Yau manifolds. In this subsection we will review two of such constructions. First of all, following refs. \cite{Maldacena:2000mw, Arean:2008az}, we will consider the case of D3-branes wrapping a two-cycle of a Calabi-Yau twofold, which is dual to a two-dimensional gauge theory with ${\cal N}=(4,4)$ supersymmetry. Secondly, we will review the similar construction of refs. \cite{Maldacena:2000mw, Divecchia, Ramallo:2008ew} of the gravity dual of three-dimensional ${\cal N}=4$ gauge theories from D4-branes wrapping two-cycles in a $CY_2$. Backgrounds  dual to 2d and 3d flavored theories with reduced supersymmetry have been also constructed \cite{Gaillard:2008wt,Arean:2009gc,D3D4D5}, and they will be also very briefly reviewed.

\subsubsection{Two-dimensional theories}
Let us consider the following setup for two sets of D3-branes in a Calabi-Yau cone of complex dimension two:
 \begin{center}
\begin{tabular}{|c|c|c|c|c|c|c|c|c|c|c|}
\multicolumn{3}{c}{ }&
\multicolumn{4}{c}{$\overbrace{\phantom{\qquad\qquad\qquad}}^{\text{CY}_2}$}\\
\hline
&\multicolumn{2}{|c|}{$\mathbb{R}^{1,1}$}
&\multicolumn{2}{|c|}{$S^2$}
&\multicolumn{2}{|c|}{$N_2$}
&\multicolumn{4}{|c|}{$\mathbb{R}^{4}$}\\
\hline
$N_c$\,\,\,D$3$ (color) &$-$&$-$&$\bigcirc$&$\bigcirc$&$\cdot$&$\cdot$&$\cdot$&$\cdot$&$\cdot$&$\cdot$\\
\hline
$N_f$\,\,\,D$3$ (flavor) &$-$&$-$&$\cdot$&$\cdot$&$-$&$-$&$\cdot$&$\cdot$&$\cdot$&$\cdot$\\
\hline
\end{tabular}
\label{44flavored-array}
\end{center}
where  $S^2$ represents the directions of a compact two-cycle and $N_2$ are the directions of the corresponding normal bundle. Notice also that the symbols ``$-$" and ``$\cdot"$ represent respectively unwrapped worldvolume directions and transverse directions, while a circle denotes wrapped directions. Let us parameterize  the cycle by means of two angular coordinates $(\theta, \phi)$ and let $\sigma$ be the radial coordinate of the CY cone. The ansatz for the string frame metric  which we will adopt is the following:
\bear
&&ds_{st}^2\,=\,H^{-{1\over 2}}\,\,\Big[\,dx_{1,1}^2\,+\,{z\over m^2}\,
\Big(\,d\theta^2\,+\,\sin^2\theta\,d\phi^2\,\,\Big)\,\Big]\,+\,\rc\rc
&&\qquad\qquad+\,
H^{{1\over 2}}\,\,\,\Big[\,{1\over z}\,\,\Big(\,
d\sigma^2\,+\,\sigma^2\,
\Big(\,d\psi+\cos\theta d\phi\,\Big)^2\,\Big)
\,+\,
d\rho^2\,+\,\rho^2\,d\Omega_3^2\,\Big]\,\,,
\qquad\qquad
\label{D3metric}
\eear
where $m$ is a constant with units of mass which, for convenience, we will take as:
\beq
{1\over m^2}\,=\,\sqrt{4\pi g_s N_c}\,\,\alpha'\,\,.
\label{m}
\eeq
Notice that in this setup there is another radial coordinate $\rho$, which represents the distance along $\mathbb{R}^{4}$, the directions 
orthogonal to both the D3-brane worldvolume and the CY cone. Moreover,  $d\Omega_3^2$ is the metric of a unit three-sphere. Furthermore, the function $z$ (which controls the size of the cycle) and   the warp factor  $H$ should be considered as functions of the two radial variables $(\rho, \sigma)$: $H\,=\,H(\rho,\sigma)$, $z\,=\,z(\rho,\sigma)$.

As in any background created by D3-branes, our solution should be endowed with 
a self-dual RR five-form $F_5$, that we write as:
\beq
F_5\,=\,{\cal F}_5\,+\,{}^*{\cal F}_5\,\,.
\label{F5}
\eeq
The presence of $N_f$ flavor D3-branes induces a violation of the Bianchi identity of $F_5$. Indeed, the WZ term of the flavor brane action contains the term $\sum_{N_f}\,\int_{{\cal M}_4}\,\hat C_4$ that acts as a source for this violation. Actually,  the smearing procedure amounts to performing the following substitution in this term:
\beq
\sum_{N_f}\,\int_{{\cal M}_4}\,\hat C_4\,\rightarrow\,\int_{{\cal M}_{10}}\,
\Omega\wedge C_4\,\,,
\eeq
where $\Omega$ is a six-form proportional to the volume  form of the  space transverse to the worldvolume of the flavor brane. The modified Bianchi identity takes the form 
$dF_5=2\kappa_{10}^2\,T_3\,\Omega$. As in the four-dimensional example discussed in  subsection \ref{secN2}, we shall locate the flavor branes at a particular value $\rho=\rho_Q$ of the $\rho$ coordinate (the mass of the matter fields is just $m_Q=\rho_Q/(2\pi\alpha')$). Moreover, we will smear the $N_f$ D3-branes along the angular directions $(\theta, \phi)$ of the cycle as well as along the external three-sphere.   The corresponding smearing form is:
\beq
\Omega\,=\,-{N_f\over 8\pi^3}\,\,\,\delta(\rho-\rho_Q)\,d\rho\wedge \omega_3\wedge\omega_2\,\,,
\label{Omega}
\eeq
with $\omega_2=\sin\theta d\theta\wedge d\phi$ and $\omega_3$ is the volume element of the external $S^3$ with line element $d\Omega_3^2$ (the minus sign in (\ref{Omega}) is due to the orientation of the worldvolume required by supersymmetry).  It is clear  that the modified Bianchi identity in this case is:
\beq
dF_5\,=\,-2\pi\,g_s\,(\alpha')^2\,N_f\,\delta(\rho-\rho_Q)\,d\rho\wedge \omega_3\wedge\omega_2\,\,.
\label{newBianchi2d}
\eeq
Accordingly,  let us represent $ F_5$ as in (\ref{F5}) with ${\cal F}_5$ being given by:
\beq
{\cal F}_5\,=\,f_5\,-\,2\pi\,g_s\,(\alpha')^2\,N_f\,\Theta(\rho-\rho_Q)\,\omega_3\wedge \omega_2\,\,,
\label{calF5-flavored}
\eeq
with  $f_5$ such that  $df_5=0$. We shall represent $f_5$ in terms of a potential ${\cal C}_4$ as $f_5=d{\cal C}_4$, where  ${\cal C}_4$ is given by the ansatz:
\beq
{\cal C}_4\,=\,g\,\omega_3\,\wedge\,(d\psi+\cos\theta \,d\phi)\,\,,
\qquad g=g(\rho,\sigma)\,\,.
\label{CalC4}
\eeq
Proceeding  as in subsection \ref{secN2}, one gets in this case the following set of BPS equations:
\bear
&&m^2\,\big[\,g\,-\,2\pi\,g_s\,(\alpha')^2\,N_f\,\Theta(\rho-\rho_Q)\,\big]\,=\,\rho^3\,z'\,\,,\rc\rc
&&m^2\,H\,=\,{z\dot z\over \sigma}\,\,,
\qquad\qquad\qquad
g'\,=\,-\sigma\,\rho^3\,\dot H\,\,,\rc\rc
&&\dot g\,=\,{\sigma \rho^3\over z}\,\,H'\,-\,{\sigma\over z^2}\,H\,m^2\,
\big[\,g\,-\,2\pi\,g_s\,(\alpha')^2\,N_f\,\Theta(\rho-\rho_Q)\,\big]
\,\,,
\label{flavored-BPSsystem}
\eear
where the prime and the dot have the same meaning as in (\ref{prime&dot}). The fulfillment of (\ref{flavored-BPSsystem}) ensures the preservation of eight supersymmetries by the background, which corresponds to ${\cal N}=(4,4)$ SUSY of the dual gauge theory.  Moreover, one can prove that $z(\rho,\sigma)$ satisfies the following PDE:
\beq
\rho\,z\,(\dot z\,-\,\sigma\,\ddot z\,)\,=\,\sigma\,(\, \rho\dot z^2\,+\,\rho z''\,+\,3 z'\,)\,+\,
{N_f\over 2 N_c}\,{\sigma\over m^2\,\rho^2}\,\,\delta (\rho-\rho_Q)\,\,.
\label{flavoredPDE}
\eeq
In the unflavored case $N_f=0$, the BPS system (\ref{flavored-BPSsystem}) (and the PDE equation (\ref{flavoredPDE})) can be solved analytically \cite{Arean:2008az} by constructing the solution in five-dimensional gauged supergravity and by uplifting it to ten dimensions \cite{Maldacena:2000mw}. After a suitable change of variables one can show \cite{Arean:2008az} that the metric and RR five-form of this gauged supergravity solution can be written as in our ansatz. In the general flavored case one has to apply numerical techniques.  However, as in the four-dimensional case, one only needs to know the solution for $\sigma=0$ in order to get the behavior of the gauge coupling. Indeed, by means of a probe calculation one can check \cite{Arean:2008az} that the supersymmetric locus of a color D3-brane occurs precisely at $\sigma=0$ and that the gauge coupling is related to $z(\rho, \sigma=0)$ by means of the relation:
 \beq
 {1\over g^2_{YM}(\rho)}\,=\,{z(\rho, \sigma=0)\over m^2 g_s}\,\,.
 \label{gYM-gravity}
 \eeq
It follows from the system (\ref{flavored-BPSsystem}) that $g(\rho, \sigma=0)$ is constant. Actually, by using a flux quantization condition similar to the one employed for the 4d case, one can verify that $g(\rho, \sigma=0)=1/m^4$, where $m$ is the constant defined in (\ref{m}). By using this result in the first equation in (\ref{flavored-BPSsystem}) one readily integrates $z(\rho, \sigma=0)$. By imposing continuity of the solution at $\rho=\rho_Q$, one gets:
\beq
 z(\rho, 0)\,=\,z_{*}\,-\,{\pi m^2 g_s\,(\alpha')^2\over \rho_Q^2}\,N_f\,
 \Theta(\rho-\rho_Q)\,-\,
 {2\pi\, m^2\,g_s\,(\alpha')^2\over \rho^2}
 \,\Big[\,N_c\,-\,{N_f\over 2}\, \Theta(\rho-\rho_Q)\,\Big]\,\,,
 \label{z-sigma0-flavored}
 \eeq
where $z_{*}$ is a constant of integration. Plugging this result in (\ref{z-sigma0-flavored}), and assuming that the energy scale $\mu$ is related to the holographic coordinate $\rho$ as $\rho=2\pi\alpha'\,\mu$, one gets:
\beq
 {1\over g_{YM}^2(\mu)}\,=\,
 {1\over g^2_{YM}}\,\Big(\,1\,-\,{g^2_{YM}\over 2\pi\mu^2}\,\big(\,N_c\,-\,{N_f(\mu)\over 2}\,\big)\Big)\,\,,
 \label{44gymrunning-gravity}
 \eeq
where $N_f(\mu)$ is again the number of flavors with mass smaller that the scale $\mu$ and $g_{YM}$  is the bare UV Yang-Mills coupling. The dependence on the scale $\mu$ of the Yang-Mills coupling displayed in (\ref{44gymrunning-gravity}) matches precisely the one in field theory, which constitutes a non-trivial test of the gravity result.

Backgrounds dual to 2d theories with ${\cal N}=(2,2)$ SUSY can be obtained by wrapping D5-branes along a four-cycle of a Calabi-Yau threefold \cite{Arean:2009gc}. An alternative construction, which improves the UV behavior of the solution, involves D3-branes wrapping a two-cycle of a $CY_3$ \cite{Maldacena:2000mw, D3D4D5}. One can further reduce the amount of supersymmetry by considering a D5-brane wrapping a four-cycle of a manifold of $G_2$ holonomy, which leads to a dual of an ${\cal N}=(1,1)$ supersymmetric gauge theory. In all these cases the flavor branes are extented along some of the non-compact normal directions of the cycle wrapped by the color branes and the corresponding backreacted solutions can be obtained numerically and are similar to the one reviewed here. 

\subsubsection{Three-dimensional theories}

A similar analysis can be carried out to obtain the gravity dual of ${\cal N}=4$ three-dimensional gauge theories. In this case one must consider  flavor and color D4-branes wrapping two-cycles according to the array:
\begin{center}
\begin{tabular}{|c|c|c|c|c|c|c|c|c|c|c|}
\multicolumn{4}{c}{ }&
\multicolumn{4}{c}{$\overbrace{\phantom{\qquad\qquad\qquad}}^{\text{CY}_2}$}\\
\hline
&\multicolumn{3}{|c|}{$\mathbb{R}^{1,2}$}
&\multicolumn{2}{|c|}{$S^2$}
&\multicolumn{2}{|c|}{$N_2$}
&\multicolumn{3}{|c|}{$\mathbb{R}^{3}$}\\
\hline
$N_c$\,\,\,D$4$ (color) &$-$&$-$&$-$&$\bigcirc$&$\bigcirc$&$\cdot$&$\cdot$&$\cdot$&$\cdot$&$\cdot$\\
\hline
$N_f$\,\,\,D$4$ (flavor) &$-$&$-$&$-$&$\cdot$&$\cdot$&$-$&$-$&$\cdot$&$\cdot$&$\cdot$\\
\hline
\end{tabular}
\label{44flavored-arrayD4}
\end{center}
The concrete ansatz for the ten-dimensional string frame metric we will adopt in this case is very similar to the 2d and 4d cases studied above, namely:
\begin{eqnarray}
&&ds^2=e^{2\Phi}\left[dx^2_{1,2}+{z \over m^2}\,\left(d\tilde{\theta}^2+\sin^2\tilde{\theta}d\tilde{\phi}^2\right)\right]\,+\,\nonumber\\
&&+e^{-2\Phi}\left[\frac{1}{z}\left(d\sigma^2+\sigma^2\left(d\psi+\cos\tilde{\theta}d\tilde{\phi}\right)^2\right)+
d\rho^2+\rho^2\left(d\theta^2+\sin^2\theta d\phi^2\right)
\right] \ ,
\label{metric-ansatz}
\end{eqnarray}
where $\Phi=\Phi(\rho,\sigma)$ is the dilaton and the constant $m$ is now given by:
\beq
{1\over m^3}\,=\,8\pi g_s\,N_c\,(\alpha')^{{3\over 2}}\,\,.
\label{R}
\eeq
As before, $z=z(\rho,\sigma)$ and the background should include an RR form whose Bianchi identity is violated due to the presence of flavor branes. For D4-branes the appropriate RR form is a four-form $F_4$.  If we locate the flavor branes at a fixed distance $\rho=\rho_Q$ in the transverse $\mathbb{R}^{3}$ and we smear them along their orthogonal angular directions, the modified Bianchi identity is:
\begin{equation}
dF_4\,=\,2\,\kappa_{10}^2 \, T_4\,\Omega\,=\,\frac{N_f}{2 N_c}\,\,\frac{1}{8 m^3}\,\delta(\rho-\rho_Q)
\,d\rho\wedge\, \omega_2 \, \wedge \, \tilde{\omega}_2 \, ,
\label{newBianchi3d}
\end{equation}
where $\omega_2$ and $\tilde{\omega}_2$ are the volume forms of the unit $(\theta, \phi)$ and $(\tilde\theta, \tilde\phi)$ two-spheres. Let us solve (\ref{newBianchi3d}) by means of  the following ansatz:
\begin{equation} \label{F4-flavor}
F_4\,=\, dC_3\,+\,
\frac{N_f}{2 N_c}\,\,\frac{1}{8\,m^3}\,\,\Theta(\rho-\rho_Q)
\,\,\tilde{\omega}_2 \, \wedge \, \omega_2 \, ,
\end{equation}
where $C_3$ is the following potential depending on the function $g(\rho,\sigma)$:
\be
C_3\,=\, - \, g\,\omega_2 \, \wedge \, (d\psi+\cos\tilde{\theta} \,d\tilde{\phi}) \, .
\label{C3}
\ee
By imposing that the system preserves eight supersymmetries, we arrive at the following system of BPS equations:
\begin{eqnarray}
 \label{flavored-BPSsystemD4}
&&g\,+\,{N_f \over 2\,N_c}\,{1\over
8\,m^3}\,\Theta(\rho-\rho_Q)\,=-\ \,{\rho^2\,{z}' \over m^2} \ , \quad \quad\qquad
e^{-4\Phi}\,\sigma\,=\,{1\over m^2}\,\, z\dot{z} \ , \rc
\\
&& g'\,=\,-4\,\sigma\,\rho^2\,e^{-4\phi}\dot{\Phi} \ , \rc\rc
&&\dot g\,=\,\,-m^{2}\,\sigma\,{
z}^{-2}\,e^{-4\Phi}\,\left[\,g\,+\,{N_f \over 2\,N_c}\,{1
\over 8\,m^3}\,\Theta(\rho-\rho_Q)\,\right]\,+\,4\,\sigma\,\rho^2\,{
z}^{-1}\,e^{-4\Phi}\,\Phi' \, .  \nonumber
\end{eqnarray}
Again, one can combine the different equations in (\ref{flavored-BPSsystemD4}) and get a single second-order PDE for $z(\rho,\sigma)$, namely:
\be 
\rho^2\,{z}\,\left(\,\dot{{z}}\,-\,\sigma\, \ddot{{
z}}\,\right)\,=\,\rho\,\sigma\,\left(\,\rho\,\dot{{z}}^2\,+\, \rho\,{
z}''\,+\, 2 {z}'\,\right)\,+\sigma\,{N_f \over 2\,N_c}\,{1
\over 8\,m^3}\,\delta(\rho-\rho_Q). 
\label{PDE-z-flavored} 
\ee
As in the 4d and 2d cases, eqs. (\ref{flavored-BPSsystemD4}) and (\ref{PDE-z-flavored}) can be solved analytically when $N_f=0$ by using gauged supergravity \cite{Maldacena:2000mw,Divecchia}. In the general flavored case one can get analytically the form of the solution for $\sigma=0$ \cite{Ramallo:2008ew}. Indeed, one can verify from (\ref{flavored-BPSsystemD4}) and the corresponding flux quantization condition 
that $g(\rho,\sigma=0)=-1/(8m^3)$ and that $ z(\rho, 0)$ is:
\beq
 z(\rho, 0)\,=\, z_{*}\,- \,{1\over 8\,\rho_Q m }
\,{N_f\over 2\,N_c}\,\Theta(\rho-\rho_Q)\,-\,{1\over 8m\,\rho}
\left[\,1\,-\,{N_f\over 2\,N_c}\,\Theta(\rho-\rho_Q)\,\right]
\, ,\label{z-sigma0-flavorD4}
\eeq
with $ z_{*}$ being a constant. Moreover, by means of a probe calculation one readily verifies that $\sigma=0$ is the SUSY locus of the color D4-branes and that the relation between the YM coupling and $ z(\rho, 0)$ is 
$ g^{-2}_{YM}(\rho)=z(\rho, 0)/(2\pi g_s \sqrt{\alpha'}\,m^2)$. Using this result and the radius-energy relation $\rho=2\pi\alpha'\mu$, one can convert (\ref{z-sigma0-flavorD4}) into the following equation for the running of the YM coupling of the 3d theories:
\be 
{1\over g_{YM}^2(\mu)}\,=\, {1\over
g^2_{YM}}\,\Big[\,1\,-\,{g^2_{YM} \over
4\pi\mu}\,\Big(\,N_c-{N_f(\mu)\over 2}\,\Big) \,\,\Big] \ ,
\label{gYM-QFT-D4} 
\ee
which again matches the corresponding field theory result. 

A gravity dual of ${\cal N}=2$ three-dimensional gauge theory based on D5-branes wrapping a three-cycle was found in \cite{Gomis:2001aa,Gauntlett:2001ur}. The addition of flavor to this background along the lines discussed here is carried out in \cite{Gaillard:2008wt,D3D4D5}. Alternatively,  for this same amount of supersymmetry one can consider D4-branes wrapping a two-cycle of a Calabi-Yau threefold \cite{Maldacena:2000mw,D3D4D5}.  

\setcounter{equation}{0}
\section{A Mathematical Viewpoint}
\label{mathviewpointzz}

In the approach we have followed up to now in this review on how to add 
unquenched  flavor, we  considered a family of equivalent embeddings of the flavor branes. This family can be generated by acting with  the isometries of the background on a 
fiducial representative embedding.  When the number $N_f$ of flavor branes is large,  considering the set of branes as  a continuous distribution is a good approximation.  We then computed the RR charge density generated by the branes, \ie\ the smearing form $\Omega$,  by explicitly performing the average over the set of embeddings and, subsequently,  we have studied the deformation induced  on the metric and forms due to the backreaction. 

 The outcome of this microscopic approach is a system of supergravity plus delocalized sources preserving some amount of supersymmetry. It turns out that, in this process, very interesting mathematical structures emerge. The reason for this is the fact that the supersymmetric sources that we are using satisfy a  calibration condition. As a consequence, one can use the methods of modern geometry to find backgrounds with smeared flavors in a systematic way \cite{Gaillard:2008wt}. In this procedure one does not deal with the set of embeddings and, for this reason, we will refer to it as the macroscopic approach, as opposed to the microscopic approach reviewed in previous sections. The goal is computing (or at least constraining) the smearing form $\Omega$ by using the same type of  technology  as the one employed in the analysis of flux compactifications of string theory (see refs. \cite{Gutowski:1999tu}-\cite{Koerber:2006hh}, \cite{Koerber:2007hd}).

The central object in this geometric approach is the 
so-called `calibration form' ${\cal K}$.  For D$p$-branes ${\cal K}$ is a 
$(p+1)$-form, which can be represented in a vielbein basis as:
\beq
{\cal K}\,=\,{1\over (p+1)!}\,{\cal K}_{a_1\cdots a_{p+1}}\,e^{a_1\,\cdots a_{p+1}}\,\,,
\eeq
with $e^{a_1\,\cdots a_{p+1}}=e^{a_1}\wedge\cdots \wedge e^{a_{p+1}}$. The
different components ${\cal K}_{a_1\cdots a_{p+1}}$  are given by fermionic bilinears of the type:
\beq
{\cal K}_{a_1\cdots a_{p+1}}\,=\,\,\epsilon^{\dagger}\,\tau\,\Gamma_{a_1\,\cdots a_{p+1}}\,\epsilon\,\,,
\label{K-bilinear}
\eeq
where $\epsilon$ are Killing spinors of the background, conveniently normalized,  and 
$\tau$ is a constant matrix which (in the type IIB theory) is $\tau=\tau_3 ^{{p-3\over 2}}\,i\tau_2$, where $\tau_2$ and $\tau_3$ are Pauli matrices and the spinor $\epsilon$ is represented as a two-dimensional vector of Majorana-Weyl spinors ($\tau$ is the same matrix that appears in the expression of the kappa symmetry matrix $\Gamma_{\kappa}$  of a D$p$-brane when all worldvolume fluxes are switched off). The form ${\cal K}$ can be used to characterize $(p+1)$-dimensional surfaces in the ten-dimensional geometry. A $(p+1)$-dimensional surface ${\cal M}_{p+1}$  is said to be calibrated by  ${\cal K}$ if its pullback to   ${\cal M}_{p+1}$ is equal to the induced volume form on  ${\cal M}_{p+1}$, namely:
\beq
\hat {\cal K}\,=\,\sqrt{-\det \hat g}\,\,\,\,d^{p+1}\,\xi\,\,,
\label{cal-con}
\eeq
where the $\xi$'s are a set of local coordinates of  ${\cal M}_{p+1}$. 
When there are no NSNS fluxes or worldvolume gauge fields the calibration 
condition (\ref{cal-con}) characterizes the supersymmetric embeddings of D$p$-branes (this can be easily established by using kappa symmetry). Actually, a D$p$-brane whose worldvolume  ${\cal M}_{p+1}$ is calibrated by ${\cal K}$ is electrically charged with respect to an RR (p+2)-form field strength $F_{p+2}$ and, in the Einstein frame, $F_{p+2}$ is related to ${\cal K}$ as:
\beq
F_{p+2}\,=\,d\Big(\,e^{{p-3\over 4}\phi}\,\,{\cal K}\,\Big)\,\,.
\label{F-K}
\eeq
Eq. (\ref{F-K}) is  a consequence of supersymmetry \cite{Gutowski:1999tu} and, actually, in our backreacted backgrounds it follows from the system of BPS equations. Moreover, as a consequence of (\ref{cal-con}), the action of a localized embedding of a D$p$-brane (without NSNS flux and with worldvolume gauge fields switched off) can be written as:
\beq
S_{Dp}^{loc}\,=\,-T_p\,\int_{{\cal M}_{p+1}}\,\,
\Big[\,e^{{p-3\over 4}\phi}\,\hat  {\cal K}\,-\,\hat C_{p+1}\,\Big]\,\,.
\label{loc-actionDp}
\eeq
Following our prescription, the smeared version of the brane action is obtained by performing the wedge product with $\Omega$ of the $(p+1)$-form inside the brackets in (\ref{loc-actionDp})  and by integrating the result over the full ten-dimensional spacetime:
\beq
S_{Dp}^{smeared}\,=\,-T_p\,\int_{{\cal M}_{10}}\,\,
\Big[\,e^{{p-3\over 4}\phi}\,  {\cal K}\,-\, C_{p+1}\,\Big]\wedge \Omega\,\,.
\label{smeared-Dp}
\eeq
Let us now define the $(8-p)$-form $F_{8-p}$, under which the D$p$-brane is magnetically charged, as:
\beq
F_{8-p}\,=\,\pm e^{{3-p\over 2}\phi}\,\,{}^*F_{p+2}\,\,,
\label{F(8-p)-F(p+2)}
\eeq
where the sign depends on the particular value of $p$ and on the conventions used. As in the examples studied in previous sections, the D$p$-brane modifies the Bianchi identity of $F_{8-p}$, namely:
\beq
dF_{8-p}\,=\,\pm 2\kappa_{10}^2\,T_p\,\Omega\,\,.
\label{Bianchi-Dp}
\eeq
Eq. (\ref{Bianchi-Dp}) establishes a crucial 
connection  between the smearing form $\Omega$ and 
the calibration form ${\cal K}$. Indeed, by using (\ref{F-K}) and (\ref{F(8-p)-F(p+2)}), the right-hand side of (\ref{Bianchi-Dp})  can be written in terms of ${\cal K}$ and its exterior derivative. Moreover, from the inspection of the smeared brane action (\ref{smeared-Dp}), one concludes that $\Omega$ can be regarded as a kind of orthogonal complement
(the Poincare dual) of ${\cal K}$ in ${\cal M}_{10}$. Interestingly, 
the possible calibration forms ${\cal K}$ in a 
manifold are known and are related to 
its supersymmetric cycles and G-structures. 
In the case of a  manifold preserving minimal SUSY in 4d,  ${\cal K}$ can 
be written in terms of powers 
of the K\"ahler form and of the holomorphic volume form. Thus, geometry and topology constrain the form of the charge  density distribution of supersymmetric configurations  and, actually, one could adopt the point of view in which the expression of $\Omega$ is determined from these constraints without explicitly performing the average over the family of embeddings. This macroscopic approach was followed in refs. \cite{Gaillard:2008wt,Gaillard:2009kz,Arean:2009gc, D3D4D5} for some particular brane setups.  

To finish this section let us detail the implementation  of these mathematical concepts in the case discussed in section \ref{AdS5X5}, namely the D3-D7 system. From now on we will assume that the metric, dilaton and forms are given by the expressions written in (\ref{metrictzero}) and (\ref{dilaton-forms}). It is convenient to define the following two-form:
\beq
{\cal J}\,=\,h^{{1\over 2}}\,\,\big[\,e^{2g}\,J_{KE}\,+\,e^{2f}\,d\rho\wedge (d\tau+A_{KE})\,\big]\,\,,
\eeq
which is such that $h^{-{1\over 2}}\,{\cal J}$ is the K\"ahler form of the transverse 6d space. Actually,  one can immediately verify  that $d\big[\,h^{-{1\over 2}}\,{\cal J}\,\big]=0$ as a consequence of the BPS equation for $g$ in (\ref{BSPsysKW}). By explicitly computing the fermion bilinear in (\ref{K-bilinear}) and by using the projections satisfied by the Killing spinor of the flavored $AdS_5\times S^5$ background, one gets that the calibration form ${\cal K}$ in this case is given by:
\beq
 {\cal K}\,=\,{1\over 2}\,{\rm Vol}(M_{1,3})\wedge {\cal J}\wedge {\cal J}\,\,,
 \eeq
with ${\rm Vol}(M_{1,3})=h^{-1}\,d^4x$ being the volume form of the Minkowski part of the space. 
Using the fact that $dA_{KE}=2J_{KE}$, one gets:
\beq
d\Big( e^{\phi}\, {\cal K}\,\Big)\,=\,{1\over 2}\,\,e^{2g+\phi}\,\,
\Big[\,\big(4g'+\phi')\,e^{2g}\,-\, 4 e^{2f}\,\Big]\,d^4x\,\wedge 
J_{KE}\wedge J_{KE}\wedge d\rho\,=\,F_9\,\,,
\label{F9-K}
\eeq
where, in the last step, we have used the condition (\ref{F-K}) for $p=7$. Let us now verify that the value of $F_9$ obtained in (\ref{F9-K}) is consistent with the expression for $F_1$ written in our ansatz (\ref{dilaton-forms}) and, thus, with the $\Omega$ displayed in  eq. (\ref{gsomega}). Taking into account that the volume form of the KE space is ${1\over 2} J_{KE}\wedge J_{KE}$, one can easily compute the Hodge dual of $F_1$ and get the following result for $F_9$:
\beq
F_9\,=\,-e^{2\phi}\,{}^*F_1\,=\,{Q_f\over 2}\,p(\rho)\,e^{4g+2\phi}\,
d^4x\,\wedge J_{KE}\wedge J_{KE}\wedge d\rho\,\,.
\label{F9-F1}
\eeq
The expressions (\ref{F9-K}) and (\ref{F9-F1}) for $F_9$ coincide if the following relation holds:
\beq
Q_f\,p(\rho)\,=\,e^{-\phi}\,\big[\,4g'+\phi'\,-\,4\,e^{2f-2g}\,\big]\,\,.
\label{Qf-relation}
\eeq
One can easily check that (\ref{Qf-relation}) is a consequence of the BPS system (\ref{BSPsysKW}).

\section{Discussion}
\label{outlook}

In hindsight, we can say that the program of finding
solutions dual to theories with unquenched fundamentals
with smeared flavor branes has been quite successful.
As expected, it simplifies matters both when looking for the background solution
and when discussing the physics they encode.

We have presented a series of example of solutions of ten-dimensional type 
IIA or type IIB supergravity coupled to D-brane sources. The philosophy and
methods used in the different cases are quite similar. Finding a consistent 
solution requires solving at the same time the closed string degrees of freedom
(namely, finding solutions of the generalised Einstein equations in the presence
of sources) and the open string degrees of freedom (namely, checking that the D-brane
embeddings which generate the mass and charge source density are indeed solutions of
the background).
Supersymmetric solutions are easier to deal with and indeed, preserving SUSY
simplifies enormously the technical work. 
It is rather remarkable that sometimes such complicated coupled systems can be
(at least almost) completely integrated and the solutions given in a simple closed
form (in particular in sections \ref{AdS5X5}, \ref{KS}; 
for the other sections, 
profuse numerical integration was necessary).
However, supersymmetry is not mandatory for the construction and we
have presented non-supersymmetric black hole solutions.

The solutions are conjectured to be dual to theories with 
unquenched quarks. Since we have always dealt with the particular case of smearing
the flavor branes over the transverse directions, we have built  duals of a very
particular class of such unquenched theories.
We have used the solutions to discuss many physical features of the different set-ups.
Many crosschecks of field theory expectations have been discussed. Just to mention
a few instances, the
running of the gauge coupling in different theories, the behaviour of the cascade in
section \ref{KS} or  the direct computation  of the first flavor contribution
to the entropy of the 
D3-D7 plasma (section \ref{D3D7plasma}) which was previously known from an indirect method
(namely, from first computing the free energy) \cite{myers}.
All this asserts that the dualities discussed in this review are on firm ground.
Due to obvious space constraints, we have not been able to include all the material
that may deserve to be reviewed, but we hope that we have given enough references to the
original literature.

\vskip.5cm

It is worth recapitulating
 about the presence of singularities in the different solutions
discussed. First, in all the cases presented 
there are IR curvature singularities when all of the flavor branes reach the bottom of 
the geometry, see the heuristic picture of section \ref{heuristiczzz}.
They pertain to the kind of singularities usually called {\it good}.
In fact, we have shown explicitly in the examples of sections \ref{AdS5X5} and
\ref{KS} how adding (even an infinitesimal) quark mass leads to regular backgrounds
(the analogous generalization for the set-ups of sections \ref{sec:D5D5}, \ref{sec:2+1}
 remains an interesting
open question). Moreover, heating up the theories can result in the formation of a black 
hole horizon behind which the IR singularity is hidden, see section \ref{D3D7plasma}
and eq. (\ref{bhflav}) for examples.

On the other hand, the solutions of sections \ref{AdS5X5} and \ref{KS} are singular in the UV
(an effect connected to having flavor D7-branes)
since the dilaton diverges at a finite position of the radial variable.
 This is expected, since
it is the consequence of the Landau pole of the dual theory (more precisely,
in the  case of
section \ref{KS} it is a  duality wall). Despite
the singularity, we have shown that it is possible to consistently compute IR observables
as long as the IR scale is well separated from the pathological UV. 
Clear examples are the meson spectrum (section 
\ref{sec:screening}) and the black hole thermodynamical properties
(section \ref{sec:thermo}). 
In the D5D5 setups of sections \ref{sec:D5D5}, \ref{sec:2+1}, the dilaton diverges linearly
in the UV, signalling a little string theory-like UV completion of the dual field theory.
We want to stress here that this already happens in the unflavored solutions and thus
is not problem associated to the backreaction.

Finally, all the solutions in section \ref{models2} have a singularity in the IR.
This singularity is not associated to the flavors as it is already present in the
flavorless solutions and, at least in some cases, can be resolved by the
worldsheet CFT \cite{Hori:2002cd}. On top of that,
for the same models, typically,
when $N_f$ is sufficiently large, a Landau pole is generated and, jointly, a UV 
singularity appears in the geometry.

\vskip.5cm

We end this discussion with two clarifications. 
\begin{itemize}

\item
We have repeatedly
 stressed that our main goal is to build duals
to theories in which $N_c$ and $N_f$ are of the same order. Nevertheless, for the set-ups
discussed in  section \ref{AdS5X5}, which include the 
specially interesting
flavored $AdS_5 \times S^5$ case,
 $N_f \ll N_c$ is needed, see section \ref{sec:rangeval} (similar comments
apply to section \ref{KS}). This is because, starting with a conformal theory, the 
introduction of extra matter generates an UV pathology, namely a Landau pole. 
Then, roughly speaking, in order to have a meaningful IR, it has to be well separated
from the pathological region, enforcing the number of flavors not to be too large.
However, backreaction effects and, accordingly, the effect of unquenched quarks,
can still be computed as an expansion in 
$N_f/N_c$.
On the other hand, for the models in sections \ref{sec:D5D5}, \ref{sec:2+1},
\ref{models2}, this restriction is not present and, indeed, it makes sense
to talk about solutions with $N_f \sim N_c$. In fact, this is imperative for
instance when discussing Seiberg-like dualities as in section \ref{sec:Seib}.

\item
Since  the  (DBI) action is used to model the D-brane
sources, one could be wary for the following reason: the effective string coupling
on a stack of $N_f$ D-branes is $g_s N_f \sim N_f / N_c$ and this should be small
for the DBI to be a good approximation \cite{Callan:1986bc}, whereas there is not a good effective description
for strong string coupling. However, this caveat is
circumvented because we do not deal with 
stacks of localized flavor branes: due to the smearing, the
typical distance between any pair of flavor branes is of the order of the 
size of the transverse space, which is typically large. As a result, the flavor symmetry
is usually broken to $U(1)^{N_f}$ and the effective open string coupling 
remains small.
As already pointed out, this amounts to keeping
 ``one window graphs'' in the Veneziano expansion \cite{Bigazzi:2008zt}.
 
\end{itemize}

\subsubsection*{Outlook}

There are still many open questions that deserve to be addressed within the framework presented
in this review. 
We briefly mention a few examples of
possible future projects.
They comprise  both making further progress in studying the models here
presented and building new solutions that could be useful in exploring the consequences
of the formalism for different physical points. 
Along the first of these lines, it would be nice to generalise the solutions of section \ref{sec:D5D5} 
to the massive quark case in order to remove the IR singularity. Also, we expect  the black hole
solution of section \ref{D3D7plasma} to encode interesting physical information. For instance, one
could consider massive embeddings in the search of a first order phase 
transition similar
to those in \cite{Kruczenski:2003uq},\cite{Kirsch:2004km},\cite{myers}.
The peculiarity of the back-reacted setting
would be, conceivably, that the area of the horizon would undergo a finite jump 
at the transition.
Along the second line, a back-reacted D4-D6 solution building on the model of 
\cite{Kruczenski:2003uq} could be useful in discussing QCD-like properties.
Another conceivable program is to look for a solution, which, as in
\cite{Chen:2009kx}, may correspond to a color-flavor locking phase. The 
study of fluctuations in these backgrounds, that will also contain 
fluctuations of the fields in the flavor branes, with a view on 
understanding holographic renormalization would be a highly interesting 
result.

Aside from this, it would be nice to find solutions (with backreacted 
flavor branes) that contain an $AdS_5$ factor. The study of conformal 
anomalies there may give interesting results.

As stressed in the introduction, finding the kind of solutions discussed
here, including the 
D-brane backreaction,
has an interest on their own, independently of AdS/CFT.
It would  be nice to understand whether they
 may turn out to be useful for different physical applications.
For instance, for models of inflation built with D3-D7 systems on the conifold
(see \cite{Baumann:2010sx} for recent progress in this direction), the analysis of section \ref{KS}
could have some relevance.

\section*{Acknowledgments}

This review is based on work done with several 
collaborators in the last few years. 
We thank all of them for  their insights and the many discussions 
during the course of those collaborations. 
Aside from them, we are specially grateful to D. Are\'an,
A. Armoni, F. Bigazzi,  E. Conde, A. 
Cotrone, S. 
Cremonesi, J. 
Gaillard, D. Martelli, I. Papadimitriou, J. Shock, 
J. Tarr\'\i o 
and D. Zoakos for a critical 
reading of the manuscript and the many useful comments, discussions and 
remarks they made.  
The research of A.P is
supported by grants FPA2007-66665C02-02 and DURSI
2009 SGR 168, and by the CPAN CSD2007-00042 project of the
Consolider-Ingenio
2010 program. The  work of AVR was funded in part by MEC and  FEDER  under grant
FPA2008-01838,  by the Spanish Consolider-Ingenio 2010 Programme CPAN (CSD2007-00042) and by Xunta de Galicia (Conseller\'\i a de Educaci\'on and grant INCITE09 206 121 PR).



\begin{thebibliography}{99}
\bibitem{Maldacena:1997re}
  J.~M.~Maldacena,
  ``The large N limit of superconformal field theories and 
supergravity'',
  Adv.\ Theor.\ Math.\ Phys.\  {\bf 2}, 231 (1998)
  [Int.\ J.\ Theor.\ Phys.\  {\bf 38}, 1113 (1999)];
  hep-th/9711200.

\bibitem{Gubser:1998bc}
  S.~S.~Gubser, I.~R.~Klebanov and A.~M.~Polyakov,
  ``Gauge theory correlators from non-critical string theory'',
  Phys.\ Lett.\ B {\bf 428}, 105 (1998);
  hep-th/9802109.
  
\bibitem{Witten:1998qj}
  E.~Witten,
   ``Anti-de Sitter space and holography'',
  Adv.\ Theor.\ Math.\ Phys.\  {\bf 2}, 253 (1998);
  hep-th/9802150.




\bibitem{Aharony:1999ti}
  O.~Aharony, S.~S.~Gubser, J.~M.~Maldacena, H.~Ooguri and Y.~Oz,
  ``Large N field theories, string theory and gravity,''
  Phys.\ Rept.\  {\bf 323}, 183 (2000)
  [arXiv:hep-th/9905111].



\bibitem{Karch:2002sh}
  A.~Karch and E.~Katz,
  ``Adding flavor to AdS/CFT'',
  JHEP {\bf 0206}, 043 (2002)
  [arXiv:hep-th/0205236]. 

\bibitem{Kruczenski:2003be}
  M.~Kruczenski, D.~Mateos, R.~C.~Myers and D.~J.~Winters,
  ``Meson spectroscopy in AdS/CFT with flavour,''
  JHEP {\bf 0307}, 049 (2003)
  [arXiv:hep-th/0304032].





\bibitem{Erdmenger:2007cm}
  J.~Erdmenger, N.~Evans, I.~Kirsch and E.~Threlfall,
  ``Mesons in Gauge/Gravity Duals - A Review,''
  Eur.\ Phys.\ J.\  A {\bf 35}, 81 (2008)
  [arXiv:0711.4467 [hep-th]].



\bibitem{Banks:1981nn}
  T.~Banks and A.~Zaks,
  ``On The Phase Structure Of Vector-Like Gauge Theories With Massless
  Fermions,''
  Nucl.\ Phys.\  B {\bf 196}, 189 (1982).

\bibitem{Seiberg:1994pq}
  N.~Seiberg,
  ``Electric - magnetic duality in supersymmetric nonAbelian gauge 
theories,''
  Nucl.\ Phys.\  B {\bf 435}, 129 (1995)
  [arXiv:hep-th/9411149].






\bibitem{Bigazzi:2005md}
  F.~Bigazzi, R.~Casero, A.~L.~Cotrone, E.~Kiritsis and A.~Paredes,
  ``Non-critical holography and four-dimensional CFT's with fundamentals,''
  JHEP {\bf 0510}, 012 (2005)
  [arXiv:hep-th/0505140].



\bibitem{Casero:2006pt}
  R.~Casero, C.~Nunez and A.~Paredes,
  ``Towards the string dual of N = 1 SQCD-like theories,''
  Phys.\ Rev.\  D {\bf 73}, 086005 (2006)
  [arXiv:hep-th/0602027].






\bibitem{Klebanov:2000hb}
  I.~R.~Klebanov and M.~J.~Strassler,
  ``Supergravity and a confining gauge theory: Duality cascades and
  chiSB-resolution of naked singularities,''
  JHEP {\bf 0008}, 052 (2000)
  [arXiv:hep-th/0007191].






\bibitem{'t Hooft:1973jz}
  G.~'t Hooft,
  ``A planar diagram theory for strong interactions,''
  Nucl.\ Phys.\  B {\bf 72}, 461 (1974).


\bibitem{Veneziano:1976wm}
  G.~Veneziano,
  ``Some Aspects Of A Unified Approach To Gauge, Dual And Gribov 
Theories,''
  Nucl.\ Phys.\  B {\bf 117}, 519 (1976).






\bibitem{Capella:1992yb}
  A.~Capella, U.~Sukhatme, C.~I.~Tan and J.~Tran Thanh Van,
  ``Dual parton model,''
  Phys.\ Rept.\  {\bf 236}, 225 (1994).








\bibitem{Kruczenski:2003uq}
  M.~Kruczenski, D.~Mateos, R.~C.~Myers and D.~J.~Winters,
  ``Towards a holographic dual of large-N(c) QCD,''
  JHEP {\bf 0405}, 041 (2004)
  [arXiv:hep-th/0311270].



\bibitem{Babington:2003vm}
  J.~Babington, J.~Erdmenger, N.~J.~Evans, Z.~Guralnik and I.~Kirsch,
  ``Chiral symmetry breaking and pions in non-supersymmetric gauge/gravity
  duals,''
  Phys.\ Rev.\  D {\bf 69}, 066007 (2004)
  [arXiv:hep-th/0306018].


\bibitem{Nunez:2003cf}
C.~Nunez, A.~Paredes and A.~V.~Ramallo,
``Flavoring the gravity dual of N = 1 Yang-Mills with probes'',
JHEP {\bf 0312}, 024 (2003);
hep-th/0311201.
 
\bibitem{Sakai:2004cn}
  T.~Sakai and S.~Sugimoto,
  ``Low energy hadron physics in holographic QCD,''
  Prog.\ Theor.\ Phys.\  {\bf 113}, 843 (2005)
  [arXiv:hep-th/0412141].
\bibitem{Sakai:2005yt}
  T.~Sakai and S.~Sugimoto,
  ``More on a holographic dual of QCD,''
  Prog.\ Theor.\ Phys.\  {\bf 114}, 1083 (2005)
  [arXiv:hep-th/0507073].



\bibitem{Itzhaki:1998dd}
  N.~Itzhaki, J.~M.~Maldacena, J.~Sonnenschein and S.~Yankielowicz,
 ``Supergravity and the large N limit of theories with sixteen
 supercharges'',
  Phys.\ Rev.\ D {\bf 58}, 046004 (1998);
  hep-th/9802042.
  
  
  \bibitem{Boonstra:1998mp}
  H.~J.~Boonstra, K.~Skenderis and P.~K.~Townsend,
  ``The domain wall/QFT correspondence,''
  JHEP {\bf 9901}, 003 (1999)
  [arXiv:hep-th/9807137].


\bibitem{cherkishash}
 S.~A.~Cherkis and A.~Hashimoto,
  ``Supergravity solution of intersecting branes and AdS/CFT with flavor,''
  JHEP {\bf 0211}, 036 (2002)
  [arXiv:hep-th/0210105].


\bibitem{HoyosBadajoz:2008fw}
  C.~Hoyos-Badajoz, C.~Nunez and I.~Papadimitriou,
  ``Comments on the String dual to N=1 SQCD,''
  Phys.\ Rev.\  D {\bf 78}, 086005 (2008)
  [arXiv:0807.3039 [hep-th]].




\bibitem{Bigazzi:2008zt}
  F.~Bigazzi, A.~L.~Cotrone and A.~Paredes,
  ``Klebanov-Witten theory with massive dynamical flavors,''
  JHEP {\bf 0809}, 048 (2008)
  [arXiv:0807.0298 [hep-th]].

\bibitem{Freedman:2003ax}
  D.~Z.~Freedman, C.~Nunez, M.~Schnabl and K.~Skenderis,
  ``Fake Supergravity and Domain Wall Stability,''
  Phys.\ Rev.\  D {\bf 69}, 104027 (2004)
  [arXiv:hep-th/0312055].


\bibitem{Martucci:2005rb}
  L.~Martucci, J.~Rosseel, D.~Van den Bleeken and A.~Van Proeyen,
  ``Dirac actions for D-branes on backgrounds with fluxes,''
  Class.\ Quant.\ Grav.\  {\bf 22}, 2745 (2005)
  [arXiv:hep-th/0504041].




\bibitem{Cederwall:1996ri}
  M.~Cederwall, A.~von Gussich, B.~E.~W.~Nilsson, P.~Sundell and 
A.~Westerberg,
  ``The Dirichlet super-p-branes in ten-dimensional type IIA and IIB
  supergravity,''
  Nucl.\ Phys.\  B {\bf 490}, 179 (1997)
  [arXiv:hep-th/9611159].
NUPHA,B490,179;
  E.~Bergshoeff and P.~K.~Townsend,
  ``Super D-branes,''
  Nucl.\ Phys.\  B {\bf 490} (1997) 145
  [arXiv:hep-th/9611173].



\bibitem{Koerber:2007hd}
  P.~Koerber and D.~Tsimpis,
  ``Supersymmetric sources, integrability and generalized-structure
  compactifications,''
  JHEP {\bf 0708}, 082 (2007)
  [arXiv:0706.1244 [hep-th]].


\bibitem{Grana:2006kf}
  M.~Grana, R.~Minasian, M.~Petrini and A.~Tomasiello,
  ``A scan for new N=1 vacua on twisted tori,''
  JHEP {\bf 0705}, 031 (2007)
  [arXiv:hep-th/0609124].
  

\bibitem{DeWolfe:2008zy}
  O.~DeWolfe, S.~Kachru and M.~Mulligan,
  ``A Gravity Dual of Metastable Dynamical Supersymmetry Breaking,''
  Phys.\ Rev.\  D {\bf 77}, 065011 (2008)
  [arXiv:0801.1520 [hep-th]].


\bibitem{Gaillard:2008wt}
  J.~Gaillard and J.~Schmude,
  ``On the geometry of string duals with backreacting flavors,''
  JHEP {\bf 0901}, 079 (2009)
  [arXiv:0811.3646 [hep-th]].




\bibitem{Kehagias:1998gn}
  A.~Kehagias,
  ``New type IIB vacua and their F-theory interpretation,''
  Phys.\ Lett.\  B {\bf 435}, 337 (1998)
  [arXiv:hep-th/9805131].



\bibitem{Aharony:1998xz}
  O.~Aharony, A.~Fayyazuddin and J.~M.~Maldacena,
  ``The large N limit of N = 2,1 field theories from three-branes in
  F-theory,''
  JHEP {\bf 9807}, 013 (1998)
  [arXiv:hep-th/9806159].



\bibitem{Grana:2001xn}
  M.~Grana and J.~Polchinski,
  ``Gauge / gravity duals with holomorphic dilaton,''
  Phys.\ Rev.\  D {\bf 65}, 126005 (2002)
  [arXiv:hep-th/0106014].



\bibitem{Bertolini:2001qa}
  M.~Bertolini, P.~Di Vecchia, M.~Frau, A.~Lerda and R.~Marotta,
  ``N = 2 gauge theories on systems of fractional D3/D7 branes,''
  Nucl.\ Phys.\  B {\bf 621}, 157 (2002)
  [arXiv:hep-th/0107057].



\bibitem{Bertolini:2002xu}
  M.~Bertolini, P.~Di Vecchia, M.~Frau, A.~Lerda and R.~Marotta,
  ``More anomalies from fractional branes,''
  Phys.\ Lett.\  B {\bf 540}, 104 (2002)
  [arXiv:hep-th/0202195].




\bibitem{Burrington:2004id}
  B.~A.~Burrington, J.~T.~Liu, L.~A.~Pando Zayas and D.~Vaman,
  ``Holographic duals of flavored N = 1 super Yang-Mills: Beyond the probe
  approximation,''
  JHEP {\bf 0502}, 022 (2005)
  [arXiv:hep-th/0406207].




\bibitem{Liu:2004ru}
  J.~T.~Liu, D.~Vaman and W.~Y.~Wen,
  ``Bubbling 1/4 BPS solutions in type IIB and supergravity reductions on  S**n
  x S**n,''
  Nucl.\ Phys.\  B {\bf 739}, 285 (2006)
  [arXiv:hep-th/0412043].




\bibitem{Kirsch:2005uy}
  I.~Kirsch and D.~Vaman,
  ``The D3/D7 background and flavor dependence of Regge trajectories,''
  Phys.\ Rev.\  D {\bf 72}, 026007 (2005)
  [arXiv:hep-th/0505164].




\bibitem{Ouyang:2003df}
  P.~Ouyang,
  ``Holomorphic D7-branes and flavored N = 1 gauge theories,''
  Nucl.\ Phys.\  B {\bf 699}, 207 (2004)
  [arXiv:hep-th/0311084].
  


\bibitem{Mia:2009wj}
  M.~Mia, K.~Dasgupta, C.~Gale and S.~Jeon,
  ``Five Easy Pieces: The Dynamics of Quarks in Strongly Coupled Plasmas,''
  arXiv:0902.1540 [hep-th].
  ``The Double Life of Thermal QCD,''
  arXiv:0902.2216 [hep-th].






\bibitem{Itzhaki:1998uz}
  N.~Itzhaki, A.~A.~Tseytlin and S.~Yankielowicz,
  ``Supergravity solutions for branes localized within branes,''
  Phys.\ Lett.\  B {\bf 432}, 298 (1998)
  [arXiv:hep-th/9803103].




\bibitem{Cherkis:2002ir}
O.~Pelc and R.~Siebelink,
  ``The D2-D6 system and a fibered AdS geometry,''
  Nucl.\ Phys.\  B {\bf 558}, 127 (1999)
  [arXiv:hep-th/9902045].
 




\bibitem{Erdmenger:2004dk}
  J.~Erdmenger and I.~Kirsch,
  ``Mesons in gauge / gravity dual with large number of fundamental fields,''
  JHEP {\bf 0412}, 025 (2004)
  [arXiv:hep-th/0408113].


\bibitem{GomezReino:2004pw}
  M.~Gomez-Reino, S.~G.~Naculich and H.~Schnitzer,
  ``Thermodynamics of the localized D2-D6 system,''
  Nucl.\ Phys.\  B {\bf 713}, 263 (2005)
  [arXiv:hep-th/0412015].
  
\bibitem{gaiottoetal}
  S.~Hohenegger and I.~Kirsch,
  ``A note on the holography of Chern-Simons matter theories with flavour,''
  JHEP {\bf 0904}, 129 (2009)
  [arXiv:0903.1730 [hep-th]].
  D.~Gaiotto and D.~L.~Jafferis,
  ``Notes on adding D6 branes wrapping RP3 in AdS4 x CP3,''
  arXiv:0903.2175 [hep-th].
  F.~Benini, C.~Closset and S.~Cremonesi,
  ``Chiral flavors and M2-branes at toric CY4 singularities,''
  arXiv:0911.4127 [hep-th].
  D.~L.~Jafferis,
  ``Quantum corrections to N=2 Chern-Simons theories with flavor and their AdS4
  duals,''
  arXiv:0911.4324 [hep-th].



\bibitem{Nastase:2003dd}
  H.~Nastase,
  ``On Dp-Dp+4 systems, QCD dual and phenomenology,''
  arXiv:hep-th/0305069.



\bibitem{Burrington:2007qd}
  B.~A.~Burrington, V.~S.~Kaplunovsky and J.~Sonnenschein,
  ``Localized Backreacted Flavor Branes in Holographic QCD,''
  JHEP {\bf 0802}, 001 (2008)
  [arXiv:0708.1234 [hep-th]].





\bibitem{Carlevaro:2009jx}
  L.~Carlevaro and D.~Israel,
  ``Heterotic Resolved Conifolds with Torsion, from Supergravity to CFT,''
  arXiv:0910.3190 [hep-th].



\bibitem{Klebanov:2004ya}
  I.~R.~Klebanov and J.~M.~Maldacena,
  ``Superconformal gauge theories and non-critical superstrings,''
  Int.\ J.\ Mod.\ Phys.\  A {\bf 19}, 5003 (2004)
  [arXiv:hep-th/0409133].
  
  
  


\bibitem{Fotopoulos:2005cn}
  A.~Fotopoulos, V.~Niarchos and N.~Prezas,
  ``D-branes and SQCD in non-critical superstring theory,''
  JHEP {\bf 0510}, 081 (2005)
  [arXiv:hep-th/0504010].

 \bibitem{Murthy:2006xt}
  S.~Murthy and J.~Troost,
  ``D-branes in non-critical superstrings and duality in N = 1 gauge theories
  with flavor,''
  JHEP {\bf 0610}, 019 (2006)
  [arXiv:hep-th/0606203].




\bibitem{Gadde:2009dj}
  A.~Gadde, E.~Pomoni and L.~Rastelli,
  ``The Veneziano Limit of N=2 Superconformal QCD: Towards the String Dual of
  N=2 $SU(N_c)$ SYM with $N_f =2 N_c$,''
  arXiv:0912.4918 [hep-th].





\bibitem{Alishahiha:2004yv}
  M.~Alishahiha, A.~Ghodsi and A.~E.~Mosaffa,
  ``On isolated conformal fixed points and noncritical string theory,''
  JHEP {\bf 0501}, 017 (2005)
  [arXiv:hep-th/0411087].



\bibitem{Casero:2005se}
  R.~Casero, A.~Paredes and J.~Sonnenschein,
  ``Fundamental matter, meson spectroscopy and non-critical string / gauge
  duality,''
  JHEP {\bf 0601}, 127 (2006)
  [arXiv:hep-th/0510110].




\bibitem{Bertoldi:2007sf}
  G.~Bertoldi, F.~Bigazzi, A.~L.~Cotrone and J.~D.~Edelstein,
  ``Holography and Unquenched Quark-Gluon Plasmas,''
  Phys.\ Rev.\  D {\bf 76}, 065007 (2007)
  [arXiv:hep-th/0702225].




\bibitem{Gursoy:2007cb}
  U.~Gursoy and E.~Kiritsis,
  ``Exploring improved holographic theories for QCD: Part I,''
  JHEP {\bf 0802}, 032 (2008)
  [arXiv:0707.1324 [hep-th]].





\bibitem{Sin:2007ze}
  S.~J.~Sin,
  ``Gravity Back-reaction to the Baryon Density for Bulk Filling Branes,''
  JHEP {\bf 0710}, 078 (2007)
  [arXiv:0707.2719 [hep-th]].

\bibitem{Jarvinen:2009fe}
  M.~Jarvinen and F.~Sannino,
  ``Holographic Conformal Window - A Bottom Up Approach,''
  arXiv:0911.2462 [hep-ph].


\bibitem{Armoni:2008jy}
  A.~Armoni,
  ``Beyond The Quenched (or Probe Brane) Approximation in Lattice (or
  Holographic) QCD,''
 Phys.\ Rev.\  D {\bf 78}, 065017 (2008)
  [arXiv:0805.1339 [hep-th]].


\bibitem{kw}
I.~R.~Klebanov and E.~Witten, ``Superconformal field theory on threebranes
at a Calabi-Yau singularity,'' Nucl.\ Phys.\ B {\bf 536} (1998) 199
[arXiv:hep-th/9807080].



\bibitem{Benini:2006hh}
  F.~Benini, F.~Canoura, S.~Cremonesi, C.~Nunez and A.~V.~Ramallo,
  ``Unquenched flavors in the Klebanov-Witten model,''
  JHEP {\bf 0702}, 090 (2007)
  [arXiv:hep-th/0612118].




\bibitem{Benini:2007gx}
  F.~Benini, F.~Canoura, S.~Cremonesi, C.~Nunez and A.~V.~Ramallo,
  ``Backreacting Flavors in the Klebanov-Strassler Background,''
  JHEP {\bf 0709}, 109 (2007)
  [arXiv:0706.1238 [hep-th]].
  S.~Cremonesi,
  ``Unquenched flavors in the Klebanov-Strassler theory,''
  Fortsch.\ Phys.\  {\bf 56}, 950 (2008)
  [arXiv:0805.4384 [hep-th]].



  \bibitem{Lawrence:1998ja}
 A.~E.~Lawrence, N.~Nekrasov and C.~Vafa,
 ``On conformal field theories in four dimensions,''
 Nucl.\ Phys.\  B {\bf 533}, 199 (1998)
 [arXiv:hep-th/9803015].
I.~R.~Klebanov and N.~A.~Nekrasov,
``Gravity duals of fractional branes and logarithmic RG flow,''
 Nucl.\ Phys.\  B {\bf 574}, 263 (2000)
 [arXiv:hep-th/9911096].


\bibitem{Maldacena:2000mw}
  J.~M.~Maldacena and C.~Nunez,
  ``Supergravity description of field theories on curved manifolds and a no  go
  theorem,''
  Int.\ J.\ Mod.\ Phys.\  A {\bf 16}, 822 (2001)
  [arXiv:hep-th/0007018].
  
\bibitem{Gubser:2000nd}
  S.~S.~Gubser,
  Adv.\ Theor.\ Math.\ Phys.\  {\bf 4}, 679 (2000)
  [arXiv:hep-th/0002160].




\bibitem{Bigazzi:2009bk}
  F.~Bigazzi, A.~L.~Cotrone, J.~Mas, A.~Paredes, A.~V.~Ramallo and J.~Tarrio,
  ``D3-D7 Quark-Gluon Plasmas,''
 JHEP {\bf 0911}, 117 (2009)
  [arXiv:0909.2865 [hep-th]].
  


 \bibitem{fkw2}F.~Bigazzi, A.~L.~Cotrone, A.~Paredes and A.~V.~Ramallo,
 ``Non chiral dynamical flavors and screening on the conifold,''
Fortsch.\ Phys.\  {\bf 57}, 514 (2009)
 [arXiv:0810.5220 [hep-th]].



  \bibitem{unquenchedmesons}F.~Bigazzi, A.~L.~Cotrone, A.~Paredes and A.~V.~Ramallo,
 ``Screening effects on meson masses from holography,''
 JHEP {\bf 0905}, 034 (2009)
 [arXiv:0903.4747 [hep-th]].






\bibitem{BCT}
  F.~Bigazzi, A.L.~Cotrone and J. Tarr\'\i o,
  ``Hydrodynamics of fundamental matter,''
  arXiv:0912.3256 [hep-th].
 



\bibitem{Elander:2009bm}
  D.~Elander,
  ``Glueball Spectra of SQCD-like Theories,''
  arXiv:0912.1600 [hep-th].




\bibitem{Kuperstein:2004hy}
  S.~Kuperstein,
  ``Meson spectroscopy from holomorphic probes on the warped deformed
  conifold,''
  JHEP {\bf 0503}, 014 (2005)
  [arXiv:hep-th/0411097].







\bibitem{Russo:1998by}
  J.~G.~Russo and K.~Sfetsos,
  ``Rotating D3 branes and {QCD} in three dimensions,''
  Adv.\ Theor.\ Math.\ Phys.\  {\bf 3}, 131 (1999)
  [arXiv:hep-th/9901056].













\bibitem{mesonmelting}
  C.~Hoyos-Badajoz, K.~Landsteiner and S.~Montero,
  ``Holographic Meson Melting,''
  JHEP {\bf 0704}, 031 (2007)
  [arXiv:hep-th/0612169].
  R.~C.~Myers, A.~O.~Starinets and R.~M.~Thomson,
  ``Holographic spectral functions and diffusion constants for fundamental
  matter,''
  JHEP {\bf 0711}, 091 (2007)
  [arXiv:0706.0162 [hep-th]].





  \bibitem{myers}D.~Mateos, R.~C.~Myers and R.~M.~Thomson,
 ``Thermodynamics of the brane,''
 JHEP {\bf 0705}, 067 (2007)
 [arXiv:hep-th/0701132].











\bibitem{Cherman:2009tw}
 A.~Cherman, T.~D.~Cohen and A.~Nellore,
 ``A bound on the speed of sound from holography,''
 Phys.\ Rev.\  D {\bf 80}, 066003 (2009)
  [arXiv:0905.0903 [hep-th]].





\bibitem{Kovtun:2004de}
 P.~Kovtun, D.~T.~Son and A.~O.~Starinets,
 ``Viscosity in strongly interacting quantum field theories from black hole
 physics,''
 Phys.\ Rev.\ Lett.\  {\bf 94}, 111601 (2005)
 [arXiv:hep-th/0405231].



\bibitem{Mateos:2006yd}
 D.~Mateos, R.~C.~Myers and R.~M.~Thomson,
 ``Holographic viscosity of fundamental matter,''
 Phys.\ Rev.\ Lett.\  {\bf 98}, 101601 (2007)
 [arXiv:hep-th/0610184].


\bibitem{buchel}
 A.~Buchel,
 ``Bulk viscosity of gauge theory plasma at strong coupling,''
 Phys.\ Lett.\  B {\bf 663}, 286 (2008)
 [arXiv:0708.3459 [hep-th]].




\bibitem{Baier}
 R.~Baier, Y.~L.~Dokshitzer, A.~H.~Mueller, S.~Peigne and D.~Schiff,
 ``Radiative energy loss and p(T)-broadening of high energy partons in
 nuclei,''
 Nucl.\ Phys.\  B {\bf 484}, 265 (1997)
 [arXiv:hep-ph/9608322].



\bibitem{liu}H. Liu, K. Rajagopal and U. A. Wiedemann, 
``Calculating the jet quenching parameter from AdS/CFT'', 
Phys.\ Rev.\ Lett.\ {\bf 97}, 182301 (2006)
[arXiv:hep-ph/0605178].


\bibitem{aredmas}N.~Armesto, J.~D.~Edelstein and J.~Mas,
 ``Jet quenching at finite 't Hooft coupling and chemical potential from
 AdS/CFT,''
 JHEP {\bf 0609}, 039 (2006)
 [arXiv:hep-ph/0606245].






\bibitem{lrw2}H.~Liu, K.~Rajagopal and U.~A.~Wiedemann,
 ``Wilson loops in heavy ion collisions and their calculation in AdS/CFT,''
 JHEP {\bf 0703}, 066 (2007)
 [arXiv:hep-ph/0612168].









\bibitem{herzog1}
C. P. Herzog, A. Karch, P. Kovtun, C. Kozcaz and L. G. Yaffe, 
``Energy loss of a heavy quark moving through N = 4 supersymmetric Yang-Mills plasma'', 
JHEP {\bf 0607}, 013 (2006) [arXiv:hep-th/0605158].

\bibitem{gubserdrag}S.~S.~Gubser,
``Drag force in AdS/CFT,'' 
Phys.\ Rev.\  D {\bf 74}, 126005 (2006)
[arXiv:hep-th/0605182].



\bibitem{herzog2}
C.~P.~Herzog,
``Energy loss of heavy quarks from asymptotically AdS geometries,''
JHEP {\bf 0609}, 032 (2006) [arXiv:hep-th/0605191].




\bibitem{holaf}
  L.~Girardello, M.~Petrini, M.~Porrati and A.~Zaffaroni,
  ``Novel local CFT and exact results on perturbations of N = 4 super
  Yang-Mills from AdS dynamics,''
  JHEP {\bf 9812}, 022 (1998)
  [arXiv:hep-th/9810126].
  D.~Z.~Freedman, S.~S.~Gubser, K.~Pilch and N.~P.~Warner,
  ``Renormalization group flows from holography supersymmetry and a
  c-theorem,''
  Adv.\ Theor.\ Math.\ Phys.\  {\bf 3}, 363 (1999)
  [arXiv:hep-th/9904017].

\bibitem{kkm} I.~R.~Klebanov, D.~Kutasov and A.~Murugan,
  ``Entanglement as a Probe of Confinement,''
  Nucl.\ Phys.\  B {\bf 796}, 274 (2008)
  [arXiv:0709.2140 [hep-th]].


\bibitem{Witten:1998zw}
  E.~Witten,
  ``Anti-de Sitter space, thermal phase transition, and confinement in  
gauge
  theories,''
  Adv.\ Theor.\ Math.\ Phys.\  {\bf 2}, 505 (1998)
  [arXiv:hep-th/9803131].



\bibitem{Maldacena:2000yy}
  J.~M.~Maldacena and C.~Nunez,
  ``Towards the large N limit of pure N = 1 super Yang Mills'',
  Phys.\ Rev.\ Lett.\  {\bf 86}, 588 (2001);
  hep-th/0008001.

\bibitem{Witten:1994ev}
  E.~Witten,
  ``Supersymmetric Yang-Mills theory on a four manifold'',
  J.\ Math.\ Phys.\  {\bf 35}, 5101 (1994);
  hep-th/9403195.


\bibitem{Bertolini:2003iv}
  M.~Bertolini,
  ``Four Lectures On The Gauge/Gravity Correspondence,''
  Int.\ J.\ Mod.\ Phys.\  A {\bf 18}, 5647 (2003)
  [arXiv:hep-th/0303160].
  F.~Bigazzi, A.~L.~Cotrone, M.~Petrini and A.~Zaffaroni,
  ``Supergravity duals of supersymmetric four dimensional gauge 
theories,''
  Riv.\ Nuovo Cim.\  {\bf 25N12}, 1 (2002)
  [arXiv:hep-th/0303191].
  A.~Paredes,
  ``Supersymmetric solutions of supergravity from wrapped branes,''
  arXiv:hep-th/0407013.


\bibitem{Andrews:2005cv}
  R.~P.~Andrews and N.~Dorey,
  ``Spherical deconstruction'',
  Phys.\ Lett.\ B {\bf 631}, 74 (2005);
  hep-th/0505107.
  R.~P.~Andrews and N.~Dorey,
  ``Deconstruction of the Maldacena-Nunez Compactification'',
  Nucl.\ Phys.\  B {\bf 751}, 304 (2006)
  [arXiv:hep-th/0601098].



\bibitem{Casero:2007jj}
  R.~Casero, C.~Nunez and A.~Paredes,
  ``Elaborations on the String Dual to N=1 SQCD,''
  Phys.\ Rev.\  D {\bf 77}, 046003 (2008)
  [arXiv:0709.3421 [hep-th]].


\bibitem{Chamseddine:1997nm}
  A.~H.~Chamseddine and M.~S.~Volkov,
 ``Non-Abelian BPS monopoles in N = 4 gauged supergravity'',
  Phys.\ Rev.\ Lett.\  {\bf 79}, 3343 (1997);
  hep-th/9707176.




\bibitem{Nunez:2008wi}
  C.~Nunez, I.~Papadimitriou and M.~Piai,
  ``Walking Dynamics from String Duals,''
  arXiv:0812.3655 [hep-th].



\bibitem{Maldacena:2009mw}
  J.~Maldacena and D.~Martelli,
  ``The unwarped, resolved, deformed conifold: fivebranes and the baryonic
  branch of the Klebanov-Strassler theory,''
  arXiv:0906.0591 [hep-th].



\bibitem{gmnp}
J\'er\^ome Gaillard, Dario Martelli, Carlos Nunez and Ioannis 
Papadimitriou. 
In preparation.



\bibitem{Nunez:2009da}
  C.~Nunez, M.~Piai and A.~Rago,
  ``Wilson Loops in string duals of Walking and Flavored Systems,''
  arXiv:0909.0748 [hep-th].







\bibitem{Elander:2009pk}
  D.~Elander, C.~Nunez and M.~Piai,
  ``A light scalar from walking solutions in gauge-string duality,''
  arXiv:0908.2808 [hep-th].
  O.~C.~Gurdogan,
  ``Walking solutions in the string background dual to N=1 SQCD-like
  theories,''
  arXiv:0906.2429 [hep-th].


\bibitem{talaveraetal}
  A.~L.~Cotrone, J.~M.~Pons and P.~Talavera,
  ``Notes on a SQCD-like plasma dual and holographic renormalization,''
  JHEP {\bf 0711}, 034 (2007)
  [arXiv:0706.2766 [hep-th]].
  O.~Lorente-Espin and P.~Talavera,
  ``A silence black hole: Hawking radiation at the Hagedorn 
temperature,''
  JHEP {\bf 0804}, 080 (2008)
  [arXiv:0710.3833 [hep-th]].

\bibitem{Caceres:2007mu}
  E.~Caceres, R.~Flauger, M.~Ihl and T.~Wrase,
  ``New Supergravity Backgrounds Dual to N=1 SQCD-like Theories with
  $N_f=2N_c$,''
  JHEP {\bf 0803}, 020 (2008)
  [arXiv:0711.4878 [hep-th]].
  E.~Caceres, R.~Flauger and T.~Wrase,
  ``Hagedorn Systems from Backreacted Finite Temperature $N_f=2N_c$
  Backgrounds,''
  arXiv:0908.4483 [hep-th].




\bibitem{Apreda:2001qb}
  R.~Apreda, F.~Bigazzi, A.~L.~Cotrone, M.~Petrini and A.~Zaffaroni,
  ``Some comments on N = 1 gauge theories from wrapped branes,''
  Phys.\ Lett.\  B {\bf 536}, 161 (2002)
  [arXiv:hep-th/0112236].




\bibitem{DiVecchia:2002ks}
  P.~Di Vecchia, A.~Lerda and P.~Merlatti,
  ``N = 1 and N = 2 super Yang-Mills theories from wrapped branes,''
  Nucl.\ Phys.\  B {\bf 646}, 43 (2002)
  [arXiv:hep-th/0205204].
  M.~Bertolini and P.~Merlatti,
  ``A note on the dual of N = 1 super Yang-Mills theory,''
  Phys.\ Lett.\  B {\bf 556}, 80 (2003)
  [arXiv:hep-th/0211142].



\bibitem{Strassler:2005qs}
  M.~J.~Strassler,
  ``The duality cascade,''
  arXiv:hep-th/0505153.

  


\bibitem{Ashok:2007sf}
  S.~K.~Ashok, S.~Murthy and J.~Troost,
  ``D-branes in unoriented non-critical strings and duality in SO(N) and Sp(N)
  gauge theories,''
  JHEP {\bf 0706}, 047 (2007)
  [arXiv:hep-th/0703148].


\bibitem{Armoni:2008gg}
  A.~Armoni, D.~Israel, G.~Moraitis and V.~Niarchos,
  ``Non-Supersymmetric Seiberg Duality, Orientifold QCD and Non-Critical
  Strings,''
  Phys.\ Rev.\  D {\bf 77}, 105009 (2008)
  [arXiv:0801.0762 [hep-th]].


\bibitem{Maldacena:1998im}
  J.~M.~Maldacena,
  ``Wilson loops in large N field theories,''
  Phys.\ Rev.\ Lett.\  {\bf 80}, 4859 (1998)
  [arXiv:hep-th/9803002].
  S.~J.~Rey and J.~T.~Yee,
  ``Macroscopic strings as heavy quarks in large N gauge theory and  
anti-de
  Sitter supergravity,''
  Eur.\ Phys.\ J.\  C {\bf 22}, 379 (2001)
  [arXiv:hep-th/9803001].


\bibitem{Bigazzi:2008gd}
  F.~Bigazzi, A.~L.~Cotrone, C.~Nunez and A.~Paredes,
  ``Heavy quark potential with dynamical flavors: a first order 
transition,''
  Phys.\ Rev.\  D {\bf 78}, 114012 (2008)
  [arXiv:0806.1741 [hep-th]].



 
\bibitem{Bigazzi:2008qq}
 F.~Bigazzi, A.~L.~Cotrone, A.~Paredes and A.~V.~Ramallo,
 ``The Klebanov-Strassler model with massive dynamical flavors,''
 JHEP {\bf 0903}, 153 (2009) [arXiv:0812.3399 [hep-th]].

  



\bibitem{Ramallo:2008ew}
  A.~V.~Ramallo, J.~P.~Shock and D.~Zoakos,
  ``Holographic flavor in N=4 gauge theories in 3d from wrapped branes,''
  JHEP {\bf 0902}, 001 (2009)
  [arXiv:0812.1975 [hep-th]].
  



\bibitem{sfetsos}
A.~Brandhuber and K.~Sfetsos,
  ``Wilson loops from multicentre and rotating branes, mass gaps and phase
  structure in gauge theories,''
  Adv.\ Theor.\ Math.\ Phys.\  {\bf 3}, 851 (1999)
  [arXiv:hep-th/9906201].

\bibitem{vandertrans}
D.~Arean, A.~Paredes and A.~V.~Ramallo,
  ``Adding flavor to the gravity dual of non-commutative gauge theories,''
  JHEP {\bf 0508}, 017 (2005)
  [arXiv:hep-th/0505181].
S.~D.~Avramis, K.~Sfetsos and K.~Siampos,
  ``Stability of strings dual to flux tubes between static quarks in N=4 SYM,''
  Nucl.\ Phys.\  B {\bf 769}, 44 (2007)
  [arXiv:hep-th/0612139];
  S.~D.~Avramis, K.~Sfetsos and D.~Zoakos,
  ``Complex marginal deformations of D3-brane geometries, their Penrose
  limits and giant gravitons,''
  Nucl.\ Phys.\  B {\bf 787}, 55 (2007)
  [arXiv:0704.2067 [hep-th]].
 ``Stability of string configurations dual to quarkonium states in AdS/CFT,''
  Nucl.\ Phys.\  B {\bf 793}, 1 (2008)
  [arXiv:0706.2655 [hep-th]].
 

\bibitem{Chamseddine:2001hk}
  A.~H.~Chamseddine and M.~S.~Volkov,
  ``Non-Abelian vacua in D = 5, N = 4 gauged supergravity,''
  JHEP {\bf 0104}, 023 (2001)
  [arXiv:hep-th/0101202].


\bibitem{Maldacena:2001pb}
  J.~M.~Maldacena and H.~S.~Nastase,
  ``The supergravity dual of a theory with dynamical supersymmetry  breaking,''
  JHEP {\bf 0109}, 024 (2001)
  [arXiv:hep-th/0105049].


\bibitem{Schvellinger:2001ib}
  M.~Schvellinger and T.~A.~Tran,
  ``Supergravity duals of gauge field theories from SU(2) x U(1) gauged
  supergravity in five dimensions,''
  JHEP {\bf 0106}, 025 (2001)
  [arXiv:hep-th/0105019].



\bibitem{Canoura:2008at}
  F.~Canoura, P.~Merlatti and A.~V.~Ramallo,
  ``The supergravity dual of 3d supersymmetric gauge theories with 
  unquenched
  flavors,''
  JHEP {\bf 0805}, 011 (2008)
  [arXiv:0803.1475 [hep-th]].





  



\bibitem{Benini:2007kg}
  F.~Benini,
  ``A chiral cascade via backreacting D7-branes with flux,''
  JHEP {\bf 0810}, 051 (2008)
  [arXiv:0710.0374 [hep-th]].



\bibitem{Benvenuti:2005wi}
  S.~Benvenuti and A.~Hanany,
  ``Conformal manifolds for the conifold and other toric field theories,''
  JHEP {\bf 0508}, 024 (2005)
  [arXiv:hep-th/0502043].
 A.~Dymarsky, I.~R.~Klebanov and N.~Seiberg,
  ``On the moduli space of the cascading SU(M+p) x SU(p) gauge theory,''
  JHEP {\bf 0601}, 155 (2006)
  [arXiv:hep-th/0511254].






\bibitem{Strassler:1996ua}
  M.~J.~Strassler,
  ``Duality in supersymmetric field theory: General conceptual background and
  an application to real particle physics,''
{\it Prepared for International Workshop on Perspectives of Strong Coupling Gauge Theories (SCGT 96), Nagoya, Japan, 13-16 Nov 1996}






\bibitem{Fiol:2002ah}
  B.~Fiol,
  ``Duality cascades and duality walls,''
  JHEP {\bf 0207}, 058 (2002)
  [arXiv:hep-th/0205155].
  




\bibitem{Hanany}
  A.~Hanany and J.~Walcher,
  ``On duality walls in string theory,''
  JHEP {\bf 0306}, 055 (2003)
  [arXiv:hep-th/0301231].
 S.~Franco, A.~Hanany, Y.~H.~He and P.~Kazakopoulos,
  ``Duality walls, duality trees and fractional branes,''
  [arXiv:hep-th/0306092].
  




\bibitem{Levi:2005hh}
  T.~S.~Levi and P.~Ouyang,
  ``Mesons and Flavor on the Conifold,''
  Phys.\ Rev.\  D {\bf 76}, 105022 (2007)
  [arXiv:hep-th/0506021].





  
\bibitem{Klebanov:2000nc}
  I.~R.~Klebanov and A.~A.~Tseytlin,
  ``Gravity Duals of Supersymmetric SU(N) x SU(N+M) Gauge Theories,''
  Nucl.\ Phys.\  B {\bf 578}, 123 (2000)
  [arXiv:hep-th/0002159].


\bibitem{Gauntlett:2001ps}
  J.~P.~Gauntlett, N.~Kim, D.~Martelli and D.~Waldram,
  ``Wrapped fivebranes and N = 2 super Yang-Mills theory,''
  Phys.\ Rev.\  D {\bf 64}, 106008 (2001)
  [arXiv:hep-th/0106117].
F.~Bigazzi, A.~L.~Cotrone and 
A.~Zaffaroni,
  ``N = 2 gauge theories from wrapped five-branes,''
  Phys.\ Lett.\ B {\bf 519}, 269 (2001);
  hep-th/0106160.




\bibitem{Paredes:2006wb}
  A.~Paredes,
  ``On unquenched N = 2 holographic flavor,''
  JHEP {\bf 0612}, 032 (2006)
  [arXiv:hep-th/0610270].



\bibitem{Arean:2008az}
  D.~Arean, P.~Merlatti, C.~Nunez and A.~V.~Ramallo,
  ``String duals of 2-d (4,4) supersymmetric gauge theories,''
  JHEP {\bf 0812}, 054 (2008)
  [arXiv:0810.1053 [hep-th]].

    
  
  
     
 \bibitem{Divecchia}
P. Di Vecchia, H. Enger, E. Imeroni and E. Lozano-Tellechea
``Gauge theories from wrapped and fractional branes",
  Nucl.\ Phys.\  B {\bf 631}, 95 (2002)
  [arXiv:hep-th/0112126].

  
  

  
\bibitem{Arean:2009gc}
  D.~Arean, E.~Conde and A.~V.~Ramallo,
  ``Gravity duals of 2d supersymmetric gauge theories,''
  JHEP {\bf 0912}, 006 (2009)
  [arXiv:0909.3106 [hep-th]].
  
  

    

\bibitem{D3D4D5} D.~Arean, E.~Conde, A.~V.~Ramallo and D. Zoakos,
``Holographic duals of SQCD models in low dimensions", work in progress. 






\bibitem{Gomis:2001aa}
  J.~Gomis and J.~G.~Russo,
  ``D = 2+1 N = 2 Yang-Mills theory from wrapped branes,''
  JHEP {\bf 0110} (2001) 028
  [arXiv:hep-th/0109177].


\bibitem{Gauntlett:2001ur}
  J.~P.~Gauntlett, N.~Kim, D.~Martelli and D.~Waldram,
  ``Fivebranes wrapped on SLAG three-cycles and related geometry,''
  JHEP {\bf 0111} (2001) 018
  [arXiv:hep-th/0110034].


\bibitem{Gaillard:2009kz}
  J.~Gaillard and J.~Schmude,
  ``The lift of type IIA supergravity with D6 sources: M-theory with torsion,''
  arXiv:0908.0305 [hep-th].


\bibitem{Gutowski:1999tu}
  J.~Gutowski, G.~Papadopoulos and P.~K.~Townsend,
  ``Supersymmetry and generalized calibrations,''
  Phys.\ Rev.\  D {\bf 60}, 106006 (1999)
  [arXiv:hep-th/9905156].



\bibitem{Koerber:2005qi}
  P.~Koerber,
  ``Stable D-branes, calibrations and generalized Calabi-Yau geometry,''
  JHEP {\bf 0508}, 099 (2005)
  [arXiv:hep-th/0506154].
  
  
\bibitem{Martucci:2005ht}
  L.~Martucci and P.~Smyth,
  ``Supersymmetric D-branes and calibrations on general N = 1 backgrounds,''
  JHEP {\bf 0511}, 048 (2005)
  [arXiv:hep-th/0507099].
  
  
  
\bibitem{Koerber:2006hh}
  P.~Koerber and L.~Martucci,
  ``Deformations of calibrated D-branes in flux generalized complex
  JHEP {\bf 0612}, 062 (2006)
  [arXiv:hep-th/0610044].

  
  
  
  


\bibitem{Hori:2002cd}
  K.~Hori and A.~Kapustin,
  ``Worldsheet descriptions of wrapped NS five-branes,''
  JHEP {\bf 0211}, 038 (2002)
  [arXiv:hep-th/0203147].


\bibitem{Callan:1986bc}
  C.~G.~.~Callan, C.~Lovelace, C.~R.~Nappi and S.~A.~Yost,
  ``String Loop Corrections To Beta Functions,''
  Nucl.\ Phys.\  B {\bf 288}, 525 (1987).
 


\bibitem{Kirsch:2004km}
  I.~Kirsch,
  ``Generalizations of the AdS/CFT correspondence,''
  Fortsch.\ Phys.\  {\bf 52}, 727 (2004)
  [arXiv:hep-th/0406274].







\bibitem{Chen:2009kx}
  H.~Y.~Chen, K.~Hashimoto and S.~Matsuura,
  ``Towards a Holographic Model of Color-Flavor Locking Phase,''
  arXiv:0909.1296 [hep-th].


\bibitem{Baumann:2010sx}
  D.~Baumann, A.~Dymarsky, S.~Kachru, I.~R.~Klebanov and L.~McAllister,
  ``D3-brane Potentials from Fluxes in AdS/CFT,''
  arXiv:1001.5028 [hep-th].


  
  
  \end{thebibliography}
\end{document}